\pdfoutput=1
\documentclass[a4paper]{jpconf}

\usepackage{amsmath,amsfonts, amssymb, amsxtra}

\usepackage{graphicx}
\usepackage{rotating}
\usepackage{supertabular}
\usepackage{tikz}

\newcommand\spm{\mathrel{\text{\framebox[0.9\width]{$\pm$}}}}
\newcommand\smp{\mathrel{\text{\framebox[0.9\width]{$\mp$}}}}
\newcommand\cpm{\mathrel{\text{\textcircled{\makebox{$\pm$}}}}}
\newcommand\cmp{\mathrel{\text{\textcircled{\makebox{$\mp$}}}}}
\newcommand\sminus{\boxminus}
\usepackage{subfigure}

\begin{document}
\title{Spin-charge-family theory is offering next step in understanding  elementary particles  and 
fields  and correspondingly universe}

\author{Norma Susana Manko\v c Bor\v stnik}

\address{FMP, University of Ljubljana, Jadranska 19, 1000 Ljubljana,Slovenia}

\ead{norma.mankoc@fmf.uni-lj.si}

\begin{abstract}
More than 40 years ago the {\it standard model} made a successful new step in understanding
properties of fermion and boson fields. Now the next step is needed, which would explain what the 
{\it standard model} and the cosmological models just assume: {\bf a.} The origin 
of quantum numbers of massless one family members. {\bf b.} The origin of families. {\bf c.} 
The origin of the vector gauge fields. {\bf d.} The origin of the Higgses and Yukawa couplings. 
{\bf e.} The origin of the dark matter. {\bf f.} The origin of the matter-antimatter asymmetry.
{\bf g.} The origin of the dark energy. {\bf h.} And several other open problems. 
The {\it spin-charge-family} theory, a kind of the Kaluza-Klein theories in $(d=(2n-1) +1)$-%
space-time, with $d=(13+1)$ and the two kinds of the spin connection fields, which are 
the gauge fields of the two kinds of the Clifford algebra objects anti-commuting with one another, 
may provide this much needed next step. The talk presents: {\bf i.} A short presentation 
of this theory. {\bf ii.} The review over the achievements of this theory so far, with some 
not published yet achievements included. {\bf iii.} Predictions for future experiments. 
%
\end{abstract}

\section{Introduction}
\label{introduction}

The {\it standard model} made a great step in understanding properties of fermion and boson fields
 by: {\bf i.} Starting with massless fields. {\bf ii.} Assuming quantum numbers of one family of 
massless quarks and leptons and relating handedness with charges. {\bf iii.} Postulating the 
existence of several families. {\bf iv.} Postulating the existence of the vector gauge fields of 
the charges of quarks and leptons. {\bf v.} Postulating a simple action for fermions and vector
bosons under the requirement of the gauge invariance. {\bf vi.}  Postulating the existence of the 
scalar field, which breaks the weak and the hyper charges of the vacuum making fermions 
and heavy bosons massive. {\bf viii.} Postulating the Yukawa couplings.

Properties of fermions and bosons in the {\it standard model} are presented and commented on 
in~\ref{sm}.

Although the assumptions from {\bf i.- v.} are elegant, in particular the assumption that all the 
elementary fields are massless gaining masses through the interactions only, as well as the choice 
of simple actions for massless fermion and vector boson fields, yet these assumptions need the 
explanation, why has nature "decided" to make this particular choice of fermions and vector 
gauge fields and what steps to take in the evolution. 

The assumption ({\bf vi.}) that there is the massive scalar field, carrying the charges in the 
fundamental representations of the groups, while all the other bosons (vector bosons) carry 
charges in the adjoint representations of the groups, and the assumption ({\bf viii.}) that the 
Yukawa couplings take care of the fermion properties, without explaining the origin of these 
couplings, do not seem either elegant or simple. 

The experiments have confirmed  so far the existence of three families of fermions with properties 
required by the {\it standard model}, of the vector fields, which are the gauge fields of the 
charges $SU(3), SU(2)$ and $U(1)$, and of the Higgs, all in accordance with the {\it standard 
model} assumptions.

To be able to predict the outcome of future experiments the next step beyond the {\it standard 
model} is needed. 

Just adding several new fields by repeating ideas of the {\it standard model}, without explaining
the assumptions of this so far so successful model, has to my understanding a little chance to be
the right step.

The {\it spin-charge-family} theory~\cite{JMP2015,norma2014MatterAntimatter,NBled2012,JMP,%
NBled2013,norma92,norma93,norma94,pikanorma,portoroz03,norma95,gmdn07,gn,gn2013,gn2015} 
does explain all the assumptions of the 
{\it standard model}: {\bf i.} The charges of the left and of the right handed quarks and leptons 
of one family - the right handed neutrinos are in this theory regular members of each family - and 
of their antiquarks and antileptons.  {\bf ii.} The appearance and properties of families. {\bf iii.} The 
appearance and properties of the vector gauge fields of the family members charges. {\bf iv.} 
The appearance and properties of scalars fields, explaining the properties of the Higgses carrying 
charges in the fundamental representations of the groups and the Yukawa couplings.

The {\it spin-charge-family} theory is offering also the explanation for the existence of the 
phenomena not explained by the {\it standard model}: {\bf a.} For the dark matter~\cite{gn}. 
{\bf b.} For the (ordinary) matter-antimatter asymmetry~\cite{norma2014MatterAntimatter}.

This theory  predicts: {\bf a.} At the low energy regime two 
decoupled groups of four families; {\bf a.i.} The fourth~\cite{JMP2015,NBled2013,JMP,pikanorma,%
gmdn07} to the already observed three families of quarks and leptons will be measured at the
LHC~\cite{gn2013}; {\bf a.ii.} The lowest of the upper four families might be considered as
 constituting the dark 
matter~\cite{gn}.  {\bf b.}  New scalar fields with the weak and the hyper charges of the Higgs%
~\cite{JMP2015,JMP}, some of them will be measured at the LHC. {\bf c.}  New $SU(2)$ vector 
gauge fields, explaining the appearance of the hyper charge, and  its $U(1)$ gauge field. {\bf d.} 
New scalar fields, which are in the fundamental representations of the colour charge (triplets), 
explaining the ordinary matter-antimatter asymmetry and causing the proton decay.

Within this theory many consequences of the {\it standard model}, like the "miraculous" 
cancellation of the triangle anomalies, can straightforwardly be explained~(Subsect.~\ref{anomaly}).

\vspace{3mm}

\section{Short presentation of the {\it spin-charge-family} theory}
\label{SCFT}

\vspace{3mm}

This section follows a lot the similar one from Ref.~\cite{norma2014MatterAntimatter}.
It briefly presents the {\it spin-charge-family} theory~(\cite{JMP2015,norma2014MatterAntimatter,%
JMP,NBled2013}, and the references therein). The details of the theory will follow in Sect.~%
(\ref{fermionsvectorsscalarsSCFT}).

Let me start with the  {\it assumptions} on which the theory is built. Comments, following the 
assumptions, will explain the meaning of each of the assumptions.

\vspace{2mm}

\noindent
{\bf A i.} $\;\;$
Fermions ($\psi$) carry in $d=(13+1)$ as the {\it internal degrees of freedom only two kinds of 
spins}, no charges, determined by the {\it two kinds of the Clifford objects}~\footnote{There exist 
only two kinds of the  Clifford algebra objects, connected with the left and right multiplication~%
(\cite{JMP2015}, Sect.~IV. Eq.(28)).}, 
($\gamma^a$ and $\tilde{\gamma}^a$),  and {\it interact correspondingly with the two kinds of
the spin connection fields} - $\omega_{ab \alpha}$ and $\tilde{\omega}_{ab \alpha}$, (the 
{\it gauge fields} of $S^{ab}=\frac{i}{4}\,$ $(\gamma^a \gamma^b -\gamma^b 
\gamma^a)$,  the generators  of $SO(13,1)$
and  
$\tilde{S}^{ab}=\frac{i}{4}\,$ $(\tilde{\gamma}^a \tilde{\gamma}^b - 
\tilde{\gamma}^b \tilde{\gamma}^a)$,  the generators of $\widetilde{SO}(13,1)$), 
and the {\it vielbeins} $f^{\alpha}{}_{a}$.
\begin{eqnarray}
{\cal A}            \,  &=& \int \; d^dx \; E\;{\mathcal L}_{f} +  
\nonumber\\  
               & & \int \; d^dx \; E\; (\alpha \,R + \tilde{\alpha} \, \tilde{R})\,,\nonumber\\
%
{\mathcal L}_f &=& \frac{1}{2}\, (\bar{\psi} \, \gamma^a p_{0a} \psi) + h.c.\,, 
\nonumber\\
p_{0a }        &=& f^{\alpha}{}_a p_{0\alpha} + \frac{1}{2E}\, \{ p_{\alpha}, E f^{\alpha}{}_a\}_- \,, 
\nonumber\\  
   p_{0\alpha} &=&  p_{\alpha}  - 
                    \frac{1}{2}  S^{ab} \omega_{ab \alpha} - 
                    \frac{1}{2}  \tilde{S}^{ab}   \tilde{\omega}_{ab \alpha}\,,                   
\nonumber\\ 
R              &=&  \frac{1}{2} \, \{ f^{\alpha [ a} f^{\beta b ]} \;(\omega_{a b \alpha, \beta} 
- \omega_{c a \alpha}\,\omega^{c}{}_{b \beta}) \} + h.c. \;, 
\nonumber\\
\tilde{R}      &=&  \frac{1}{2} \, \{ f^{\alpha [ a} f^{\beta b ]} \;(\tilde{\omega}_{a b \alpha,\beta} - 
\tilde{\omega}_{c a \alpha} \,\tilde{\omega}^{c}{}_{b \beta})\} + h.c.\;. 
\label{wholeaction}
\end{eqnarray}
Here~\footnote{$f^{\alpha}{}_{a}$ are inverted vielbeins to 
$e^{a}{}_{\alpha}$ with the properties $e^a{}_{\alpha} f^{\alpha}{\!}_b = \delta^a{\!}_b,\; 
e^a{\!}_{\alpha} f^{\beta}{\!}_a = \delta^{\beta}_{\alpha} $, $ E = \det(e^a{\!}_{\alpha}) $. 
Latin indices  
$a,b,..,m,n,..,s,t,..$ denote a tangent space (a flat index),
while Greek indices $\alpha, \beta,..,\mu, \nu,.. \sigma,\tau, ..$ denote an Einstein 
index (a curved index). Letters  from the beginning of both the alphabets
indicate a general index ($a,b,c,..$   and $\alpha, \beta, \gamma,.. $ ), 
from the middle of both the alphabets   
the observed dimensions $0,1,2,3$ ($m,n,..$ and $\mu,\nu,..$), indices from 
the bottom of the alphabets
indicate the compactified dimensions ($s,t,..$ and $\sigma,\tau,..$). 
We assume the signature $\eta^{ab} =
diag\{1,-1,-1,\cdots,-1\}$.} 
$f^{\alpha [a} f^{\beta b]}= f^{\alpha a} f^{\beta b} - f^{\alpha b} f^{\beta a}$.
$R$ and $\tilde{R}$ are the two scalars ($R$ is a curvature).
%
\vspace{2mm}

\noindent
{\bf A ii.} $\;\;$ 
The manifold $M^{(13+1)}$ breaks first into $M^{(7+1)}$ $\times$ $M^{(6)}$ (which manifests 
as $SO(7,1)$ $\times SU(3)$ $\times U(1)$), affecting both internal degrees of freedom - the 
one represented by $\gamma^a$ and the one represented by $\tilde{\gamma}^a$. Since the 
left handed (with respect to $M^{(7+1)}$) spinors couple differently to scalar (with respect to 
$M^{(7+1)}$) fields than the right handed ones, the break can leave massless and mass
protected $2^{((7+1)/2-1)}$ families~\cite{NHD}. The rest of families get heavy
masses~\footnote{A toy model~\cite{NHD,DN012} was studied in $d=(5+1)$ with the same 
action as in Eq.~(\ref{wholeaction}). The break from $d=(5+1)$ to $d=(3+1) \times$ an almost
$S^{2}$ was studied. For a particular choice of vielbeins and for a class of spin connection fields 
the manifold $M^{(5+1)}$ breaks into $M^{(3+1)}$ times an almost $S^2$, while 
$2^{((3+1)/2-1)}$ families remain massless and mass protected. 
Equivalent assumption, although not yet proved how does it really work, is made in the 
$d=(13+1)$ case.  
 This study is in progress.}.

\vspace{2mm}

\noindent
{\bf A iii.} $\;\;$ 
There is additional breaking of symmetry: The manifold $M^{(7+1)}$ breaks further into 
$M^{(3+1)} \times$  $M^{(4)}$. 

\vspace{2mm}

\noindent
{\bf A iv.} $\;\;$ 
There is a scalar condensate (Table~\ref{Table con.}) of two right handed neutrinos with the 
family quantum numbers of the upper four families, bringing masses of the scale $\propto 10^{16}$
GeV or higher to all the vector and scalar gauge fields, which 
interact with the condensate~\cite{norma2014MatterAntimatter}.   

\vspace{2mm}

\noindent
{\bf A v.} $\;\;$ 
There are  the scalar fields with the space index $(7,8)$ carrying the weak ($\tau^{1i}$) and the 
hyper charges ($Y=\tau^{23} + \tau^{4}$, $\tau^{1i}$ and $\tau^{2i}$ are generators of the
 subgroups of 
$SO(4)$, $\tau^{4}$ and $\tau^{3i}$  are the generators of $U(1)_{II}$ and $SU(3)$, respectively,
which are subgroups of $SO(6)$), which with their nonzero vacuum expectation values change the
properties of the vacuum and break the weak charge and the hyper charge. Interacting with fermions
and with the weak and hyper bosons, they bring masses to heavy bosons and to twice four groups of 
families. Carrying no  electromagnetic ($Q=\tau^{13} + Y$) and colour ($\tau^{3i}$) charges and 
no $SO(3,1)$ spin, the scalar fields  leave the electromagnetic, colour and gravity fields in $d=(3+1)$ 
massless. 

\vspace{2mm}

{\it Comments ({\bf C}) on the assumptions} ({\bf A}):

\vspace{2mm}

{\bf C i.} The  simple starting action, Eq.(\ref{wholeaction}), of the {\it spin-charge-family} theory leads 
in the low energy regime - after the breaks of the starting symmetry -  to the effective action, which is
the  {\it standard model} action with the right handed neutrinos included, what offers the explanation for
all the {\it standard model} assumptions, as well as for the appearance of the families, of the Higgs and 
of the Yukawa couplings:

{\it C i.a.} $\;\,$ One Weyl (massless) representation of $SO(13,1)$ contains~\cite{NBled2013,%
NBled2012,JMP,pikanorma,JMP2015}, if analyzed with respect to the subgroups $SO(3,1)$, 
$SU(2)_{I}$, $SU(2)_{II}$, $SU(3)$ and $U(1)$ (Eqs.~(A1) - (A6) of Ref.~\cite{JMP2015}), all
the family members and anti-members assumed by the {\it standard model}, with the right handed 
neutrinos as the regular members of each family in addition: It contains the left handed weak 
($SU(2)_{I}$) charged and  $SU(2)_{II}$ chargeless colour triplet quarks and colourless leptons
and right handed weak chargeless and $SU(2)_{II}$ charged coloured quarks and colourless leptons,
as well as the right handed weak charged and $SU(2)_{II}$ chargeless anti-coloured (anti-triplet) 
antiquarks and (anti)colourless antileptons, and left handed weak chargeless and $SU(2)_{II}$ 
charged antiquarks and antileptons. (The anti-fermion states are reachable from the fermion states 
by the application of the discrete symmetry operator $\mathbb{C}_{{\cal N}}$ ${\cal P}_{{\cal N}}$, 
presented in Ref.~\cite{HNds}.)

{\it C i.b.} $\;\,$ Before the electroweak break all observable gauge fields  are massless: 
the gravity, the colour octet vector gauge fields  (of the group $SU(3)$ from $SO(6)$), 
the weak triplet vector gauge fields  (of the group $SU(2)_{I}$ from $SO(4)$),
and  the hyper singlet vector gauge field (a superposition of $U(1)$ from 
$SO(6)$ and the third component of $SU(2)_{II}$ triplet gauge fields). 
All are the superposition of the $f^{\alpha}{}_{c}$ $\omega_{ab \alpha}$ spin connection gauge fields. 

{\it C i.c.} $\;\,$ Before the electroweak break there are  two decoupled massless groups of four
families of quarks and leptons, in the fundamental representations of 
$\widetilde{SU}(2)_{R,\widetilde{SO}(3,1)}\times$ $ \widetilde{SU}(2)_{II,\widetilde{SO}(4)}$ 
and 
$\widetilde{SU}(2)_{L,\widetilde{SO}(3,1)}\times$ $ \widetilde{SU}(2)_{I,\widetilde{SO}(4)}$ 
groups, respectively - the subgroups of $\widetilde{SO}(3,1) $ and $\widetilde{SO}(4)$ 
(
Table~\ref{Table III.}). These eight families remain massless up to the electroweak break due to
the "mass protection mechanism": Right handed members have no
left handed partners with the same charges. 

{\it C i.d.} $\;\,$ There are scalar fields~\cite{norma2014MatterAntimatter,JMP2015}   
with the space index ($7,8$) and with respect to the space index with the weak charge and the hyper 
charge of the Higgs's scalar (Eq.~(\ref{checktau13Y})). 
They have additional quantum numbers, belonging either to one of the two groups of two triplets 
or to three singlets.\\
One group of the two triplets belong to the family groups 
 $\widetilde{SU}(2)_{R\,\widetilde{SO}(3,1)}$ and $ \widetilde{SU}(2)_{II\,\widetilde{SO}(4)}$  
and couple to the upper four families. The second group of the two triplets belong to the family
groups  $\widetilde{SU}(2)_{L\,\widetilde{SO}(3,1)}$ and $ \widetilde{SU}(2)_{I\,\widetilde{SO}(4)}$
 and couple to the lower four families. All these scalars are the superposition of $f^{\sigma}{}_{s} $ 
$\tilde{\omega}_{ab\sigma}$~\footnote{$\vec{\tilde{A}}^{\tilde{1}}_{s} = 
(\tilde{\omega}_{\tilde{5} \tilde{8}s}-  \tilde{\omega}_{\tilde{6} \tilde{7}s},\, 
\tilde{\omega}_{\tilde{5} \tilde{7}s}+  \tilde{\omega}_{\tilde{6} \tilde{8}s}, \,
\tilde{\omega}_{\tilde{5} \tilde{6}s}-
  \tilde{\omega}_{\tilde{7} \tilde{8}s})\,$,  $\vec{\tilde{A}}^{\tilde{N}_{\tilde{L}}}_{s} =
 (\tilde{\omega}_{\tilde{2} \tilde{3}s}+i\,
  \tilde{\omega}_{\tilde{0} \tilde{1}s}, \,    \tilde{\omega}_{\tilde{3} \tilde{1}s}+i\,
  \tilde{\omega}_{\tilde{0} \tilde{2}s},  \,   \tilde{\omega}_{\tilde{1} \tilde{2}s}+i\,
  \tilde{\omega}_{\tilde{0} \tilde{3}s})\,$, $ \vec{\tilde{A}}^{\tilde{2}}_{s} = 
(\tilde{\omega}_{\tilde{5} \tilde{8}s}+  \tilde{\omega}_{\tilde{6} \tilde{7}s}, \,
\tilde{\omega}_{\tilde{5} \tilde{7}s}-   \tilde{\omega}_{\tilde{6} \tilde{8}s},\,
 \tilde{\omega}_{\tilde{5} \tilde{6}s}+  \tilde{\omega}_{\tilde{7} \tilde{8}s})\,$  and
$\vec{\tilde{A}}^{\tilde{N}_{\tilde{R}}}_{s} = (\tilde{\omega}_{\tilde{2} \tilde{3}s}-i\,
  \tilde{\omega}_{\tilde{0} \tilde{1}s}, \,    \tilde{\omega}_{\tilde{3} \tilde{1}s}-i\,
  \tilde{\omega}_{\tilde{0} \tilde{2}s}, \,    \tilde{\omega}_{\tilde{1} \tilde{2}s}-i\,
  \tilde{\omega}_{\tilde{0} \tilde{3}s})\,$, where $(s\in (7,8))$~(Ref.~\cite{JMP2015}, Eq. (A8)).}.
Scalars, belonging to three singlets, are the gauge fields of  the charges ($Q,Q',Y'$)~%
\footnote{$Q: =  \tau^{13} + Y\,$,  $Q':= -Y \tan^2\vartheta_1 + \tau^{13}\,$,
$Q':= -Y \tan^2\vartheta_1 + \tau^{13}\,$,
$Y:= \tau^{4} + \tau^{23}\,$, $Y':= -\tau^{4}\tan^2\vartheta_2 + \tau^{23}\,$,
$Q: =  \tau^{13} + Y\,$
~(Ref.~\cite{JMP2015}, Eq. (A7).}   
and couple to the family members of both groups of families. They are
the superposition of $f^{\sigma}{}_{s}$ $\omega_{ab\sigma}$~\footnote{$
 A^{4}_{s} = 
g^{Q}\, Q A^{Q}_{s} + g^{Q'}\, Q' A^{Q'}_{s} + g^{Y'}\,Y'\, A^{Y'}_{s}\,$, 
$A^{4}_{s}  = - (\omega_{9\,10\,s} + \omega_{11\,12\,s} + \omega_{13\,14\,s})\,$,
$A^{13}_{s} =  (\omega_{56 s}- \omega_{78 s})\,$, 
$A^{23}_{s}=(\omega_{56 s}+ \omega_{78 s})\,$,
$A^{Q}_{s} = \sin \vartheta_{1} \,A^{13}_{s} + \cos \vartheta_{1} \,A^{Y}_{s}\,$,
$A^{Q'}_{s}  = \cos \vartheta_{1} \,A^{13}_{s} - \sin \vartheta_{1} \,A^{Y}_{s}\,$,
$A^{Y'}_{s} = \cos \vartheta_{2} \,A^{23}_{s} - \sin \vartheta_{2} \,A^{4}_{s}\,$,
with $(s\in (7,8))$~(Re.~\cite{JMP2015}, Eq. (A9)).}. 
Both kinds of scalar fields determine the fermion masses (Eq.~(\ref{M0})), offering the 
explanation for the Yukawa couplings and the heavy bosons masses.

{\it C i.e.}$\;\,$ The starting action contains also  additional $SU(2)_{II}$ (from $SO(4)$, 
Eq.~(\ref{so42})) vector 
gauge fields (one of the components contributes to the hyper charge gauge fields as explained 
above), as well as the scalar fields  with the space index $s\in (5,6)$ and 
$t\in (9,10,\dots,14)$. All these fields gain masses of the scale of the condensate
(Table~\ref{Table con.}) with which they interact. They all are 
expressible as superposition of $f^{\mu}{}_{m}\,\tilde{\omega}_{ab\mu}$. In the case of 
free fields (if no spinor source, carrying their quantum numbers, is present) both
 $f^{\mu}{}_{m}\, \omega_{ab\mu}$ and $f^{\mu}{}_{m}\, \tilde{\omega}_{ab\mu}$ are 
expressible with vielbeins, Eq. (C9) in Ref.~\cite{JMP2015}%
, correspondingly only one kind of the three gauge fields are the 
propagating fields.

\vspace{2mm}

\noindent
{\bf C ii., C iii.}: $\;\;$ There are many ways of breaking symmetries from $d=(13+1)$ to
$d=(3+1)$. The assumed breaks explain why the weak and the hyper charges are connected 
with the handedness of spinors, manifesting correspondingly the properties of the 
family members - the quarks and the leptons, left and right handed (Table~\ref{Table so13+1.})
 - and of the vector gauge fields.\\
Since the left handed members are weak charged while the right handed ones are weak 
chargeless~\footnote{ The left handed anti-members are weak chargeless while the right handed 
anti-members are weak charged.}, the family members remain massless and mass protected up to
 the electroweak break, 
when the nonzero vacuum expectation values of the scalar fields with the space index $(7,8)$
break the weak and the hyper charge symmetry. \\
Antiparticles are accessible from particles by the application of the operator 
$\mathbb{C}_{{\cal N}}$ $\cdot {\cal P}_{{\cal N}}$, as explained in Refs.~\cite{HNds,TDN}. 
This discrete symmetry operator does not contain $\tilde{\gamma}^a$'s degrees of freedom: 
To each family member there corresponds the anti-member, with the same family quantum 
number.

\vspace{2mm}

\noindent
{\bf C iv.}: $\;\;$ 
It is the condensate of two right handed neutrinos with the quantum numbers of the upper four 
families (Table~\ref{Table con.}), which makes massive all the scalar gauge fields (with the index 
($5,6,7,8$), as well as those with the index $(9,\dots,14)$) and the vector gauge fields, 
manifesting nonzero $\tau^{4}$, $\tau^{23}$, 
$\tilde{\tau}^{4}$, $\tilde{\tau}^{23}$, 
$\tilde{N}^{3}_{R}$~\footnote{$ \vec{\tau}^{1}:=\frac{1}{2}\, (S^{58}-  S^{67}, \,S^{57} 
+ S^{68}, \,S^{56}-  S^{78} )\,$,
$ \vec{\tau}^{2}:= \frac{1}{2} (S^{58}+  S^{67}, \,S^{57} - S^{68}, \,S^{56}+ 
 S^{78} )\,$, $\vec{N}_{\pm} (= \vec{N}_{(L,R)}): \,\frac{1}{2} 
(S^{23}\pm i S^{01},S^{31}\pm i S^{02}, S^{12}\pm i S^{03} )\,$,
$\vec{\tau}^{3}: = \frac{1}{2} \,\{  S^{9\;12} - S^{10\;11} \,,
  S^{9\;11} + S^{10\;12} ,\, S^{9\;10} - S^{11\;12}\,,
  S^{9\;14} -  S^{10\;13} ,\,  S^{9\;13} + S^{10\;14} \,,
  S^{11\;14} -  S^{12\;13}\,,
  S^{11\;13} +  S^{12\;14}\,, 
 \frac{1}{\sqrt{3}} ( S^{9\;10} + S^{11\;12} - 
 2 S^{13\;14})\}\,$,
$ \tau^{4}: = - \frac{1}{3}(S^{9\;10} + S^{11\;12} + S^{13\;14})\,$,
$\vec{\tilde{\tau}}^{1}:=\frac{1}{2} (\tilde{S}^{58}-  \tilde{S}^{67}, \,\tilde{S}^{57} + 
\tilde{S}^{68}, \,\tilde{S}^{56}-  \tilde{S}^{78} )\,$,
$\vec{\tilde{\tau}}^{2}:=\frac{1}{2} (\tilde{S}^{58}+  \tilde{S}^{67}, \,\tilde{S}^{57} - 
\tilde{S}^{68}, \,\tilde{S}^{56}+  \tilde{S}^{78} )\,$,
$\vec{\tilde{N}}_{L,R}:= \,\frac{1}{2} (\tilde{S}^{23}\pm i \tilde{S}^{01},
\tilde{S}^{31}\pm i \tilde{S}^{02}, \tilde{S}^{12}\pm i \tilde{S}^{03} )\,$,
$\tilde{\tau}^{4}: = - \frac{1}{3}(\tilde{S}^{9\;10} + \tilde{S}^{11\;12} + \tilde{S}^{13\;14})\,$~%
Ref.~(\cite{JMP2015}, Eqs.~(A1-A6)).
}. 
Only the vector gauge fields of 
$Y$, $SU(3)$ and $SU(2)_{I}$ remain massless, since they do not interact with the condensate,
their corresponding quantum numbers  are zero ($Y=0$, $\tau^{3i}=0$ and $\tau^{1i}=0$).

\vspace{2mm}

\noindent
{\bf C v.}: $\;\;$ 
At the electroweak break the scalar fields with the space index $s=(7,8)$ - originating in 
$\tilde{\omega}_{abs}$
~\footnote{$\tilde{A}^{N_{L}\spm}_{\scriptscriptstyle{\stackrel{78}{(\pm)}}}= $ 
$\tilde{A}^{N_{L}1}_{\scriptscriptstyle{\stackrel{78}{(\pm)}}}\, \smp\, i $ 
$\tilde{A}^{N_{L}2}_{\scriptscriptstyle{\stackrel{78}{(\pm)}}}$,
%
$\tilde{A}^{\tilde{N}_{L}\spm}_{\scriptscriptstyle{\stackrel{78}{(\pm)}}} = 
\{(\tilde{\omega}_{\tilde{2} \tilde{3} \scriptscriptstyle{\stackrel{78}{(\pm)}}} + i \,
   \tilde{\omega}_{\tilde{0} \tilde{1} \scriptscriptstyle{\stackrel{78}{(\pm)}}} ) \smp\,i \, 
  (\tilde{\omega}_{\tilde{3} \tilde{1} \scriptscriptstyle{\stackrel{78}{(\pm)}}}  + i\,
   \tilde{\omega}_{\tilde{0} \tilde{2} \scriptscriptstyle{\stackrel{78}{(\pm)}}})\}\,$,
$\tilde{A}^{\tilde{N}_{L}3}_{\scriptscriptstyle{\stackrel{78}{(\pm)}}} = 
  (\tilde{\omega}_{\tilde{1} \tilde{2} \scriptscriptstyle{\stackrel{78}{(\pm)}}} +i \,
   \tilde{\omega}_{\tilde{0} \tilde{3} \scriptscriptstyle{\stackrel{78}{(\pm)}}})\,$,
$A^{Q}_{\scriptscriptstyle{\stackrel{78}{(\pm)}}} =
 \omega_{56 \scriptscriptstyle{\stackrel{78}{(\pm)}}} -  
(\omega_{9\,10 \scriptscriptstyle{\stackrel{78}{(\pm)}}} + 
\omega_{11\,12 \scriptscriptstyle{\stackrel{78}{(\pm)}}} + 
\omega_{13\,14 \scriptscriptstyle{\stackrel{78}{(\pm)}}})\,$, with
$
%
N_{L}^{3}\,\tilde{A}^{\tilde{N}_{L}\spm}_{\scriptscriptstyle{\stackrel{78}{(\pm)}}} = \spm 
\tilde{A}^{\tilde{N}_{L}\spm}_{\scriptscriptstyle{\stackrel{78}{(\pm)}}}\,$,
$N_{L}^{3}\,\tilde{A}^{\tilde{N}_{L}3}_{\scriptscriptstyle{\stackrel{78}{(\pm)}}}=0\,$ and
$Q \,A^{Q}_{\scriptscriptstyle{\stackrel{78}{(\pm)}}} = 0\,$~(Ref.~\cite{JMP2015}, Eq.~(22)).}, 
 as well as some superposition of
$\omega_{s' s" s}$ with the quantum numbers ($Q,Q',Y'$) (footnotes in this paper and  
Ref.~\cite{JMP2015}, Eq.~(22)), 
conserving
the electromagnetic charge - change their mutual interaction, and gaining nonzero vacuum 
expectation values change correspondingly also their masses. They contribute to the mass matrices 
of twice four families, as well as to the masses of the heavy vector bosons (to the two 
members of the weak triplet and the superposition of the third member of the weak triplet with 
the hyper vector field~(Ref.~(\cite{JMP2015
}, Eqs.~(17-20)). 

All the rest scalar fields keep masses of the scale of the condensate and are correspondingly  
unobservable in the low energy regime.

\vspace{3mm}

 \begin{table}
 \begin{center}
 \begin{tabular}{c|c c c r c r r |c c c c c c c}
 \hline
 state & $S^{03}$& $ S^{12}$ & $\tau^{13}$& $\tau^{23}$ &$\tau^{4}$& $Y$&$Q$&$\tilde{\tau}^{13}$&
 $\tilde{\tau}^{23}$&$\tilde{\tau}^{4}$&$\tilde{Y} $& $\tilde{Q}$&$\tilde{N}_{L}^{3}$& $\tilde{N}_{R}^{3}$
 \\
 \hline
 ${\bf (|\nu_{1 R}^{VIII}>_{1}\,|\nu_{2 R}^{VIII}>_{2})}$
 & $0$& $0$& $0$& $1$ & $-1$ & $0$ & $0$&$0$&$1$&$-1$& $0$& $0$& $0$& $1$\\ 
 \hline
 $ (|\nu_{1 R}^{VIII}>_{1}|e_{2 R}^{VIII}>_{2})$
 & $0$& $0$& $0$& $0$ & $-1$ & $-1$& $-1$ & $0$ &$1$&$-1$& $0$& $0$& $0$& $1$\\ 
 $ (|e_{1 R}^{VIII}>_{1}|e_{2 R}^{VIII}>_{2})$
 & $0$& $0$& $0$& $-1$& $-1$ & $-2$& $-2$ & $0$ &$1$&$-1$& $0$& $0$& $0$& $1$\\ 
 \hline 
 \end{tabular}
 \end{center}
\caption{\label{Table con.} This table is taken from~\cite{norma2014MatterAntimatter}.
The condensate of the two right handed neutrinos $\nu_{R}$,  with the $VIII^{th}$ 
family quantum numbers, coupled to spin zero and belonging to a triplet with 
respect to the generators $\tau^{2i}$, is presented together with its two partners. 
The right handed neutrino has $Q=0=Y$. The triplet carries $\tau^4=-1$, $\tilde{\tau}^{23}=1$,
$\tilde{\tau}^{4} =-1$, $\tilde{N}_{R}^3 = 1$, $\tilde{N}_{L}^3 = 0$, $\tilde{Y}=0 $, $\tilde{Q}=0$. 
The family quantum numbers are presented in Table~\ref{Table III.}. 
}
 \end{table}

The fourth family to the observed three ones is {\it  predicted to be observed} at the LHC. Its 
properties are under consideration~\cite{gn2013,gn2015}. The {\it baryons of the stable family of 
the upper four families offer the explanation for the dark matter}~\cite{gn}. The {\it 
triplet and anti-triplet scalar fields}  contribute together with the condensate to the
 {\it matter/anti-matter asymmetry}~\cite{norma2014MatterAntimatter}.

\vspace{3mm}

\section{Quarks,  leptons and vector and scalar gauge fields in the {\it spin-charge-family} theory}
\label{fermionsvectorsscalarsSCFT}

\vspace{3mm}

I shall formally rewrite the part of the action in Eq.(\ref{wholeaction}), which determines the 
spinor degrees of freedom, in the way that we can clearly see that the action does manifest  in the low 
energy regime by the {\it standard model} required degrees of freedom of the fermions 
(Table~\ref{Table SMfermions.}), of the vector gauge fields (Table~\ref{Table SMvectors.}) and  
of the scalar gauge fields (Table~\ref{Table SMscalar.})~\cite{NBled2013,NBled2012,JMP,portoroz03,%
JMP2015,pikanorma,norma92,norma93,norma94,norma95,gmdn07,gn,gn2013}.
\begin{eqnarray}
\label{faction}
{\mathcal L}_f &=&  \bar{\psi}\gamma^{m} (p_{m}- \sum_{A,i}\; g^{Ai}\tau^{Ai} 
A^{Ai}_{m}) \psi + \nonumber\\
               & &  \{ \sum_{s=7,8}\;  \bar{\psi} \gamma^{s} p_{0s} \; \psi \} + \nonumber\\ 
& & \{ \sum_{t=5,6,9,\dots, 14}\;  \bar{\psi} \gamma^{t} p_{0t} \; \psi \}
\,, 
\end{eqnarray}
where $p_{0s} =  p_{s}  - \frac{1}{2}  S^{s' s"} \omega_{s' s" s} - 
                    \frac{1}{2}  \tilde{S}^{ab}   \tilde{\omega}_{ab s}$, 
$p_{0t}   =    p_{t}  - \frac{1}{2}  S^{t' t"} \omega_{t' t" t} - 
                    \frac{1}{2}  \tilde{S}^{ab}   \tilde{\omega}_{ab t}$,                    
with $ m \in (0,1,2,3)$, $s \in (7,8),\, (s',s") \in (5,6,7,8)$, $(a,b)$ (appearing in
 $\tilde{S}^{ab}$) run within  either $ (0,1,2,3)$ or $ (5,6,7,8)$, $t$ runs $ \in (5,\dots,14)$, 
$(t',t")$ run either $ \in  (5,6,7,8)$ or $\in (9,10,\dots,14)$. 
The spinor function $\psi$ represents all family members of all the $2^{\frac{7+1}{2}-1}=8$ 
families.

The first line of Eq.~(\ref{faction}) determines (in $d=(3+1)$) the kinematics and dynamics of
spinor (fermion) fields, coupled to the vector gauge fields. The generators $\tau^{Ai} $ of the charge 
groups are expressible  
in terms of $S^{ab}$ through the complex coefficients $c^{Ai}{ }_{ab}$ (definition of $\tau^{Ai}$ can
be found in the footnote ${}^8$, the rest in~\footnote{ 
After the electroweak break the nonconserved charges manifest $ Y:= \tau^{4} + \tau^{23}\,$, 
$Y':= -\tau^{4} \tan^2\vartheta_2 + \tau^{23}\,$,$Q':= -Y \tan^2\vartheta_1 + \tau^{13}$,
 $\theta_1$ is the electroweak angle,  breaking the weak $SU(2)_{I}$ and the hyper charge, 
$\theta_2$ is the angle of the break of $SU(2)_{II}$ from $SU(2)_{I} \times SU(2)_{II}$.
$ Q: =  \tau^{13} + Y\,$ remains the conserved charge.},
%
\begin{eqnarray}
\tau^{Ai} = \sum_{a,b} \;c^{Ai}{ }_{ab} \; S^{ab}\,,
\label{tau}
\end{eqnarray}
fulfilling the commutation relations 
\begin{eqnarray}
\{\tau^{Ai}, \tau^{Bj}\}_- = i \delta^{AB} f^{Aijk} \tau^{Ak}\,.
\label{taucom}
\end{eqnarray}
They represent the colour ($\tau^{3i}$), the weak $(\tau^{1i})$ and the hyper ($Y$) 
charges, as well as the $SU(2)_{II}$ ($\tau^{2i}$) and $U(1)_{II}$ ($\tau^{4}$) charges, 
the gauge fields of these last two groups gain masses interacting with  the 
condensate, Table~\ref{Table con.}. The condensate leaves massless, besides the colour and gravity 
gauge fields, the weak and the hyper charge vector gauge fields. The 
corresponding vector gauge fields  $A^{Ai}_{m}$ are expressible with the spin connection fields
 $\omega_{stm}$. 
\begin{eqnarray}
A^{Ai}_{m} = \sum_{s,t} \;c^{Ai}{ }_{st} \; \omega^{st}{}_{m}\,, 
\label{AAiomega}
\end{eqnarray}
%
with $(s,t)$  either in $ (5,6,7,8)$ or in
$ (9,\dots,14)$, in agreement with the assumptions {\bf A ii.} and {\bf A iii.}. I
demonstrate~\cite{JMP2015,DNBled2015} in Subsect.~\ref{vectors} 
the equivalence between  the usual Kaluza-Klein procedure leading to the vector gauge fields 
through the vielbeins and the procedure with the spin connections used by the {\it spin-charge-family}
theory. 

All the vector gauge fields, appearing in the first line of Eq.~(\ref{faction}), except 
$A^{2 \pm}_{m}$ and $A^{Y'}_{m}$ ($=\cos \vartheta_{2} \,A^{23}_{m} - 
\sin \vartheta_{2}\, A^{4}_{m}$, $Y'$  and $\tau^{4}$ are defined 
in the footnote~\footnote{$Y':= -\tau^{4} \tan^2\vartheta_2 + \tau^{23}$, $\tau^{4} = 
-\frac{1}{3}(S^{9\,10}+ S^{11\,12} + S^{13\,14})$.}),  
are massless before the electroweak break. 
$\vec{A}^{3}_{m}$ carries the colour charge $SU(3)$ (originating in $SO(6)$),
$\vec{A}^{1}_{m}$ carries the weak charge $SU(2)_{I}$ ($SU(2)_{I}$ and  $SU(2)_{II}$ 
are the subgroups of $SO(4)$) and $A^{Y}_{m}$ ($= \sin \vartheta_{2}\, A^{23}_{m} +
\cos \vartheta_{2} \,A^{4}_{m}\,$) carries 
the corresponding $U(1)$ charge ($Y=\tau^{23} + \tau^{4}$, $\tau^{4}$ originates in $SO(6)$  
and $\tau^{23}$ is the third component of the second $SU(2)_{II}$ group, $A^{4}_{m}$ and 
$\vec{A}^{2}_{m}$ are the corresponding vector gauge fields). 
The fields $A^{2 \pm}_{m}$ and $A^{Y'}_{m}$ get masses of the 
order of the condensate scale through the interaction with the condensate of the two right
handed neutrinos with the quantum numbers of one of the group of four families 
(the assumption {\bf A iv.}, Table~\ref{Table con.}). (See Ref.~\cite{JMP2015}.) 

\vspace{2mm}

\subsection{Quarks and leptons in the {\it spin-charge-family} theory}
\label{quarksleptons}

\vspace{2mm}

To offer the explanation for the origin of quantum numbers of one (anyone) family of massless 
quarks and leptons, assumed by  the {\it standard model} (Table~\ref{Table SMfermions.}), the
 {\it spin-charge-family} theory must answer the question, where do the {\it standard model} 
charges originate and why are the weak and the hyper charge connecting with the spin of quarks 
and leptons. The theory must answer also the question: Where do families of quarks and leptons 
originate?

This section demonstrates that spinors, which carry  in $d=(13+1)$ nothing but spins of two kinds, 
determined by the two kinds of the Clifford algebra objects
~(Sect.~IV. and App.~B~in~\cite{JMP2015}), explain the origin of spins and charges 
and of families: One kind of spins manifests in $d=(3+1)$ at low energies the spin and all the 
charges, connecting the spin (the handedness) and the charges~\cite{norma92,norma93,norma94,%
norma95,hn02,hn03,DKhn}. The second kind explains the origin of families.

 To explain the appearance of the electroweak break, which causes that all the families become 
massive, the properties of the scalar fields must be explained. This is done
in Subsects.~\ref{scalars}, where also the appearance of the Yukawa couplings is explained,  while 
masses of the two groups of four families, predicted by the
{\it spin-charge-family} theory, will be discussed in Subsect.~\ref{massmatrices}.

In Table~\ref{Table so13+1.} one Weyl representation of spinors in $d=(13+1)$ is presented. The 
{\it technique}~\cite{norma92,norma93,hn02,hn03}, \ref{technique}, is used, which makes
that the states themselves demonstrate properties of spinors. Besides the states also the quantum 
numbers of the members are presented with respect to the groups $SO(3,1)$, $SU(2)_{I}$, 
$SU(2)_{II}$, $U(1)_{II}$ and $SU(3)$, which are the subgroups of the group $SO(13,1)$.
One easily sees that the states of one Weyl representation include all the quarks and the leptons, 
and the antiquarks and the antileptons of one family of quarks and leptons,  with just the quantum as 
assumed by the {\it standard model}, 
Table~\ref{Table SMfermions.}. 

Table~\ref{Table so13+1.} demonstrates that left handed quarks and leptons carry the weak  
charge ($SU(2)_{I}$ with $\tau^{1i}$ as generators) and the
hyper charge ($Y=\tau^{23} + \tau^{4}$,  $\tau^{2i}$ are generators of $SU(2)_{II}$, which
is a subgroup of $SO(4)$, $\tau^{4}$  are generators of $U(1)_{II}$, which is a subgroup of 
$SO(6)$) just as required by the {\it standard model}, Table~\ref{Table SMfermions.}, while the
right handed quarks and leptons are weak chargeless, carrying the $SU(2)_{II}$ charge with 
$\tau^{2i}$ as generators, determining the hyper charges of quarks and leptons, again in
agreement with the {\it standard model}.

Quarks carry the colour charge in the fundamental representation and the "fermion charge" 
$\tau^{4} =\frac{1}{6}$, leptons are colourless carrying the  "fermion charge" $\tau^{4}
 =-\frac{1}{2}$. Correspondingly is the hyper charge of either left handed or right handed quarks 
and leptons in agreement with the {\it standard model} assumptions. 

Table~\ref{Table so13+1.} demonstrates that left handed antiquarks and antileptons are weak  
chargeless and right handed antiquarks and antileptons are weak charged. Antiquarks carry 
antitriplet charges, while leptons are anticolourless. Correspondingly fermions and anti-fermions 
carry opposite colour, hyper, electromagnetic and "fermion" charges than fermions.

Handedness is in the one Weyl representation of $SO(13,1)$ strongly related to the weak charge 
and the hyper charge. 

Left and right handed neutrinos (carrying nonzero $Y'= -\tau^{4}\tan^2\vartheta_2 +\tau^{23}$ 
quantum number) are the regular members of each family, and so are the antineutrinos. 
\begin{table}
\tiny{
\caption{The left handed ($\Gamma^{(13,1)} = -1$)  ($ = \Gamma^{(7,1)} \times \Gamma^{(6)}$) 
multiplet of spinors - the members of the $SO(13,1)$ group representation, 
manifesting the subgroup $SO(7,1)$ - of the colour charged quarks and antiquarks and the colourless 
leptons and antileptons, is presented in the massless basis using the technique presented
in~\ref{technique}. 
It contains the left handed  ($\Gamma^{(3,1)}=-1$) weak charged  ($\tau^{13}=\pm \frac{1}{2}$) 
and $SU(2)_{II}$ chargeless ($\tau^{23}=0$) 
quarks and the right handed weak chargeless and $SU(2)_{II}$ charged ($\tau^{23}=\pm \frac{1}{2}$) 
quarks of three colours  ($c^i$ $= (\tau^{33}, \tau^{38})$) 
with the "spinor" charge ($\tau^{4}=\frac{1}{6}$) 
and the colourless left handed weak charged and  right handed weak chargeless leptons
with the "spinor" charge ($\tau^{4}=-\frac{1}{2}$).  
$ S^{12}$ defines the ordinary spin 
$\pm \frac{1}{2}$. Table contains also the corresponding anti-states with opposite charges, reachable 
from the particle states by the application of the discrete symmetry operator 
$\mathbb{C}_{{\cal N}}$ ${\cal P}_{{\cal N}}$, presented in Refs.~\cite{HNds,TDN}.
%
The vacuum state, 
on which the nilpotents and projectors operate, is not shown. 
The reader can find this  Weyl representation also in Refs.~\cite{norma2014MatterAntimatter,%
portoroz03,JMP2015}. Table is separated into three parts.
}
{\begin{tabular}{r c |c | c c|c c c|c c c |r r}
\hline
i&$$&$|^a\psi_i>$&$\Gamma^{(3,1)}$&$ S^{12}$&$\Gamma^{(4)}$&
$\tau^{13}$&$\tau^{23}$&$\tau^{33}$&$\tau^{38}$&$\tau^{4}$&$Y$&$Q$\\
\hline
&& ${\rm (Anti)octet},\,\Gamma^{(1,7)} = (-1)\,1\,, \,\Gamma^{(6)} = (1)\,-1$&&&&&&& \\
&& ${\rm of \;(anti) quarks \;and \;(anti)leptons}$&&&&&&&\\
\hline\hline 
1&$ u_{R}^{c1}$&$ \stackrel{03}{(+i)}\,\stackrel{12}{(+)}|
\stackrel{56}{(+)}\,\stackrel{78}{(+)}
||\stackrel{9 \;10}{(+)}\;\;\stackrel{11\;12}{(-)}\;\;\stackrel{13\;14}{(-)} $ &1&$\frac{1}{2}$&1&0&
$\frac{1}{2}$&$\frac{1}{2}$&$\frac{1}{2\,\sqrt{3}}$&$\frac{1}{6}$&$\frac{2}{3}$&$\frac{2}{3}$\\
\hline 
2&$u_{R}^{c1}$&$\stackrel{03}{[-i]}\,\stackrel{12}{[-]}|\stackrel{56}{(+)}\,\stackrel{78}{(+)}
||\stackrel{9 \;10}{(+)}\;\;\stackrel{11\;12}{(-)}\;\;\stackrel{13\;14}{(-)}$&1&$-\frac{1}{2}$&1&0&
$\frac{1}{2}$&$\frac{1}{2}$&$\frac{1}{2\,\sqrt{3}}$&$\frac{1}{6}$&$\frac{2}{3}$&$\frac{2}{3}$\\
\hline
3&$d_{R}^{c1}$&$\stackrel{03}{(+i)}\,\stackrel{12}{(+)}|\stackrel{56}{[-]}\,\stackrel{78}{[-]}
||\stackrel{9 \;10}{(+)}\;\;\stackrel{11\;12}{(-)}\;\;\stackrel{13\;14}{(-)}$&1&$\frac{1}{2}$&1&0&
$-\frac{1}{2}$&$\frac{1}{2}$&$\frac{1}{2\,\sqrt{3}}$&$\frac{1}{6}$&$-\frac{1}{3}$&$-\frac{1}{3}$\\
\hline 
4&$ d_{R}^{c1} $&$\stackrel{03}{[-i]}\,\stackrel{12}{[-]}|
\stackrel{56}{[-]}\,\stackrel{78}{[-]}
||\stackrel{9 \;10}{(+)}\;\;\stackrel{11\;12}{(-)}\;\;\stackrel{13\;14}{(-)} $&1&$-\frac{1}{2}$&1&0&
$-\frac{1}{2}$&$\frac{1}{2}$&$\frac{1}{2\,\sqrt{3}}$&$\frac{1}{6}$&$-\frac{1}{3}$&$-\frac{1}{3}$\\
\hline
5&$d_{L}^{c1}$&$\stackrel{03}{[-i]}\,\stackrel{12}{(+)}|\stackrel{56}{[-]}\,\stackrel{78}{(+)}
||\stackrel{9 \;10}{(+)}\;\;\stackrel{11\;12}{(-)}\;\;\stackrel{13\;14}{(-)}$&-1&$\frac{1}{2}$&-1&
$-\frac{1}{2}$&0&$\frac{1}{2}$&$\frac{1}{2\,\sqrt{3}}$&$\frac{1}{6}$&$\frac{1}{6}$&$-\frac{1}{3}$\\
\hline
6&$d_{L}^{c1} $&$\stackrel{03}{(+i)}\,\stackrel{12}{[-]}|\stackrel{56}{[-]}\,\stackrel{78}{(+)}
||\stackrel{9 \;10}{(+)}\;\;\stackrel{11\;12}{(-)}\;\;\stackrel{13\;14}{(-)} $&-1&$-\frac{1}{2}$&-1&
$-\frac{1}{2}$&0&$\frac{1}{2}$&$\frac{1}{2\,\sqrt{3}}$&$\frac{1}{6}$&$\frac{1}{6}$&$-\frac{1}{3}$\\
\hline
7&$ u_{L}^{c1}$&$\stackrel{03}{[-i]}\,\stackrel{12}{(+)}|\stackrel{56}{(+)}\,\stackrel{78}{[-]}
||\stackrel{9 \;10}{(+)}\;\;\stackrel{11\;12}{(-)}\;\;\stackrel{13\;14}{(-)}$ &-1&$\frac{1}{2}$&-1&
$\frac{1}{2}$&0 &$\frac{1}{2}$&$\frac{1}{2\,\sqrt{3}}$&$\frac{1}{6}$&$\frac{1}{6}$&$\frac{2}{3}$\\
\hline
8&$u_{L}^{c1}$&$\stackrel{03}{(+i)}\,\stackrel{12}{[-]}|\stackrel{56}{(+)}\,\stackrel{78}{[-]}
||\stackrel{9 \;10}{(+)}\;\;\stackrel{11\;12}{(-)}\;\;\stackrel{13\;14}{(-)}$&-1&$-\frac{1}{2}$&-1&
$\frac{1}{2}$&0&$\frac{1}{2}$&$\frac{1}{2\,\sqrt{3}}$&$\frac{1}{6}$&$\frac{1}{6}$&$\frac{2}{3}$\\
\hline\hline
9&$ u_{R}^{c2}$&$ \stackrel{03}{(+i)}\,\stackrel{12}{(+)}|
\stackrel{56}{(+)}\,\stackrel{78}{(+)}
||\stackrel{9 \;10}{[-]}\;\;\stackrel{11\;12}{[+]}\;\;\stackrel{13\;14}{(-)} $ &1&$\frac{1}{2}$&1&0&
$\frac{1}{2}$&$-\frac{1}{2}$&$\frac{1}{2\,\sqrt{3}}$&$\frac{1}{6}$&$\frac{2}{3}$&$\frac{2}{3}$\\
\hline 
10&$u_{R}^{c2}$&$\stackrel{03}{[-i]}\,\stackrel{12}{[-]}|\stackrel{56}{(+)}\,\stackrel{78}{(+)}
||\stackrel{9 \;10}{[-]}\;\;\stackrel{11\;12}{[+]}\;\;\stackrel{13\;14}{(-)}$&1&$-\frac{1}{2}$&1&0&
$\frac{1}{2}$&$-\frac{1}{2}$&$\frac{1}{2\,\sqrt{3}}$&$\frac{1}{6}$&$\frac{2}{3}$&$\frac{2}{3}$\\
\hline
$\cdots$&&&&&&&&&&&&\\
\hline 
\hline
\end{tabular}}
\label{Table so13+1.}
}
\end{table}

\begin{table}
\tiny{
\caption{Continuation of Table\ref{Table so13+1.}}
{\begin{tabular}{r c |c | c c|c c c|c c c |r r}
\hline
i&$$&$|^a\psi_i>$&$\Gamma^{(3,1)}$&$ S^{12}$&$\Gamma^{(4)}$&
$\tau^{13}$&$\tau^{23}$&$\tau^{33}$&$\tau^{38}$&$\tau^{4}$&$Y$&$Q$\\
\hline
&& ${\rm (Anti)octet},\,\Gamma^{(1,7)} = (-1)\,1\,, \,\Gamma^{(6)} = (1)\,-1$&&&&&&& \\
&& ${\rm of \;(anti) quarks \;and \;(anti)leptons}$&&&&&&&\\
\hline\hline 
\hline\hline
17&$ u_{R}^{c3}$&$ \stackrel{03}{(+i)}\,\stackrel{12}{(+)}|
\stackrel{56}{(+)}\,\stackrel{78}{(+)}
||\stackrel{9 \;10}{[-]}\;\;\stackrel{11\;12}{(-)}\;\;\stackrel{13\;14}{[+]} $ &1&$\frac{1}{2}$&1&0&
$\frac{1}{2}$&$0$&$-\frac{1}{\sqrt{3}}$&$\frac{1}{6}$&$\frac{2}{3}$&$\frac{2}{3}$\\
\hline 
18&$u_{R}^{c3}$&$\stackrel{03}{[-i]}\,\stackrel{12}{[-]}|\stackrel{56}{(+)}\,\stackrel{78}{(+)}
||\stackrel{9 \;10}{[-]}\;\;\stackrel{11\;12}{(-)}\;\;\stackrel{13\;14}{[+]}$&1&$-\frac{1}{2}$&1&0&
$\frac{1}{2}$&$0$&$-\frac{1}{\sqrt{3}}$&$\frac{1}{6}$&$\frac{2}{3}$&$\frac{2}{3}$\\
\hline
$\cdots$&&&&&&&&&&&&\\
\hline\hline
25&$ \nu_{R}$&$ \stackrel{03}{(+i)}\,\stackrel{12}{(+)}|
\stackrel{56}{(+)}\,\stackrel{78}{(+)}
||\stackrel{9 \;10}{(+)}\;\;\stackrel{11\;12}{[+]}\;\;\stackrel{13\;14}{[+]} $ &1&$\frac{1}{2}$&1&0&
$\frac{1}{2}$&$0$&$0$&$-\frac{1}{2}$&$0$&$0$\\
\hline 
26&$\nu_{R}$&$\stackrel{03}{[-i]}\,\stackrel{12}{[-]}|\stackrel{56}{(+)}\,\stackrel{78}{(+)}
||\stackrel{9 \;10}{(+)}\;\;\stackrel{11\;12}{[+]}\;\;\stackrel{13\;14}{[+]}$&1&$-\frac{1}{2}$&1&0&
$\frac{1}{2}$ &$0$&$0$&$-\frac{1}{2}$&$0$&$0$\\
\hline
27&$e_{R}$&$\stackrel{03}{(+i)}\,\stackrel{12}{(+)}|\stackrel{56}{[-]}\,\stackrel{78}{[-]}
||\stackrel{9 \;10}{(+)}\;\;\stackrel{11\;12}{[+]}\;\;\stackrel{13\;14}{[+]}$&1&$\frac{1}{2}$&1&0&
$-\frac{1}{2}$&$0$&$0$&$-\frac{1}{2}$&$-1$&$-1$\\
\hline 
28&$ e_{R} $&$\stackrel{03}{[-i]}\,\stackrel{12}{[-]}|
\stackrel{56}{[-]}\,\stackrel{78}{[-]}
||\stackrel{9 \;10}{(+)}\;\;\stackrel{11\;12}{[+]}\;\;\stackrel{13\;14}{[+]} $&1&$-\frac{1}{2}$&1&0&
$-\frac{1}{2}$&$0$&$0$&$-\frac{1}{2}$&$-1$&$-1$\\
\hline
29&$e_{L}$&$\stackrel{03}{[-i]}\,\stackrel{12}{(+)}|\stackrel{56}{[-]}\,\stackrel{78}{(+)}
||\stackrel{9 \;10}{(+)}\;\;\stackrel{11\;12}{[+]}\;\;\stackrel{13\;14}{[+]}$&-1&$\frac{1}{2}$&-1&
$-\frac{1}{2}$&0&$0$&$0$&$-\frac{1}{2}$&$-\frac{1}{2}$&$-1$\\
\hline
30&$e_{L} $&$\stackrel{03}{(+i)}\,\stackrel{12}{[-]}|\stackrel{56}{[-]}\,\stackrel{78}{(+)}
||\stackrel{9 \;10}{(+)}\;\;\stackrel{11\;12}{[+]}\;\;\stackrel{13\;14}{[+]} $&-1&$-\frac{1}{2}$&-1&
$-\frac{1}{2}$&0&$0$&$0$&$-\frac{1}{2}$&$-\frac{1}{2}$&$-1$\\
\hline
31&$ \nu_{L}$&$\stackrel{03}{[-i]}\,\stackrel{12}{(+)}|\stackrel{56}{(+)}\,\stackrel{78}{[-]}
||\stackrel{9 \;10}{(+)}\;\;\stackrel{11\;12}{[+]}\;\;\stackrel{13\;14}{[+]}$ &-1&$\frac{1}{2}$&-1&
$\frac{1}{2}$&0 &$0$&$0$&$-\frac{1}{2}$&$-\frac{1}{2}$&$0$\\
\hline
32&$\nu_{L}$&$\stackrel{03}{(+i)}\,\stackrel{12}{[-]}|\stackrel{56}{(+)}\,\stackrel{78}{[-]}
||\stackrel{9 \;10}{(+)}\;\;\stackrel{11\;12}{[+]}\;\;\stackrel{13\;14}{[+]}$&-1&$-\frac{1}{2}$&-1&
$\frac{1}{2}$&0&$0$&$0$&$-\frac{1}{2}$&$-\frac{1}{2}$&$0$\\
\hline\hline
\end{tabular}}
\label{Table so13+1.a}
}
\end{table}

\begin{table}
\tiny{
\caption{Continuation of Table\ref{Table so13+1.}}
{\begin{tabular}{r c |c | c c|c c c|c c c |r r}
\hline
i&$$&$|^a\psi_i>$&$\Gamma^{(3,1)}$&$ S^{12}$&$\Gamma^{(4)}$&
$\tau^{13}$&$\tau^{23}$&$\tau^{33}$&$\tau^{38}$&$\tau^{4}$&$Y$&$Q$\\
\hline
&& ${\rm (Anti)octet},\,\Gamma^{(1,7)} = (-1)\,1\,, \,\Gamma^{(6)} = (1)\,-1$&&&&&&& \\
&& ${\rm of \;(anti) quarks \;and \;(anti)leptons}$&&&&&&&\\
\hline\hline 
33&$ \bar{d}_{L}^{\bar{c1}}$&$ \stackrel{03}{[-i]}\,\stackrel{12}{(+)}|
\stackrel{56}{(+)}\,\stackrel{78}{(+)}
||\stackrel{9 \;10}{[-]}\;\;\stackrel{11\;12}{[+]}\;\;\stackrel{13\;14}{[+]} $ &-1&$\frac{1}{2}$&1&0&
$\frac{1}{2}$&$-\frac{1}{2}$&$-\frac{1}{2\,\sqrt{3}}$&$-\frac{1}{6}$&$\frac{1}{3}$&$\frac{1}{3}$\\
\hline 
34&$\bar{d}_{L}^{\bar{c1}}$&$\stackrel{03}{(+i)}\,\stackrel{12}{[-]}|\stackrel{56}{(+)}\,\stackrel{78}{(+)}
||\stackrel{9 \;10}{[-]}\;\;\stackrel{11\;12}{[+]}\;\;\stackrel{13\;14}{[+]}$&-1&$-\frac{1}{2}$&1&0&
$\frac{1}{2}$&$-\frac{1}{2}$&$-\frac{1}{2\,\sqrt{3}}$&$-\frac{1}{6}$&$\frac{1}{3}$&$\frac{1}{3}$\\
\hline
35&$\bar{u}_{L}^{\bar{c1}}$&$\stackrel{03}{[-i]}\,\stackrel{12}{(+)}|\stackrel{56}{[-]}\,\stackrel{78}{[-]}
||\stackrel{9 \;10}{[-]}\;\;\stackrel{11\;12}{[+]}\;\;\stackrel{13\;14}{[+]}$&-1&$\frac{1}{2}$&1&0&
$-\frac{1}{2}$&$-\frac{1}{2}$&$-\frac{1}{2\,\sqrt{3}}$&$-\frac{1}{6}$&$-\frac{2}{3}$&$-\frac{2}{3}$\\
\hline
36&$ \bar{u}_{L}^{\bar{c1}} $&$\stackrel{03}{(+i)}\,\stackrel{12}{[-]}|
\stackrel{56}{[-]}\,\stackrel{78}{[-]}
||\stackrel{9 \;10}{[-]}\;\;\stackrel{11\;12}{[+]}\;\;\stackrel{13\;14}{[+]} $&-1&$-\frac{1}{2}$&1&0&
$-\frac{1}{2}$&$-\frac{1}{2}$&$-\frac{1}{2\,\sqrt{3}}$&$-\frac{1}{6}$&$-\frac{2}{3}$&$-\frac{2}{3}$\\
\hline
37&$\bar{d}_{R}^{\bar{c1}}$&$\stackrel{03}{(+i)}\,\stackrel{12}{(+)}|\stackrel{56}{(+)}\,\stackrel{78}{[-]}
||\stackrel{9 \;10}{[-]}\;\;\stackrel{11\;12}{[+]}\;\;\stackrel{13\;14}{[+]}$&1&$\frac{1}{2}$&-1&
$\frac{1}{2}$&0&$-\frac{1}{2}$&$-\frac{1}{2\,\sqrt{3}}$&$-\frac{1}{6}$&$-\frac{1}{6}$&$\frac{1}{3}$\\
\hline
38&$\bar{d}_{R}^{\bar{c1}} $&$\stackrel{03}{[-i]}\,\stackrel{12}{[-]}|\stackrel{56}{(+)}\,\stackrel{78}{[-]}
||\stackrel{9 \;10}{[-]}\;\;\stackrel{11\;12}{[+]}\;\;\stackrel{13\;14}{[+]} $&1&$-\frac{1}{2}$&-1&
$\frac{1}{2}$&0&$-\frac{1}{2}$&$-\frac{1}{2\,\sqrt{3}}$&$-\frac{1}{6}$&$-\frac{1}{6}$&$\frac{1}{3}$\\
\hline
39&$ \bar{u}_{R}^{\bar{c1}}$&$\stackrel{03}{(+i)}\,\stackrel{12}{(+)}|\stackrel{56}{[-]}\,\stackrel{78}{(+)}
||\stackrel{9 \;10}{[-]}\;\;\stackrel{11\;12}{[+]}\;\;\stackrel{13\;14}{[+]}$ &1&$\frac{1}{2}$&-1&
$-\frac{1}{2}$&0 &$-\frac{1}{2}$&$-\frac{1}{2\,\sqrt{3}}$&$-\frac{1}{6}$&$-\frac{1}{6}$&$-\frac{2}{3}$\\
\hline
40&$\bar{u}_{R}^{\bar{c1}}$&$\stackrel{03}{[-i]}\,\stackrel{12}{[-]}|\stackrel{56}{[-]}\,\stackrel{78}{(+)}
||\stackrel{9 \;10}{[-]}\;\;\stackrel{11\;12}{[+]}\;\;\stackrel{13\;14}{[+]}$&1&$-\frac{1}{2}$&-1&
$-\frac{1}{2}$&0&$-\frac{1}{2}$&$-\frac{1}{2\,\sqrt{3}}$&$-\frac{1}{6}$&$-\frac{1}{6}$&$-\frac{2}{3}$\\
\hline\hline
41&$ \bar{d}_{L}^{\bar{c2}}$&$ \stackrel{03}{[-i]}\,\stackrel{12}{(+)}|
\stackrel{56}{(+)}\,\stackrel{78}{(+)}
||\stackrel{9 \;10}{(+)}\;\;\stackrel{11\;12}{(-)}\;\;\stackrel{13\;14}{[+]} $ &-1&$\frac{1}{2}$&1&0&
$\frac{1}{2}$&$\frac{1}{2}$&$-\frac{1}{2\,\sqrt{3}}$&$-\frac{1}{6}$&$\frac{1}{3}$&$\frac{1}{3}$\\
\hline 
$\cdots$ &&&&&&&&&&&& \\
\hline\hline
49&$ \bar{d}_{L}^{\bar{c3}}$&$ \stackrel{03}{[-i]}\,\stackrel{12}{(+)}|
\stackrel{56}{(+)}\,\stackrel{78}{(+)}
||\stackrel{9 \;10}{(+)}\;\;\stackrel{11\;12}{[+]}\;\;\stackrel{13\;14}{(-)} $ &-1&$\frac{1}{2}$&1&0&
$\frac{1}{2}$&$0$&$\frac{1}{\sqrt{3}}$&$-\frac{1}{6}$&$\frac{1}{3}$&$\frac{1}{3}$\\
\hline 
$\cdots$ &&&&&&&&&&&& \\
\hline\hline
57&$ \bar{e}_{L}$&$ \stackrel{03}{[-i]}\,\stackrel{12}{(+)}|
\stackrel{56}{(+)}\,\stackrel{78}{(+)}
||\stackrel{9 \;10}{[-]}\;\;\stackrel{11\;12}{(-)}\;\;\stackrel{13\;14}{(-)} $ &-1&$\frac{1}{2}$&1&0&
$\frac{1}{2}$&$0$&$0$&$\frac{1}{2}$&$1$&$1$\\
\hline 
58&$\bar{e}_{L}$&$\stackrel{03}{(+i)}\,\stackrel{12}{[-]}|\stackrel{56}{(+)}\,\stackrel{78}{(+)}
||\stackrel{9 \;10}{[-]}\;\;\stackrel{11\;12}{(-)}\;\;\stackrel{13\;14}{(-)}$&-1&$-\frac{1}{2}$&1&0&
$\frac{1}{2}$ &$0$&$0$&$\frac{1}{2}$&$1$&$1$\\
\hline
59&$\bar{\nu}_{L}$&$\stackrel{03}{[-i]}\,\stackrel{12}{(+)}|\stackrel{56}{[-]}\,\stackrel{78}{[-]}
||\stackrel{9 \;10}{[-]}\;\;\stackrel{11\;12}{(-)}\;\;\stackrel{13\;14}{(-)}$&-1&$\frac{1}{2}$&1&0&
$-\frac{1}{2}$&$0$&$0$&$\frac{1}{2}$&$0$&$0$\\
\hline 
60&$ \bar{\nu}_{L} $&$\stackrel{03}{(+i)}\,\stackrel{12}{[-]}|
\stackrel{56}{[-]}\,\stackrel{78}{[-]}
||\stackrel{9 \;10}{[-]}\;\;\stackrel{11\;12}{(-)}\;\;\stackrel{13\;14}{(-)} $&-1&$-\frac{1}{2}$&1&0&
$-\frac{1}{2}$&$0$&$0$&$\frac{1}{2}$&$0$&$0$\\
\hline
61&$\bar{\nu}_{R}$&$\stackrel{03}{(+i)}\,\stackrel{12}{(+)}|\stackrel{56}{[-]}\,\stackrel{78}{(+)}
||\stackrel{9 \;10}{[-]}\;\;\stackrel{11\;12}{(-)}\;\;\stackrel{13\;14}{(-)}$&1&$\frac{1}{2}$&-1&
$-\frac{1}{2}$&0&$0$&$0$&$\frac{1}{2}$&$\frac{1}{2}$&$0$\\
\hline
62&$\bar{\nu}_{R} $&$\stackrel{03}{[-i]}\,\stackrel{12}{[-]}|\stackrel{56}{[-]}\,\stackrel{78}{(+)}
||\stackrel{9 \;10}{[-]}\;\;\stackrel{11\;12}{(-)}\;\;\stackrel{13\;14}{(-)} $&1&$-\frac{1}{2}$&-1&
$-\frac{1}{2}$&0&$0$&$0$&$\frac{1}{2}$&$\frac{1}{2}$&$0$\\
\hline
63&$ \bar{e}_{R}$&$\stackrel{03}{(+i)}\,\stackrel{12}{(+)}|\stackrel{56}{(+)}\,\stackrel{78}{[-]}
||\stackrel{9 \;10}{[-]}\;\;\stackrel{11\;12}{(-)}\;\;\stackrel{13\;14}{(-)}$ &1&$\frac{1}{2}$&-1&
$\frac{1}{2}$&0 &$0$&$0$&$\frac{1}{2}$&$\frac{1}{2}$&$1$\\
\hline
64&$\bar{e}_{R}$&$\stackrel{03}{[-i]}\,\stackrel{12}{[-]}|\stackrel{56}{(+)}\,\stackrel{78}{[-]}
||\stackrel{9 \;10}{[-]}\;\;\stackrel{11\;12}{(-)}\;\;\stackrel{13\;14}{(-)}$&1&$-\frac{1}{2}$&-1&
$\frac{1}{2}$&0&$0$&$0$&$\frac{1}{2}$&$\frac{1}{2}$&$1$\\
\hline 
\end{tabular}}
\label{Table so13+1.b}}
\end{table}
%

Let me call attention to the reader that the term $\gamma^0\, \stackrel{78}{(-)} \, $
$ \tau^{Ai}$ $A^{Ai}_{\scriptscriptstyle{\stackrel{78}{(-)}}}$, where $ \tau^{Ai}$ represent the
superposition of either $S^{ab}$ or $\tilde{S}^{ab}$ and 
$A^{Ai}_{\scriptscriptstyle{\stackrel{78}{(-)}}}$ represent correspondingly the superposition of either 
the scalar fields $\omega_{ab s}$ or the scalar fields $\tilde{\omega}_{ab s}$, $s\in (7,8)$, as 
presented in Subsect.~\ref{scalars},  Eq.~(\ref{eigentau1tau2}), 
transforms the right handed $u_{R}^{c1}$ quark from the first line of Tables~\ref{Table so13+1.}%
--\ref{Table so13+1.b} 
into the left handed $u_{L}^{c1}$ quark from the seventh line of the same table~\footnote{This
transformation of the right handed family members into the corresponding left handed partners can 
easily be calculated by using Eqs.~(\ref{graphbinoms}, \ref{snmb:gammatildegamma}, 
\ref{plusminus}).}, which can, due to the properties of the scalar fields (Eq.~(\ref{checktau13Y})),
be interpreted also in the {\it standard model} way, namely, that
$A^{Ai}_{\scriptscriptstyle{\stackrel{78}{(-)}}}$ "dress" $u_{R}^{c1}$ giving it the weak and the
hyper charge of the left handed $u_{L}^{c1}$ quark, while $\gamma^0$ changes handedness. 
Equivalently happens to $\nu_{R}$ from the $25^{th}$ line, which transforms under the action of  
$\gamma^0\, \stackrel{78}{(-)}$ $ \tau^{Ai}$  $A^{Ai}_{\scriptscriptstyle{\stackrel{78}{(-)}}}$, 
into  $\nu_{L}$ from the $31^{th}$ line.

The operator $\gamma^0\,\stackrel{78}{(+)}$ $\tau^{Ai}$ 
$ A^{Ai}_{\scriptscriptstyle{\stackrel{78}{(+)}}}$ transforms $d_{R}^{c1}$ from the third line of 
Tables~\ref{Table so13+1.}--\ref{Table so13+1.b}  into $d_{L}^{c1}$ from the  fifth line of this
 table, or $e_{R}$ 
from the $27^{th}$ line into $e_{L}$ from the $29^{th}$ line, where 
$ A^{Ai}_{\scriptscriptstyle{\stackrel{78}{(+)}}}$  belong to the scalar fields from
Eq.~(\ref{commonAi}).

The operator $\tau^{Ai}$, if representing the first three operators in
Eq.~(\ref{commonAi}),  (only) multiplies the right handed family member with its eigenvalue.

The term $\gamma^0\,\stackrel{78}{(\mp)}$ 
$\tau^{Ai}$ $A^{Ai}_{\scriptscriptstyle{\stackrel{78}{(\mp)}}}$ of the action
(Eqs.~(\ref{wholeaction}, \ref{faction})) determines the Yukawa 
couplings~(\ref{scalars}, \ref{massmatrices}).


Since spinors (fermions) carry besides the family members quantum numbers also the family 
quantum numbers, determined by $\tilde{S}^{ab}= \frac{i}{4} (\tilde{\gamma}^{a} 
\tilde{\gamma}^{b} - \tilde{\gamma}^{b}\tilde{\gamma}^{a})$, there are correspondingly
$2^{(7+1)/2-1} =8$ families~\cite{JMP2015}, which split into two groups of families, each
 manifesting the ($\widetilde{SU}(2)_{\widetilde{SO}(3,1)}$ $\times \widetilde{SU}(2)_{\widetilde{SO}(4)}$
$\times U(1)$) symmetry.

The  eight families of the right  handed $u_{1R}$ quark (the first member of the eight-plet of quarks 
from Tables~\ref{Table so13+1.} -- \ref{Table so13+1.b}) and of the right handed  $\nu_{1R}$ 
leptons  (the first member of the eight-plet of leptons from Tables~\ref{Table so13+1.} -- 
\ref{Table so13+1.b}) are presented as an example in Table~\ref{Table III.}~\cite{JMP}. 

All the other members of any of the eight families of quarks or leptons follow  from any member 
of a particular family by the application of the  operators $N^{\pm}_{R,L}$ and  $\tau^{(2,1)\pm}$ 
on this particular member.  

%
\begin{table}
\begin{tiny}
\caption{\label{Table III.} 
Eight families of the right handed $u^{c1}_{R}$ (\ref{Table so13+1.}--\ref{Table so13+1.b}) 
quark with the spin $\frac{1}{2}$, colour charge $(\tau^{33}=1/2$, $\tau^{38}=1/(2\sqrt{3})$, 
and of  the colourless right handed neutrino $\nu_{R}$ of spin $\frac{1}{2}$ 
are presented in the  left and in the right column, respectively.
They belong to two groups of four families, one ($I$) is a doublet with respect to 
($\vec{\tilde{N}}_{L}$ and  $\vec{\tilde{\tau}}^{(1)}$) and  a singlet with respect to 
($\vec{\tilde{N}}_{R}$ and  $\vec{\tilde{\tau}}^{(2)}$), the other ($II$) is a singlet with respect to 
($\vec{\tilde{N}}_{L}$ and  $\vec{\tilde{\tau}}^{(1)}$) and  a doublet with with respect to 
($\vec{\tilde{N}}_{R}$ and  $\vec{\tilde{\tau}}^{(2)}$).
All the families of each of the two groups follow from the starting one by the application of  
one of the two operators ($\tilde{N}^{\pm}_{R,L}$, $\tilde{\tau}^{(2,1)\pm}$), 
Eq.~(\ref{plusminus}), respectively.  The generators 
($N^{\pm}_{R,L} $, $\tau^{(2,1)\pm}$) (Eq.~(\ref{plusminus}))
transform $u_{1R}$ of one family to all the members of the same family of the same colour. 
The same generators transform equivalently the right handed   neutrino $\nu_{1R}$  to all the 
colourless members of the same family. The table is taken from Ref.~\cite{norma2014MatterAntimatter}
}
 {\begin{tabular}{r|c|c|c|c|c c c c c}
 \hline
 &&&&&$\tilde{\tau}^{13}$&$\tilde{\tau}^{23}$&$\tilde{N}_{L}^{3}$&$\tilde{N}_{R}^{3}$&$\tilde{\tau}^{4}$\\
 \hline
 $I$&$u^{c1}_{R\,1}$&
   $ \stackrel{03}{(+i)}\,\stackrel{12}{[+]}|\stackrel{56}{[+]}\,\stackrel{78}{(+)} ||
   \stackrel{9 \;10}{(+)}\;\;\stackrel{11\;12}{(-)}\;\;\stackrel{13\;14}{(-)}$ & 
   $\nu_{R\,2}$&
   $ \stackrel{03}{(+i)}\,\stackrel{12}{[+]}|\stackrel{56}{[+]}\,\stackrel{78}{(+)} ||
   \stackrel{9 \;10}{(+)}\;\;\stackrel{11\;12}{[+]}\;\;\stackrel{13\;14}{[+]}$ 
  &$-\frac{1}{2}$&$0$&$-\frac{1}{2}$&$0$&$-\frac{1}{2}$ 
 \\
  $I$&$u^{c1}_{R\,2}$&
   $ \stackrel{03}{[+i]}\,\stackrel{12}{(+)}|\stackrel{56}{[+]}\,\stackrel{78}{(+)} ||
   \stackrel{9 \;10}{(+)}\;\;\stackrel{11\;12}{(-)}\;\;\stackrel{13\;14}{(-)}$ & 
   $\nu_{R\,2}$&
   $ \stackrel{03}{[+i]}\,\stackrel{12}{(+)}|\stackrel{56}{[+]}\,\stackrel{78}{(+)} ||
   \stackrel{9 \;10}{(+)}\;\;\stackrel{11\;12}{[+]}\;\;\stackrel{13\;14}{[+]}$ 
  &$-\frac{1}{2}$&$0$&$\frac{1}{2}$&$0$&$-\frac{1}{2}$
 \\
  $I$&$u^{c1}_{R\,3}$&
   $ \stackrel{03}{(+i)}\,\stackrel{12}{[+]}|\stackrel{56}{(+)}\,\stackrel{78}{[+]} ||
   \stackrel{9 \;10}{(+)}\;\;\stackrel{11\;12}{(-)}\;\;\stackrel{13\;14}{(-)}$ & 
   $\nu_{R\,3}$&
   $ \stackrel{03}{(+i)}\,\stackrel{12}{[+]}|\stackrel{56}{(+)}\,\stackrel{78}{[+]} ||
   \stackrel{9 \;10}{(+)}\;\;\stackrel{11\;12}{[+]}\;\;\stackrel{13\;14}{[+]}$ 
  &$\frac{1}{2}$&$0$&$-\frac{1}{2}$&$0$&$-\frac{1}{2}$
 \\
 $I$&$u^{c1}_{R\,4}$&
  $ \stackrel{03}{[+i]}\,\stackrel{12}{(+)}|\stackrel{56}{(+)}\,\stackrel{78}{[+]} ||
  \stackrel{9 \;10}{(+)}\;\;\stackrel{11\;12}{(-)}\;\;\stackrel{13\;14}{(-)}$ & 
  $\nu_{R\,4}$&
  $ \stackrel{03}{[+i]}\,\stackrel{12}{(+)}|\stackrel{56}{(+)}\,\stackrel{78}{[+]} ||
  \stackrel{9 \;10}{(+)}\;\;\stackrel{11\;12}{[+]}\;\;\stackrel{13\;14}{[+]}$ 
  &$\frac{1}{2}$&$0$&$\frac{1}{2}$&$0$&$-\frac{1}{2}$
  \\
  \hline
  $II$& $u^{c1}_{R\,5}$&
        $ \stackrel{03}{[+i]}\,\stackrel{12}{[+]}|\stackrel{56}{[+]}\,\stackrel{78}{[+]}||
        \stackrel{9 \;10}{(+)}\;\;\stackrel{11\;12}{(-)}\;\;\stackrel{13\;14}{(-)}$ & 
        $\nu_{R\,5}$&
        $ \stackrel{03}{[+i]}\,\stackrel{12}{[+]}|\stackrel{56}{[+]}\,\stackrel{78}{[+]}|| 
        \stackrel{9 \;10}{(+)}\;\;\stackrel{11\;12}{[+]}\;\;\stackrel{13\;14}{[+]}$ 
        &$0$&$-\frac{1}{2}$&$0$&$-\frac{1}{2}$&$-\frac{1}{2}$
 \\ 
  $II$& $u^{c1}_{R\,6}$&
      $ \stackrel{03}{(+i)}\,\stackrel{12}{(+)}|\stackrel{56}{[+]}\,\stackrel{78}{[+]}||
      \stackrel{9 \;10}{(+)}\;\;\stackrel{11\;12}{(-)}\;\;\stackrel{13\;14}{(-)}$ & 
      $\nu_{R\,6}$&
      $ \stackrel{03}{(+i)}\,\stackrel{12}{(+)}|\stackrel{56}{[+]}\,\stackrel{78}{[+]}|| 
      \stackrel{9 \;10}{(+)}\;\;\stackrel{11\;12}{[+]}\;\;\stackrel{13\;14}{[+]}$ 
      &$0$&$-\frac{1}{2}$&$0$&$\frac{1}{2}$&$-\frac{1}{2}$
 \\ 
 $II$& $u^{c1}_{R\,7}$&
 $ \stackrel{03}{[+i]}\,\stackrel{12}{[+]}|\stackrel{56}{(+)}\,\stackrel{78}{(+)}||
 \stackrel{9 \;10}{(+)}\;\;\stackrel{11\;12}{(-)}\;\;\stackrel{13\;14}{(-)}$ & 
      $\nu_{R\,7}$&
      $ \stackrel{03}{[+i]}\,\stackrel{12}{[+]}|\stackrel{56}{(+)}\,\stackrel{78}{(+)}|| 
      \stackrel{9 \;10}{(+)}\;\;\stackrel{11\;12}{[+]}\;\;\stackrel{13\;14}{[+]}$ 
    &$0$&$\frac{1}{2}$&$0$&$-\frac{1}{2}$&$-\frac{1}{2}$
  \\
   $II$& $u^{c1}_{R\,8}$&
    $ \stackrel{03}{(+i)}\,\stackrel{12}{(+)}|\stackrel{56}{(+)}\,\stackrel{78}{(+)}||
    \stackrel{9 \;10}{(+)}\;\;\stackrel{11\;12}{(-)}\;\;\stackrel{13\;14}{(-)}$ & 
    $\nu_{R\,8}$&
    $ \stackrel{03}{(+i)}\,\stackrel{12}{(+)}|\stackrel{56}{(+)}\,\stackrel{78}{(+)}|| 
    \stackrel{9 \;10}{(+)}\;\;\stackrel{11\;12}{[+]}\;\;\stackrel{13\;14}{[+]}$ 
    &$0$&$\frac{1}{2}$&$0$&$\frac{1}{2}$&$-\frac{1}{3}$
 \\ 
 \hline 
 \end{tabular}
}
\end{tiny}
%
\end{table}
%

The eight families separate into two groups of four families: One group  contains  doublets with respect 
to $\vec{\tilde{N}}_{R}$ and  $\vec{\tilde{\tau}}^{2}$, these families are singlets with respect to 
$\vec{\tilde{N}}_{L}$ and  $\vec{\tilde{\tau}}^{1}$.
Another group of four families contains  doublets 
with respect to  $\vec{\tilde{N}}_{L}$ and  $\vec{\tilde{\tau}}^{1}$, these families are singlets 
with respect to  $\vec{\tilde{N}}_{R}$ and  $\vec{\tilde{\tau}}^{2}$. 

If $\tau^{Ai}$ represents the last four operators of Eq.~(\ref{commonAi}) in Subsect.~\ref{scalars}, 
the operators 
$\gamma^0\,\stackrel{78}{(\mp)}$ $\tau^{Ai}$ 
$ A^{Ai}_{\scriptscriptstyle{\stackrel{78}{(\mp)}}}\;$ (($\mp$) for $(u_{R},\nu_{R})$ and 
$(d_{R},e_{R})$, respectively)    
transform the right handed family member of one family into the left handed partner of another 
family within the same group of four families, since these four operators manifest the symmetry
twice ($\widetilde{SU}(2)_{\widetilde{SO}(3,1)}\times \widetilde{SU}(2)_{\widetilde{SO}(4)}$). 
One group of four families carries  the family quantum numbers ($\vec{\tilde{\tau}}^{1}$, 
$\vec{\tilde{N}}_{L}$), the other group of four families carries the family quantum numbers
 ($\vec{\tilde{\tau}}^{2}$, $\vec{\tilde{N}}_{R}$).

The contribution of the scalar fields to masses of fermions and to the Yukawa couplings will be 
discussed in Subsects.~(\ref{massmatrices}, \ref{scalars}), respectively.

At each break of symmetry  fermions can gain masses~\cite{witten} of the order of the scale 
of the condensate. In the 
Refs.~\cite{NHD,DN012} we discuss possible conditions under which fermion remain massless 
at the break. This discussion concerns in our case the break of $d=(2(2n+1)-1,1)$, for $n=9$ 
or larger, down to $d=(13+1)$ and from $d=(13+1)$ to $d=(7+1)$ before the symmetry between 
spinors and antispinors is broken. After that the massless fermions are mass protected, since the left 
handed and right handed members differ in the weak and hyper charges, until the weak and the 
hyper charges are no longer conserved quantum numbers.

Let me point out here that there are scalar fields, the gauge scalars  of  $\vec{\tilde{N}}_{R}$ 
and  $\vec{\tilde{\tau}}^{2}$, which couple only to the four families  which are doublets with 
respect to these two groups, while the scalar fields, which are the gauge scalars  of  $\vec{\tilde{N}}_{L}$ and  
$\vec{\tilde{\tau}}^{1}$, couple only to the four families  which are doublets with respect to 
these last two groups. 
Each of the two kinds of scalar contribute after the electroweak transition to their own group of 
four families, while the scalar gauge fields of ($Q, Q', Y'$) couple to family members of all eight 
families. 


\vspace{2mm}

\subsection{Vector gauge fields in the {\it spin-charge-family} theory}
\label{vectors}

\vspace{2mm}

In the starting action of Eq.~(\ref{wholeaction}) of the {\it spin-charge-family} theory all the gauge 
fields are the gravitational ones: the vielbeins and the spin connections of two kinds.  Both kinds of the 
spin connection fields are uniquely determined by the vielbeins (Ref.~\cite{JMP2015}, Eqs.~((30)-(32), 
(C9))), if  there are no spinor sources present. Spinors (fermions) interact with the vielbeins
and the two kinds of the spin connection fields. After the break of the starting symmetry $SO(13,1)$
the starting action manifests at low energies  the effective action.  Eq.~(\ref{faction}) represents 
the effective action for fermions in $d=(3+1)$ interacting with the vector gauge fields, which are the 
superposition of the spin connection gauge fields with the vector index $m$ ($m=(0,1,2,3)$) -
$A^{Ai}_{m} = \sum_{s,t} $ $c^{Ai st}\, \omega_{stm}$ - and the scalar gauge fields, which are 
the superposition of the spin connection gauge fields of both kinds  
$\omega_{abs}$'s and $\tilde{\omega}_{abs}$'s   with the scalar index $s$  ($s\ge5$) - 
$A^{(Q,Q',Y')}_{s} = \sum_{s,t} $ $c^{(Q,Q",Y') st}\, \omega_{stm}$ (for ($Q,Q',Y'$), 
respectively), $\tilde{A}^{\tilde{A}i)}_{s} = \sum_{s,t} $$\tilde{c}^{\tilde{A}i st}\, 
\tilde{\omega}_{stm}$.

I comment in this section that the vector gauge fields in the {\it spin-charge-family}  theory appear 
equivalently either from vielbeins $f^{\sigma}{}_{m}$ - like it is usually proceeded in the 
Kaluza-Klein-like  theories~\cite{zelenaknjiga,mil} (with which the {\it spin-charge-family} theory
has many things in common) - or from the spin connection fields $\sum_{s,t}  c^{Ai}{}_{st}\,
\omega^{st}{}_{m}$. This is indeed known for a long 
time~\cite{mil,zelenaknjiga,norma2004}. 

This section reviews Refs.~\cite{JMP2015,DNBled2015,DNproof}, where this
equivalence is demonstrated  when the spaces of $d\ge5$ have the metric tensor $g_{\sigma \tau} 
= \eta_{\sigma \tau}\, f^{-2}$, where ($x^{\sigma}$, $x^{\tau}$) determine the coordinates of
 the (almost~\cite{NHD}) compactified space, $\eta_{\sigma \tau}$ is the diagonal matrix in
 this space and $f$ is any  scalar function of these coordinates.

Let the space with $s\ge 5$ have the symmetry allowing the infinitesimal 
transformations of the kind
\begin{eqnarray}
\label{deltaxsigmagen}
x'^{\mu} &=& x^{\mu}\,, \quad 
 x'^{\sigma} = 
 x^{\sigma} - i  \, \sum_{A,i, s,t} \varepsilon^{Ai} (x^{\mu})\, c_{Ai}{}^{st }M_{st} \,
x^{\sigma}\,,
\end{eqnarray}
%
%
where $M^{st}= S^{st} + L^{st}$, $L^{st}=x^s p^t-x^t p^s$, $S^{st}$ concern internal 
degrees of freedom of boson and fermion fields,  $\{M^{st}, 
M^{s't'}\}_{-} = i (\eta^{s t'} M^{ts'} + \eta^{ts'} M^{st'} - \eta^{s s'} M^{tt'} - 
\eta^{t t'} M^{ss'})$. 
From Eq.~(\ref{deltaxsigmagen}) it follows 
\begin{eqnarray}
\label{deltaxsigma1}
- i 
\sum_{s,t} c_{Ai}{}^{st }M_{st} \,
x^{\sigma}  &=&  E^{\sigma}_{Ai} = \sum_{s,t}  c_{Ai}{}^{st }\, (x_{s}\, f^{\sigma}{}_{t} - 
x_{t}\, f^{\sigma}{}_{s})\,,\nonumber\\
\sum_{s,t}  c_{Ai}{}^{st }\, M_{st}{}^{\sigma}: &=& i E^{\sigma}_{Ai}\,,
\end{eqnarray}
and correspondingly: $\tau_{Ai}=  E^{\sigma}_{Ai} p_{\sigma}$, where $\tau^{Ai}=
\sum_{s,t} c_{Ai}{}^{st} \,M_{st} $ with the commutation relations $\{\tau_{Ai},\tau_{Bj}\}_{-}=$
$ i \delta^{A B}\, f^{Aijk} \, \tau_{Ak}$, $f^{Aijk} $ are the structure constants of the symmetry
group $A$. 
One derives, when taking into account  Eq.~(\ref{deltaxsigma1}) and the commutation relations 
among generators of the infinitesimal transformations $\tau_{Ai}$ the equation for the Killing vectors 
$E^{\sigma}_{Ai}$~\cite{mil}
\begin{eqnarray}
\label{comE}
E^{\sigma}_{Ai} p_{\sigma}  E^{\tau}_{Bj} p_{\tau} - 
E^{\sigma}_{Bj} p_{\sigma}  E^{\tau}_{Ai} p_{\tau}
 &=&  i  \delta^{A B}\, f^{Aijk} \, E^{\tau}_{Ak} p_{\tau}\,,  
\end{eqnarray}
and the Killing equation
\begin{eqnarray}
\label{Killing}
&&D_{\sigma} E_{\tau Ai}  + D_{\tau} E_{\sigma Bj} =0\,,\nonumber\\
&&D_{\sigma} E_{\tau Ai} = \partial_{\sigma}  E_{\tau Ai}  - \Gamma^{\tau'}_{\tau \sigma}
 E_{\tau' Ai}\,. 
\end{eqnarray}
Let the corresponding background field ($g_{\alpha \beta} = e^{a}{}_{\alpha} \,e^{a}{}_{\beta}$) 
be
\begin{eqnarray}
e^{a}{}_{\alpha}=
\begin{pmatrix} \delta^{m}{}_{\mu} & e^{m}{}_{\sigma}=0 \\
 e^{s}{}_{\mu} & e^s{}_{\sigma} 
\end{pmatrix}
\,, \quad 
f^{\alpha}{}_{a} =
\begin{pmatrix} \delta^{\mu}{}_{m}  & f^{\sigma}{}_{m} \\
0= f^{\mu}{}_{s} & f^{\sigma}{}_{s}\,,  
\end{pmatrix}
\,,
\label{fe}
\end{eqnarray}
so that  the background field in $ d=(3+1)$ is flat.
 From $e^{a}{}_{\mu}f^{\sigma}{}_{a}=$
$\delta^{\sigma}_{\mu}=0 $ it follows
\begin{eqnarray}
\label{fe1}
e^{s}{}_{\mu} &=& -\delta^{m}_{\mu} e^{s}{}_{\sigma} f^{\sigma}{}_{m} \,.  
\end{eqnarray}
This leads to
\begin{equation}
g_{\alpha \beta} =
\begin{pmatrix}
 \eta_{m n} +  f^{\sigma}{}_{m}  f^{\tau}{}_{n}  
e^{s}{}_{\sigma} e_{s \tau}\;\;\; & -f^{\tau}{}_{m} e^{s}{}_{\tau} e_{s\sigma}\\
-f^{\tau}{}_{n} e^s{}_{\tau} e_{s \sigma} & e^s{}_\sigma e_{s\tau}
\end{pmatrix}\,, 
\label{gmdown}
\end{equation}
and
\begin{equation}
g^{\alpha \beta} =
\begin{pmatrix}
 \eta^{m n}\quad \quad\quad & f^{\sigma}{}_{m}\\
f^{\sigma}{}_{m} \quad \quad\quad & f^{\sigma}{}_{s}  f^{\tau s} +
 f^{\sigma}{}_{m}  f^{\tau m}
\end{pmatrix}\,.
\label{gmup}
\end{equation}
We have: $ \Gamma^{\tau'}_{\tau \sigma} =\frac{1}{2}  g^{\tau' \sigma'} (g_{\sigma \sigma'},_{\tau}
+ g_{\tau \sigma'},_{\sigma} - g_{\sigma \tau},_{\sigma'})$.

Let us  make a choice for the vielbein
\begin{eqnarray}
\label{fmagen} 
f^{\sigma}{}_{m}&=& \sum_{A}\,\vec{\tau}^{A\sigma}\, \vec{\cal{A}}^{A}_{m} \,, 
\end{eqnarray}
where we expect $\vec{\cal{A}}^{A}_{m} $ that they manifest in $d=(3+1)$ as the gauge fields 
of the charges $\tau^{Ai}$. 
To prove this we must compare the gauge fields $A^{A}_{m}= c_{Ai}{}^{st} \omega_{stm}$,
appearing in Eq.~(\ref{faction}), with the gauge fields $\vec{\cal{A}}^{A}_{m}$. 

 If there are no fermions present then the  vector gauge fields of the family members charges 
and of the family charges -  $\omega_{abm}$ and $\tilde{\omega}_{abm}$, respectively - are 
uniquely expressible with the vielbeins~\cite{norma2014MatterAntimatter,JMP2015,DNproof}.
We are interested in the vector gauge fields in $d=(3+1)$, for which we find
\begin{eqnarray}
\label{omegaabe1m}
\omega_{stm} &=&  \frac{1}{2E} \{   f^{\sigma}{}_{m}\,[e_{t \sigma}\partial_\tau
(Ef^{\tau}{}_{s})  - e_{s\sigma}\,\partial_\tau (Ef^{\tau}{}_{t})]
  \nonumber\\
                  & & + e_{s\sigma} \partial_\tau [E (f^{\sigma}{}_{m}
 f^{\tau}{}_{t} - f^{\tau}{}_{m}  f^{\sigma}{}_{t})]- e_{t\sigma} \partial_\tau [E
 (f^{\sigma}{}_{m}  f^{\tau}{}_{s} - f^{\tau}{}_{m}  f^{\sigma}{}_{s})] \}\,.
  \end{eqnarray}
We must show that  if we calculate $c_{Ai}{}^{st}\, \omega_{stm}$ by taking  $\omega_{stm}$ 
from Eq.~(\ref{omegaabe1m}), and if we put  on the right hand side of this equation the vielbeins 
$f^{\sigma}{}_{m} = \sum_{A}\,\vec{\tau}^{A\sigma}\, \vec{\cal{A}}^{A}_{m} $ from 
Eq.~(\ref{fmagen}), we  must end up with the equality relation
\begin{eqnarray}
\label{AcalA} 
A^{Ai}_{m}&=&{\cal A}^{Ai}_{m} \,. 
\end{eqnarray}
It is not difficult to check that Eq.~(\ref{AcalA})  follows, if we take into account that 
$e^{s}{}_{\mu}=$ $ - \delta^{m}_{\mu}$ and make a choice of the symmetry of space $(d-4)$: 
$f^{\sigma}{}_{s} =f \, \delta^{\sigma}_{s}$ (Ref.~\cite{JMP2015}, Sect.~ II.).

Calculating from Eqs.~(\ref{gmdown}, \ref{gmup}) the Riemann curvature $R^{(d)} $ in 
$d$-dimensional space  by taking into account that  ($d=(3+1)$) space is flat, one obtains
\begin{eqnarray}
\label{actionvg}
R^{(d)} &=& R^{(d-4)} - \frac{1}{4} g_{\sigma \tau} E^{\sigma}{}_{A i}
 E^{\tau}{}_{A' i'} \, F^{Ai}{}_{mn} F_{A' i'}{}^{mn}\,,\nonumber\\
F^{Ai}{}_{mn} &=& \partial_{m} A ^{Ai}_{n}- \partial_{n} A ^{Ai}_{m}
- i f^{Aijk} \, A^{Aj}_{m} \,A^{Ak}_{n}\,,\nonumber\\
A^{Ai}_{m}&=& \sum_{s t} \,c^{Ai}{}_{st}\, \omega^{st}{}_{m}\,,\nonumber\\
\tau^{Ai} &=&  \sum_{s t}\, c^{Ai st} \,M_{st}\,.
\end{eqnarray}
The integration of the action $ \int \,E \,d^{4} x d^{(d-4)} x\,R^{(d)}$ over $(d-4)$ space
(in which it turns out that only even functions of coordinates $x^{\sigma}$ give nonzero 
contributions) leads to the well known effective action for the vector gauge fields in $d=(3+1)$ 
space: $\int \,d^{4} x \, 
\{ - \frac{1}{4}  F^{A i}{}_{\mu \nu} \, F^{A i\mu \nu}\,\} $. 

The quadratic form for the vector gauge fields,  Eq.~(\ref{actionvg}), in $d=(3+1)$, obtained from 
the curvature $R^{(d)}$ can be found in many text book~\cite{mil}.  In Ref.~(\cite{MatejPavsic}, 
Sect.~5.3) the Lagrange function for the gauge  vector fields is derived by using the Clifford algebra 
space. The author allows besides the curvature $R$ also its quadratic form $R^2$ (Eq.~(240)).

\vspace{2mm}

\subsection{Scalar fields in the {\it spin-charge-family} theory explain the origin of the Higgs
and Yukawa couplings }
\label{scalars}

\vspace{2mm}

In the {\it spin-charge-family} theory the spin connection fields of both kinds, $\omega_{abs}$ and
$\tilde{\omega}_{abs}$, carrying the space index $s=(7,8)$, explain the Higgs and the Yukawa 
couplings of the {\it standard model}.  They all belong to the weak charge doublets~(\cite{JMP2015,%
norma2014MatterAntimatter} and references therein), as will be demonstrated in this section.

After gaining nonzero vacuum expectation values these scalar fields break the weak and the hyper 
charges of the vacuum (the assumption {\bf A v.} and comments {\bf C v.})  making all the fermions, 
massive due to the interaction with the vacuum.  Also the heavy bosons gain masses, while interacting
with the vacuum.

The gauge scalar fields with the space index $s>8$ contribute to the 
matter-antimatter asymmetry in the universe~\cite{norma2014MatterAntimatter}.

This section follows mainly the equivalent sections in Refs.~\cite{norma2014MatterAntimatter,%
JMP2015}.

It turns out~\cite{norma2014MatterAntimatter} that all scalars (the gauge fields with the space index 
$s \ge 5$) of the action (Eq.~(\ref{wholeaction})) carry charges in fundamental 
representations due to the space index: They are either doublets (Table~\ref{Table doublets.}), 
$s=(5,6,7,8)$, or triplets (Table~\ref{Table bosons.},  and Sect.~ II,~Table~ I in 
Ref.~\cite{norma2014MatterAntimatter}), $s=(9,10,..,13,14)$. 
Scalars with the space indices $s \in (7,8)$ and $s \in (5,6)$ are  
the $SU(2)$ {\em doublets} (Table~\ref{Table doublets.}) with respect to this degree of freedom. 

All scalars carry additional quantum numbers:  Besides the quantum numbers determined by the 
space index $s$ they carry also the quantum numbers $Ai$, Eq.~(\ref{tau}), the states of which 
belong to the adjoint representations. They originate in $S^{ab}$ or $\tilde{S}^{ab}$, 
Eq.~(\ref{sabtildesab}), $S^{ab} =\frac{i}{4} (\gamma^a\, \gamma^b - \gamma^b\, 
\gamma^a)\,$, $\tilde{S}^{ab} = \,\frac{i}{4} (\tilde{\gamma}^a\, \tilde{\gamma}^b - 
\tilde{\gamma}^b\, \tilde{\gamma}^a)$, the gauge fields of which  are $\omega_{abs}$ and 
$\tilde{\omega}_{abs}$, respectively. 
$S^{ab}$ determine family members spin and charges, $\tilde{S}^{ab}$ determine family charges.

The  infinitesimal generators ${\cal S}^{ab}$, which apply on the spin connections 
$\omega_{bd e}$ ($= f^{\alpha}{}_{e}\, $ 
$\omega_{bd \alpha}$) and $\tilde{\omega}_{\tilde{b} \tilde{d} e}$ ($= f^{\alpha}{}_{e}\,$ 
$\tilde{\omega}_{\tilde{b} \tilde{d} \alpha}$), on either the space index $e$ or any of the
indices $(b,d,\tilde{b},\tilde{d})$, operates as follows 
\begin{eqnarray}
\label{bosonspin0}
{\cal S}^{ab} \, A^{d\dots e \dots g} &=& i \,(\eta^{ae} \,A^{d\dots b \dots g} - 
\eta^{be}\,A^{d\dots a \dots g} )\,,
\end{eqnarray}
(see Section~IV. and Appendix~B in Ref.~\cite{JMP2015}).

The expressions for the infinitesimal operators of the subgroups of the starting group 
(presented in Eq.~(\ref{tau}) and the footnote ${}^{8}$ and determined by 
the coefficients 
$c^{Ai}{}_{ab}$ in Eq.~(\ref{tau}))  are the same for all three kinds of degrees of freedom,
 $S^{ab}$,  $\tilde{S}^{ab}$ or  ${\cal S}^{ab}$.  Correspondingly the 
commutation relations are also the same. 

 At the electroweak break all the scalar fields with the space index $(7,8)$, those which belong
to one of twice two triplets carrying the family quantum numbers ($\tilde{\tau}^{\tilde{A}i}$) 
 and those which belong to one of the three singlets carrying the family members quantum
numbers ($Q,Q',Y'$), Eq.~(\ref{commonAi}), 
start to self interact, gaining nonzero vacuum expectation values and breaking the weak charge, 
the hyper charge and the family charges.

Let me introduce  a  common notation  $ A^{Ai}_{s}$ for all the scalar fields with $s=(7,8)$,
independently of whether  they originate in $\omega_{abs}$ - in this case $ Ai$ 
$=(Q$,$Q',Y'$) - or in $\tilde{\omega}_{\tilde{a}\tilde{b}s}$ - in this case all the
family quantum numbers of all eight families contribute. 
\begin{eqnarray}
\label{commonAi}
 A^{Ai}_{s} &{\rm represents}& (\,A^{Q}_{s}\,,A^{Q'}_{s}\,, A^{Y'}_{s}\,, 
 \vec{\tilde{A}}^{\tilde{1}}_{s}\,, 
 \vec{\tilde{A}}^{\tilde{N}_{\tilde{L}}}_{s}\,, \vec{\tilde{A}}^{\tilde{2}}_{s}\,, 
 \vec{\tilde{A}}^{\tilde{N}_{\tilde{R}}}_{s}\,)\,,\nonumber\\
\tau^{Ai} &{\rm represents}& (Q,\,Q',\,Y', \,\vec{\tilde{\tau}}^{1},\, \vec{\tilde{N}}_{L},\,
\vec{\tilde{\tau}}^{2},\,\vec{\tilde{N}}_{R})\,.
\end{eqnarray}
Here $\tau^{Ai}$ represent all the operators, which apply on the spinor states.
These scalars, the gauge scalar fields of the generators $\tau^{Ai}$ and $\tilde{\tau}^{Ai}$, 
are expressible in terms of the spin connection fields (Ref.~\cite{JMP2015}, Eqs.~(10, 22, A8, A9)).

Let me demonstrate~\cite{JMP2015} that all the scalar fields with the space index $(7,8)$ carry with
respect to this space index the weak and the hyper charge ($\mp \frac{1}{2}$, 
$\pm \frac{1}{2}$), respectively. This means that all these scalars have properties as required
for the Higgs in the {\it standard model}. 

Let me make a choice of the superposition of the scalar fields so that they are eigenstates of 
$\tau^{13}= \frac{1}{2}({\cal S}^{56} - {\cal S}^{78})$ (Eq~(\ref{tau}) and footnotes 
at one page before).  I rewrite for this purpose the second line of Eq.~(\ref{faction}) 
as follows (the momentum $p_{s}$ is left out~\footnote{It is expected that solutions with 
nonzero momenta lead to higher masses of fermion fields in $d=(3+1)$~\cite{NHD,DN012}.
We pay correspondingly no attention to the momentum $p_{s}\,, s\in(5,\dots,8)$, when having in 
mind the lowest energy solutions, manifesting at low energies.})
\begin{eqnarray}
\label{eigentau1tau2}
 & &\sum_{s=(7,8), A,i}\, \bar{\psi} \,\gamma^s\, ( - \tau^{Ai} \,A^{Ai}_{s}\,)\,\psi =
\nonumber\\
 & &- \bar{\psi}\,\{\,\stackrel{78}{(+)}\, \tau^{Ai} \,(A^{Ai}_{7} - i   
 \,A^{Ai}_{8})\, + \stackrel{78}{(-)}(\tau^{Ai} \,(A^{Ai}_{7} + i \,A^{Ai}_{8})\,\}\,\psi\,,
 \nonumber\\
 & &\stackrel{78}{(\pm)} = \frac{1}{2}\, (\gamma^{7} \pm \,i \, \gamma^{8}\,)\,,\quad
 A^{Ai}_{\scriptscriptstyle{\stackrel{78}{(\pm)}}}: = (A^{Ai}_7 \,\mp i\, A^{Ai}_8)\,,
\end{eqnarray}
with the summation over $A$ and $i$ performed, since $A^{Ai}_s$ represent the scalar fields 
($A^{Q}_{s}$, $A^{Q'}_{s}$, $A^{Y'}_{s}$, 
$\tilde{A}^{\tilde{4}}_{s}$, 
$\vec{\tilde{A}}^{\tilde{1}}_{s}$, $\vec{\tilde{A}}^{\tilde{2}}_{s}$,
 $\vec{\tilde{A}}^{\tilde{N}_{R}}_{s}$ and  $\vec{\tilde{A}}^{\tilde{N}_{L}}_{s}$).

The application of the operators $Y$ ($Y= \tau^{23} +\tau^{4}$, $\tau^{23} =
 \frac{1}{2} ({\cal S}^{56} +{\cal S}^{78})$, $\tau^{4}=$ $ -\frac{1}{3} ({\cal S}^{9\,10}
 +  {\cal S}^{11\,12} +{\cal S}^{13\,14})$), $Q$ ($Q= \tau^{13} +Y $ 
and  $ \tau^{13}$ ($\tau^{13} = \frac{1}{2} ({\cal S}^{56} - {\cal S}^{78})$)  
on the fields ($A^{Ai}_{7}\mp i\,A^{Ai}_{8})$ gives  (${\cal S}^{ab}$ is defined in 
Eq.~(\ref{bosonspin0})) 

\begin{eqnarray}
\label{checktau13Y}
\tau^{13}\,(A^{Ai}_7 \,\mp i\, A^{Ai}_8)&=& \pm \,\frac{1}{2}\,(A^{Ai}_7 \,
\mp i\, A^{Ai}_8)\,,\nonumber\\
Y\,(A^{Ai}_7 \,\mp i\, A^{Ai}_8)&=& \mp \,\frac{1}{2}\,(A^{Ai}_7 \,
\mp i\, A^{Ai}_8)\,,\nonumber\\
Q\,(A^{Ai}_7 \,\mp i\, A^{Ai}_8)&=& 0\,.
\end{eqnarray}
Since  $ \tau^{4}$, $Y$, $\tau^{13}$ and 
$\tau^{1 +},  \tau^{1 -}$ give zero if 
applied on  ($A^{Q}_{s}$, $A^{Q'}_{s}$ and $A^{Y'}_{s}$) with  respect to the quantum numbers 
($Q, Q', Y'$),  and since $Y$ and $\tau^{13}$ commute with the family quantum numbers, one 
sees that the scalar fields $A^{Ai}_{s}$ ( =($A^{Q}_{s}$, $A^{Y}_{s}$, $A^{Y'}_{s}$, 
$\tilde{A}^{\tilde{4}}_{s}$, $\tilde{A}^{\tilde{Q}}_{s}$, $\vec{\tilde{A}}^{\tilde{1}}_{s}$, 
$\vec{\tilde{A}}^{\tilde{2}}_{s}$, $\vec{\tilde{A}}^{\tilde{N}_{R}}_{s}$, 
$\vec{\tilde{A}}^{\tilde{N}_{L}}_{s}$)), rewritten as  %
$A^{Ai}_{\scriptscriptstyle{\stackrel{78}{(\pm)}}} $ $= (A^{Ai}_7 \,\mp i\, A^{Ai}_8)\,$,
%
are eigenstates of $\tau^{13}$ and $Y$, having the quantum numbers of the {\it standard model} 
Higgs' scalar.

These superposition of $A^{Ai}_{\scriptscriptstyle{\stackrel{78}{(\pm)}}}$ are presented in 
Table~\ref{Table doublets.} as two doublets with respect to the weak charge 
${\cal \tau}^{13}$,  with the eigenvalue of ${\cal \tau}^{23}$  (the second 
$SU(2)_{II}$ charge) 
equal to either $-\frac{1}{2}$ or $+\frac{1}{2}$, respectively. 
\begin{table}
\caption{
The two scalar weak doublets, one with $ {\cal \tau}^{23}=- \frac{1}{2}$  and the other with 
$ {\cal \tau}^{23}=+ \frac{1}{2}$, both with the "spinor" quantum number ${\cal \tau}^{4}$ 
$=0$, are presented. 
In this table all the scalar fields carry besides the quantum numbers determined by the space index 
also  the quantum numbers $A$ and $i$ from Eq.~(\ref{commonAi}).}
 \begin{center}
{ \begin{tabular}{c|c| c c c c r}
 \hline
 &state & ${\cal \tau}^{13}$& $ {\cal \tau}^{23}$ & spin& ${\cal \tau}^{4}$& $ Q$\\
 \hline
 $A^{Ai}_{\scriptscriptstyle{\stackrel{78}{(-)}}}$ & $A^{Ai}_{7}+iA^{Ai}_{8}$& $+
\frac{1}{2}$& 
 $-\frac{1}{2}$& 0&0& 0\\
 $A^{Ai}_{\scriptscriptstyle{\stackrel{56}{(-)}}}$ & $A^{Ai}_{5}+iA^{Ai}_{6}$& 
$-\frac{1}{2}$& 
 $-\frac{1}{2}$& 0&0& -1\\
 \hline 
$A^{Ai}_{\scriptscriptstyle{\stackrel{78}{(+)}}}$ & $A^{Ai}_{7}-iA^{Ai}_{8}$& 
$-\frac{1}{2}$& 
$+\frac{1}{2}$& 0&0& 0\\
$A^{Ai}_{\scriptscriptstyle{\stackrel{56}{(+)}}}$ & $A^{Ai}_{5}-iA^{Ai}_{6}$& $+
\frac{1}{2}$& 
$+\frac{1}{2}$& 0& 0&+1\\ 
\hline
\end{tabular}
}
 \end{center}
\label{Table doublets.}
 \end{table}

The operators ${\cal \tau}^{1\spm} = {\cal \tau}^{11}\pm i {\cal \tau}^{12} $
\begin{eqnarray}
\label{YQ}
\tau^{1\spm} &=& \frac{1}{2} [({\cal S}^{58}- {\cal S}^{67})\,\smp \,i\,
({\cal S}^{57}+ {\cal S}^{68})]\,, 
\end{eqnarray}
transform one member of a doublet from Table~\ref{Table doublets.} into another member of 
the same doublet, keeping $ \tau^{23}$  ($= \frac{1}{2}\,({\cal S}^{56}+ {\cal S}^{78})$) 
unchanged, clarifying the above statement.

It is shown in Ref.~(\cite{JMP2015}, Eq.~(22)) that the scalar fields 
$A^{Ai}_{\scriptscriptstyle{\stackrel{78}{(\pm)}}}$  
are either {\it triplets} as the gauge fields of the {\it family quantum numbers} 
($\vec{\tilde{N}}_{R}, \,$ $\vec{\tilde{N}}_{L},\,$ $ \vec{\tilde{\tau}}^{2},\,$ 
$\vec{\tilde{\tau}}^{1}$)
or they are singlets as the gauge fields of 
$Q=\tau^{13}+Y, \,Q'= -\tan^{2}\vartheta_{1} Y$ $ + \tau^{13} $ and
 $Y' = -\tan^2 \vartheta_{2} \tau^{4} + \tau^{23}$.

One finds 
\begin{eqnarray}
\label{checktildeNL3Q}
\tilde{N}_{L}^{3}\,\tilde{A}^{\tilde{N}_{L} \spm}_{\scriptscriptstyle{\stackrel{78}{(\pm)}}} &=&
\spm   \tilde{A}^{\tilde{N}_{L}\spm}_{\scriptscriptstyle{\stackrel{78}{(\pm)}}}\,,\quad
\tilde{N}_{L}^{3}\,\tilde{A}^{\tilde{N}_{L}3}_{\scriptscriptstyle{\stackrel{78}{(\pm)}}}=0\,,\nonumber\\
Q \,A^{Q}_{\scriptscriptstyle{\stackrel{78}{(\pm)}}} &=&0\,.
\end{eqnarray}
with $ Q={\cal S}^{56} + {\cal \tau}^{4}= {\cal S}^{56} -\frac{1}{3}({\cal S}^{9\,10}+
{\cal S}^{11\,12} + {\cal S}^{13\,14})$, and with ${\cal \tau}^{4}$ defined in the footnote 
${}^{8}$ (if one replaces $S^{ab}$ by ${\cal S}^{ab}$ from Eq.~(\ref{bosonspin0})). 

Similarly one finds properties with respect to the $Ai$ quantum numbers for all the scalar fields
$A^{Ai}_{\scriptscriptstyle{\stackrel{78}{(\pm)}}}$. 

After the appearance of the condensate (Table~\ref{Table con.}), which breaks the
 $SU(2)_{II}$ symmetry (bringing masses to all the scalar fields), the weak charge  
$\vec{\tau}^{1}$ and the hyper charge $Y$ remain the conserved
 charges~\footnote{It is $\tau^{23}$ which determines the hyper charge $Y$ 
($ Y= S^{23} + \tau^{4}$) of these scalar fields, since $\tau^{4}$, if applies on the 
scalar index of these scalar fields, gives zero, according to equations in the footnote ${}^{8}$.}.

The nonzero vacuum expectation values of the scalar fields of Eq.~(\ref{commonAi}) break the 
mass protection mechanism of quarks and leptons and determine correspondingly the mass 
matrices (Eq.~(\ref{M0})) of the two groups of quarks and leptons.  

Obviously the scalar fields in the {\it spin-charge-family} theory have all the properties of the 
Higgs. I show below and in Subsect.~\ref{massmatrices} that these scalar fields explain also 
the Yukawa couplings of the {\it standard model}.

All  other scalar fields: $A^{Ai}_{s}\,, s\in(5,6)$ and  $A^{Ai}_{t}\,, t \in(9, \dots,14)$
have masses of the order of the condensate scale and contribute to matter-antimatter 
asymmetry~\cite{norma2014MatterAntimatter}.

\vspace{2mm}

{\it Effective action for  scalar fields with the space index $(7,8)$}
%

\vspace{2mm}
  
It would be possible, at  least in principle, to derive the low energy effective action for scalars from 
the starting action (Eq.~\ref{wholeaction}) by guessing the boundary conditions, under which the 
universe evolved, since all  the scalar fields, as well as their Lagrange density, are included in the 
starting action. This is  an extremely demanding project.

In what follows the effective Lagrange density for the scalar fields is assumed to be changed from 
the Lagrange density before the electroweak break 
${\cal L}_{s} =E\, \{(p_{m} \,A^{Ai}_{s})^{\dagger} \, (p^{m} \,A^{Ai}_{s}) -
(m'_{Ai})^2 \, A^{Ai\dagger}_{s} \,A^{Ai}_{s}\}$  to 
\begin{eqnarray}
\label{interactingphi}
{\cal L}_{sg} &=& E\,\sum_{A,i} \,\{(p_{m} \,A^{Ai}_{s})^{\dagger} \,(p^{m} \,
A^{Ai}_{s}) - (-\lambda^{Ai} + (m'_{Ai})^2)) \,A^{Ai \dagger}_{s} A^{Ai}_{s} 
\nonumber\\ 
&+& \sum_{B,j}\, \Lambda^{Ai Bj}\,A^{Ai \dagger}_{s} A^{Ai}_{s} \;
A^{Bj \dagger}_{s} A^{Bj}_{s}\}\,,
\end{eqnarray}
where $-\lambda^{Ai} +  m'^{2}_{Ai}= m^{2}_{Ai}$  and $m_{Ai}$ manifests as the mass of
the $A^{Ai}_{s}$ scalar.

The Lagrange density leads to the coupled equations of motion for many scalar fields with, 
in this assumption, harmonic interactions. It requires a lot of effort to extract the dependence 
of the eigen modes  on the parameters of the Lagrange density to see the influence of the 
parameters on the properties of fermions. This work has not yet been done. First attempts are in 
progress.

\vspace{2mm}

{\it Yukawa couplings in the {\it spin-charge-family} theory}

\vspace{2mm}

Let $\psi^{\alpha}_{(L,R)}$ denote massless and $\Psi^{\alpha}_{(L,R)}$  massive four vectors 
for each family member  $\alpha= (u_{L,R}, d_{L,R}, \nu_{L,R}, e_{L,R})$, let say for the group 
of four families among which there are the observed three families, after taking into account 
loop corrections in all orders.  
\begin{eqnarray}
\label{psim}
\psi^{\alpha}_{(L,R)} &=& V^{\alpha}_{(L,R)} \,\Psi^{\alpha}_{(L,R)} \,,
\end{eqnarray}
and let
$(\psi^{\alpha \,k}_{(L,R)}\,$,  $\,\Psi^{\alpha\, k}_{(L,R)} )$ 
be any component of the four vectors, massless and massive, respectively.
On the tree level  we have 
$\psi^{\alpha}_{(L,R)}=V^{\alpha}_{(o)}\:
\Psi^{\alpha \,(o)}_{(L,R)}$ 
and
\begin{equation}
\label{treenotation}
          < \psi^{\alpha}_{L}|\gamma^0 \, {\cal M}^{\alpha}_{(o)}\,
 |\psi^{\alpha}_{R}> = < \Psi^{\alpha \,(o)}_{L}|\gamma^0 \,V^{\alpha\,
 \dagger}_{(o)}\, {\cal M}^{\alpha}_{(o)}\,V^{\alpha}_{(o)}\,|\Psi^{\alpha}_{R \,(o)}>,
\end{equation}
with 
${\cal M}^{\alpha}_{(o)k\, k'}=\sum_{A,i}\, (-g^{Ai} \, v_{Ai\, \mp})\,\, C^{\alpha}_{k\,k'}\,$. 
Here $g^{Ai} \, v_{Ai\, \mp}$ represent the nonzero vacuum expectation values of the scalar  fields.
In this case the coefficients $ C^{\alpha}_{k\,k'}$ are determined by the mass matrices, Eq.(\ref{M0}),
on the tree level. It then follows
\begin{eqnarray}
\label{Phipsi}
\overline{\Psi}^{\alpha}\,V^{\alpha \dagger}_{(o)}\, {\cal M}^{\alpha}_{(o)}\,V^{\alpha }_{(o)}\:
 \Psi^{\alpha} &=& \overline{\Psi}^{\alpha}\,{\rm diag}(m^{\alpha}_{(o)1}\,,\cdots\,,
m^{\alpha}_{(o)4})\,
 \Psi^{\alpha}\,, \nonumber\\
V^{\alpha \dagger}_{(o)}\, {\cal M}^{\alpha}_{(o)}\,V^{\alpha }_{(o)}&=&
 \Phi^{\alpha}_{\Psi(o)}\,.
\end{eqnarray}
On the tree level the coupling constants $m^{\alpha}_{(o)k}$ (in some units) of the dynamical scalar 
fields $\Phi^{\alpha}_{\Psi(o) \,k}$ - the superposition of $A^{Ai}_{s}$ - to 
the family member  $\Psi^{\alpha \,k}$ belonging to the $k^{th}$ family are  equal to
\begin{eqnarray}
\label{Phipsiex}
 (\Phi^{\alpha}_{\Psi(o)})_{k\,k'}\,\Psi^{\alpha\,k'} &=& \delta_{k\,k'}
 \,m^{\alpha}_{(o)k}\,\Psi^{\alpha\,k}\,. 
\end{eqnarray}

The superposition of scalar fields $(\Phi^{\alpha}_{\Psi(o)})$, which couple to fermions
and depend on the quantum numbers $\alpha$ and  $k$, are in general   different 
from the superposition, 
which are their mass eigenstates.  
Each family member $\alpha$ of each massive family $k$ couples in  general 
to different superposition of scalar fields.

It turns out that  mass matrices of both - quarks and leptons -  
behave in a very similar way. No additional neutrinos, offering a "sea-saw" mechanism, are needed.
All this is already included in the starting action.

\vspace{2mm}

\subsection{The condensate in the {\it spin-charge-family} theory}
\label{condensate}

\vspace{2mm}

The appearance of the condensate of two right handed neutrinos with properties presented in
Table~\ref{Table con.} is in this paper assumed so that in the low energy regime the 
{\it spin-charge-family} theory leads to the effective action, explaining the assumptions of the 
{\it standard model} and consequently the observed phenomena. 
The condensate should appear during the expansion of the universe, due to particular boundary 
conditions and the conditions in the universe in the time of the appearance of the condensate. 
This study has not yet been done.

The condensate, presented on Table~\ref{Table con.}, does not influence the colour, the weak and 
the hyper charges  ($\vec{\tau}^{3}$, $\vec{\tau^{1}}, Y$, respectively) of the corresponding 
gauge fields. Since these vector gauge fields don't  interact with  the condensate,
the colour, the weak and the hyper charges remain the conserved quantities up to the electroweak 
phase transition.

The condensate changes the properties of the scalar fields, which are before the appearance of the
condensate massless scalar gauge fields. Interaction with the condensate makes all the scalar fields
massive.

After the electroweak break, when the scalar fields with the space index $s=(7,8)$ - those with the 
family quantum numbers $(\tilde{N}^{i}_{(L,R)}, \tilde{\tau}^{(1,2) i})$ and those with the family 
members quantum numbers ($Q,Q',Y'$) - start to strongly self interact~(Eq.~(\ref{interactingphi})), 
gaining nonzero vacuum expectation values (Eq.~(\ref{interactingphi})) and correspondingly changing 
 their own masses as well as the properties of the vacuum, so that the weak charge and the hyper 
charge are no longer conserved quantities. The only conserved charges are then the colour and the 
electromagnetic charges.

\vspace{3mm}

\section{Summary of the {\it spin-charge-family} theory achievements so far}
\label{achievements}

\vspace{3mm}

To understand better the history of our universe the explanation of the {\it standard model} 
assumptions is certainly needed. It is also needed to know the number of families in the low energy 
regime and to understand the appearance of phenomena like the existence of the dark matter,  the 
matter-antimatter asymmetry and the dark energy.

I have demonstrated so far, that the {\it spin-charge-family} theory, starting with the simple action
in $d=(13+1)$ for  fermions (carrying only two kinds of spins, no charges)  and for the gauge 
fields  to which fermions are coupled (vielbeins and two kinds of  spin connection fields)   
offers the explanation for all the assumptions of the {\it standard model}:\\
{\bf a.} $\;\,$ The theory explains all the properties of the family members - quarks and leptons,
left and right handed, relating handedness and charges, and their right and left handed antiquarks 
and antileptons.\\ 
{\bf b.} $\;\,$ It explains the appearance and properties of the families of family members. \\
{\bf c.} $\;\,$ It explains the existence of the gauge vector fields of the family members charges. \\
{\bf d.} $\;\,$ It explains the appearance and properties of the scalar field (the Higgs) and the 
Yukawa couplings.\\

The {\it spin-charge-family} theory predicts that there are at the low energy regime two decoupled 
groups of four families of quarks and leptons, what means that besides the observed three there is
the fourth not yet observed family of quarks and leptons.

The existence of two decoupled groups of four families also means that the stable of the 
upper four families must also be observed. In Subsect.~\ref{darkmatter}~\cite{gn} the possibility is 
discussed that the stable of the upper four families constitutes the dark matter.

In this section I overview the {\it spin-charge-family} theory achievements, explaining:\\
{\bf i.} $\;\,$ The properties of the lower four families, the three of which have already been
observed, as they follow from the properties of the scalar fields of this theory, 
Subsect.~\ref{massmatrices},  \\
{\bf i. a.} presenting the results of the calculations (not yet published),\\
{\bf i. b.}  discussing whether or not present experiments speak or not against the existence of the
fourth family (not yet published), in particular I shall comment the contribution of the fourth family to
the production of the Higgs in the quark-fusion process, Subsect.~\ref{fourthfamily}, the topics which 
seem to speak the most against the existence of the fourth family.\\
{\bf ii.} $\;\,$
The fact that this theory easily explains the "miraculous" cancellation of the triangle anomalies in the
{\it standard model}, Subsect.~\ref{anomaly} (also this topics is not yet published).\\
{\bf iii.} $\;\,$ The existence of the dark matter, Subsect.~\ref{darkmatter}.\\
{\bf iv.} $\;\,$ The explanation for the matter-antimatter asymmetry, Subsect.~\ref{matterantimatter}.

\vspace{2mm}

\subsection{Masses of the lower four families of quarks and leptons in the {\it spin-charge-family}
 theory~\cite{gmdn07,gn2013,gn2015}}
\label{massmatrices}

\vspace{2mm}

This subsection is a short report of the not yet published results of Ref.~\cite{gn2015} (published in the
Proceedings).

There are two groups of four families. The mass matrix of each family member of each of the 
group of four families demonstrates  in the 
massless basis the $U(1)\times \widetilde{SU}(2)\times$ $\widetilde{SU}(2)$ symmetry
(each of the two $\widetilde{SU}(2)$ is a subgroup, one of $\widetilde{SO}(3,1)$ and the 
other of $\widetilde{SO}(4)$). 

The scalars with the family quantum numbers split the eight families into twice four families.  
To the masses of the lower four families the scalar fields, which are the gauge fields of 
$\vec{\tilde{N}}_{L}$ and  $\vec{\tilde{\tau}}^{1}$ contribute. To the masses of the upper 
four families the gauge fields of $\vec{\tilde{N}}_{R}$ and  $\vec{\tilde{\tau}}^{2}$ contribute. 
The scalars with the family members quantum numbers $(Q,Q',Y')$ contribute  to the masses of 
the lower and upper four families. 

I discuss here properties of quarks and leptons of the lower four families, Eq.~(\ref{M0}).

Let $\psi_{i}$, $i=(1,2,3,4)$, denote the massless basis for a particular family member 
$\alpha$. And let us denote the two kinds of the operators, which transform the basis 
vectors into one another as
\begin{eqnarray}
\label{taunl}
\tilde{N}^{i}_{L}\,,\,i=(1,2,3)\,, \quad \quad \tilde{\tau}^{i}_{L}\,,\,i=(1,2,3)\,.
\end{eqnarray}
One finds
\begin{eqnarray}
\label{taunlonpsi}
&&\tilde{N}^{3}_{L}\, (\psi_1, \psi_2,\psi_3,\psi_4)= \frac{1}{2} (-\psi_1,\; \psi_2,-\psi_3,\;\psi_4)\,,\nonumber\\
&&\tilde{N}^{+}_{L}\, (\psi_1, \psi_2,\psi_3,\psi_4)=\;\,  (\psi_2, \;\;0,\;\,\psi_4,\;\;0)\,,\nonumber\\
&&\tilde{N}^{-}_{L}\, (\psi_1, \psi_2,\psi_3,\psi_4)=\;\,  (0\;\;, \;\psi_1,\;\;0,\;\psi_3)\,,\nonumber\\
&&\tilde{\tau}^{13}\,\; (\psi_1, \psi_2,\psi_3,\psi_4)= \frac{1}{2} (-\psi_1, -\psi_2,\;\psi_3,\;\psi_4)\,, \nonumber\\
&&\tilde{\tau}^{1+}\,\; (\psi_1, \psi_2,\psi_3,\psi_4)=\;\, (\psi_3,\; \psi_4,\;\;0,\;\;0)\,,\nonumber\\ 
&&\tilde{\tau}^{1-}\,\; (\psi_1, \psi_2,\psi_3,\psi_4)=\;\, (\;\;0, \;\;0,\;\,\psi_1,\,\;\psi_2) \,.
\end{eqnarray}
This is indeed what the two $SU(2)$ operators in the {\it spin-charge-family} theory do on the lower 
four families. The gauge scalar fields 
$A^{Ai}_{\scriptscriptstyle{\stackrel{78}{(\pm)}}},\, (A,i) =[\tilde{N}^{i}_{L}, i=({\pm},3), 
\tilde{\tau}^{1i}, i=({\pm,3)}]$, 
Eqs.~(\ref{eigentau1tau2}, \ref{commonAi}), of these operators determine 
the off diagonal and diagonal matrix elements 
after the electroweak phase transition in which scalar fields gain nonzero vacuum expectation values. 

In addition to these two kinds of $SU(2)$ scalars there are three $U(1)$ scalars, which distinguish 
among the family members, contributing on the tree level the same diagonal matrix elements for 
all the families.

In loop corrections in all orders the symmetry of mass matrices remains unchanged, while the 
three $U(1)$ scalars 
manifest in off diagonal elements as well.  

 All the scalars, the two triplets and the three singlets, are doublets with respect to the weak 
charge, contributing to the weak and the hyper charge of the fermions so that they transform 
the right handed members into the left handed onces  with the phases presented in 
Table~\ref{Table so13+1.}.
\begin{small}
 \begin{equation}
 \label{M0}
 {\cal M}^{\alpha} = \begin{pmatrix} 
 - a_1 - a & e     &   d & b\\ 
 e     & - a_2 - a &   b & d\\ 
 d     & b     & a_2 - a & e\\
 b     &  d    & e   & a_1 - a
 \end{pmatrix}^{\alpha}\,.
 \end{equation}
 \end{small}

Although any accurate $3\times 3$ submatrix of the $4 \times 4$ unitary matrix determines the 
$4 \times 4$ matrix uniquely, neither the quark  nor (in particular) the lepton $3\times 3$ mixing
matrix are measured accurately enough that it would be possible to determine 
three complex phases of the $4 \times 4$ mixing matrix as well as the mixing matrix elements 
of the fourth family members to the lower three. 

 We therefore assumed in our calculations~\cite{gmdn07,gn2013,gn2015} that the mass matrices 
are symmetric 
and real. Correspondingly the mixing matrices are orthogonal. We fitted the $6$ free parameters 
of each quark mass matrix, Eq.~(\ref{M0}),  to twice three,  that is $6$, measured quark masses, 
and to the $6$ (from the experimental data extracted) parameters of the 
corresponding $4 \times 4$ mixing matrix.

While the experimental accuracy of the quark masses of the lower three families does not influence 
the calculated mass matrices considerably, it turned out that the experimental accuracy of the 
$3\times 3$ quark mixing matrix is not good enough to trustworthy determine the mass intervals 
for the fourth family quarks.

Taking into account our calculations,  in which we fit parameters of Eq.~(\ref{M0}) to the 
experimental data for masses and mixing matrices  for quarks and the meson decays evaluations
 in the literature, as well as our own evaluations, we estimated that the fourth family quarks masses 
might be above $1$ TeV. 
Choosing the masses of the fourth family quarks we were able  not only to calculate the fourth
family matrix elements to the lower three families, but also predict towards which values will the
matrix elements of the $3\times 3$  submatrix move in more accurate 
experiments~\cite{gn2015}.

The two fitted mass matrices, Ref.~(\cite{gn2015}, Eqs.~(23, 27)) lead to masses of 
Eq.~(\ref{mdudnew21})  for the choice of $M_{u_{4}}/{\rm MeV/c^2} = 700\,000=
M_{d_{4}}/{\rm MeV/c^2}$
 \begin{eqnarray}
 \label{mdudnew21}
  {\bf M}^{u}/{\rm MeV/c^2} &=& (1.3 , 620.0 , 172\,000. ,  700\,000.)\,,\nonumber\\ 
  {\bf M}^{d}/{\rm MeV/c^2} &=& (2.88508 , 55.024 ,  2\,899.99 ,   700\,000.)\,, 
  \end{eqnarray}
and to masses of 
Eq.~(\ref{mdudnew22})  for the choice of $M_{u_{4}}/{\rm MeV/c^2} =1\,200\,000=
M_{d_{4}}/{\rm MeV/c^2}$
\begin{eqnarray}
 \label{mdudnew22}
  {\bf M}^{u}/{\rm MeV/c^2} &=& (1.3 , 620.0 , 172\,000. ,   1\,200\,000.)\,,\nonumber\\ 
  {\bf M}^{d}/{\rm MeV/c^2} &=& (2.88508 , 55.024 ,  2\,899.99 ,   1\,200\,000.)\,. 
  \end{eqnarray}

They lead to the $4 \times 4$ mixing matrix 
in which we fit two kinds of the experimental - the old data ($exp_o$) and the new data ($exp_n$) - 
each used in calculations  for  the choice  $m_{u_4}= m_{d_4}=700$ GeV  ($old_{1}, new_{1}$) 
and $m_{u_4}= m_{d_4}=1\,200$ GeV ($old_{2}, new_{2}$), Eq.~(\ref{vudoldnewexp}) 
     \begin{equation}
      \label{vudoldnewexp}
      |V_{(ud)}|= \begin{pmatrix}
      \hline
     exp_o &   0.97425 \pm 0.00022    &  0.2252 \pm 0.0009    &  0.00415 \pm 0.00049     \\
     exp_n  &    0.97425 \pm 0.00022    &  0.2253 \pm 0.0008    &  0.00413 \pm 0.00049   \\
     \hline
     old_1 &   0.97423                &  0.22531              &  0.003\\
     old_2 &   0.97425                &  0.22536              &  0.00301\\
     new_1  &    0.97423                &  0.22531              &  0.00299\\  
     new_2  &    0.97423                &  0.22538              &  0.00299 \\ 
     \hline 
     exp_o &   0.230   \pm 0.011      &  1.006  \pm 0.023     &  0.0409  \pm 0.0011     \\
     exp_n  &  0.225   \pm 0.008      &  0.986  \pm 0.016     &  0.0411  \pm 0.0013    \\
    \hline
    old_1  &   0.22526                &  0.97338              &  0.042 \\
    old_2  &   0.22534                &  0.97336              &  0.04239\\
    new_1  &  0.22534                &  0.97335              &  0.04245\\  
    new_2  &  0.22531                &  0.97336              &  0.04248\\ 
    \hline
    exp_o  &   0.0084  \pm 0.0006     &  0.0429 \pm 0.0026    &  0.89    \pm 0.07      \\ 
    exp_n  &  0.0084  \pm 0.0006     &  0.0400 \pm 0.0027    &  1.021   \pm 0.032     \\
   \hline
    old_1  &   0.00663                &  0.04197              &  0.9991\\
    old_2  &   0.00663                &  0.04198              &  0.9991\\
    new_1  &  0.00667                &  0.04203              &  0.99909 \\  
    new_2  &  0.00667                &  0.04206              &  0.99909   
     \end{pmatrix}\,.
     \end{equation}
 It was noticed~\cite{gn2015} that  the mass matrix elements of the  $u$ and  $d$ quarks matrices 
change by a factor of  $\approx 1.5$ in the average, when the masses of the fourth family members 
grow from $700$ GeV to $1\,200$ GeV. 
The two mass matrices become more "democratic" (the matrix elements become closer to each other). 
Although the mixing matrix elements of the $3\times 3$ submatrix   of the $4\times 4$ matrix 
 do not  change  a lot 
with the masses of the fourth family  quarks,  they do  change so that they agree better with the 
newer~\cite{datanew} than with the older~\cite{dataold}  experimental values.

From the above results it follows:\\
i.$\;\,\,$  The prediction 
             of the calculated mixing matrix elements, obtained 
             by fitting the symmetry of the mass matrices (Eq.~(\ref{M0})) to the experimental 
             data~\cite{dataold},  was confirmed by more accurate experimental data~\cite{datanew}.
             In all cases are the calculated $3 \times 3$ matrix elements closer to the new experimental 
             values than to the old experimental values. \\ 
ii.$\;\,\,$ The fourth family masses change the mass matrices considerably, while their influence on 
             the $3 \times 3$ submatrix of the $4 \times 4$ mixing matrix is much weaker. \\
iii.$\;\;\,$ We expect 
             that more accurate experiments will bring a slightly smaller values for 
             ($V_{u_1 d_1}$, $V_{u_1 d_3}$, $V_{u_3 d_3}$), smaller ($V_{u_2 d_2}$, $V_{u_3 d_1}$), 
             ($V_{u_1 d_2}$,  $V_{u_2 d_1}$) will slightly grow  and ($V_{u_2 d_3}$)  $V_{u_3 d_2}$ will 
             grow. \\
iv.$\,\,$  The matrix elements $V_{u_i d_4}$ and $V_{u_4 d_i}$ change considerably with the mass
            of the fourth family members, and they differ quite a lot also when using new instead of the old 
            experimental data for the mixing matrix.\\
 v.$\,\,$ Fitting (twice $6$) free parameters of the mass matrices to the new experimental 
            data~\cite{datanew}  gives smaller uncertainty in fitting procedure 
            than when fitting to the old experimental data~\cite{dataold},
            while  the masses of the fourth family members do not influence the uncertainty of 
            the calculations considerably. Only very accurate mixing matrix elements would allow 
            to determine fourth family quarks masses more accurately.\\ 
vi.$\,\,$ It is difficult to predict the interval for the masses of the fourth family members since the 
            choice of the fourth family quark masses does not 
            appreciably influence either the fitting procedure or the obtained $3 \times 3$ mixing 
            submatrix, and also not the accuracy of the masses of the three lower families. 
            Other experimental data, like decays of mesons, speaks for masses of the fourth family 
            quarks close or above $1$ TeV.\\
vii. $\,\,$ For the masses of the fourth family members above $1$ TeV the mass matrices are 
           close to the democratic matrix: The matrix elements are closer to one another the higher 
           is the mass of the fourth family member. In such a case are the fourth family masses mostly
           determined by the scalars carrying the family quantum numbers. Correspondingly  the
           masses of the $u_{4}$-quarks are closer to the   masses of the $d_{4}$-quarks.

The complex mass matrices would lead to unitary and not to orthogonal mixing matrices.
The more accurate experimental data for quarks mixing matrix would allow us to extract also 
the phases of the unitary mixing matrix, allowing us to predict the fourth family masses.

\vspace{2mm}

\subsection{Is the existence of the fourth family in agreement with the present experiments?}
\label{fourthfamily}

\vspace{2mm}

This part is following equivalent part in my contribution to the Proceedings to  the Conference on 
New Physics at the Large Hadron Collider, $29$ February -- $4$ March, {\bf 2016}, Nanyang Executive 
Centre, NTU, Singapore. The paper, a more elaborated version of this notice, is in preparation.

The {\it spin-charge-family} theory predicts the existence of the fourth family to the observed three, 
while there has been no direct observation of the fourth family quarks with the masses below $1$ TeV. 
The fourth family quarks with masses above $1$ TeV contribute according to the {\it standard model} 
(the {\it standard model} Yukawa couplings of quarks to the scalar Higgs are proportional to  
$\frac{m^{\alpha}_{4}}{v}$, where $m^{\alpha}_{4}$ is the fourth family member 
($\alpha=u,d$) mass and $v$ the vacuum expectation value of the scalar) to either the 
quark-gluon fusion production of the scalar field (the Higgs) or to the scalar field decay into two 
photons $\approx 10$ times too much in comparison with the observations. Correspondingly 
the high energy physicists do not expect the existence of the fourth family members 
at all~\cite{AhmedAliMatthiasNeubert}.

I am stressing~\cite{norma2016higgs} in this subsection that the $u_i$-quarks and $d_i$-quarks 
of an $i^{th}$ family, if they couple with the opposite sign (with respect to the "${}_{\pm}$" 
degree of freedom)   to the scalar fields carrying the family ($\tilde{A}, i$) quantum numbers%
~\footnote{The contribution of the scalar fields $\tilde{A}^{\tilde{A}i}_{\pm}\,, \,(\tilde{A},i)=
(\tilde{\tau}^{1i}, \tilde{N}^{i}_{L}$), Eq.~\ref{commonAi}), are the same for all the family 
members.} and  have the same masses, do not contribute to either the quark-gluon fusion production 
of the scalar fields with the family quantum numbers or to the decay of these scalars into two 
photons. 
Since the $u_4$-quarks and $d_4$-quarks might have similar masses (Subsect.~\ref{massmatrices})
and since their masses are for $m_{u_{4}} > 1$ TeV and  $m_{d_{4}} > 1$ TeV mostly determined
by the scalars with the family quantum numbers, the observations so far are consequently not in 
contradiction with the {\it spin-charge-family} theory prediction that there exists the fourth family
coupled to the observed three.

The  couplings of $u_i$ and $d_i$ to the scalars carrying the family members quantum numbers 
are determined besides by the corresponding couplings also by the eigenvalues of the operators 
$(Q,Q',Y')$ on the quarks states (which do distinguish between $u_i$ and $d_i$). 

The strong influence of the scalar fields carrying the family members quantum numbers to the 
masses of the lower (observed) three families manifests in the huge differences in the  masses 
of $u_i$ and $d_i$, $i=(1,2,3)$, among families ($i$) and family members ($u,d$). For the fourth
family quarks, which are more and more decoupled from the observed three families the higher 
are their masses~\cite{gn2015,gn2013}, the influence of the scalar fields carrying the family 
members quantum numbers on their masses is expected to be much weaker. 
Correspondingly the $u_4$ and $d_4$ masses become closer to each other the higher are their 
masses and the weaker are their couplings (the mixing matrix elements) to the lower three families.

If the masses of the fourth family quarks are close to each other, then $u_{4}$ and 
$d_{4}$ contribute in the quark-gluon fusion  very little to the production of the scalar field - the 
Higgs - which is mostly  superposition of the scalar fields with the family members quantum numbers,
what is in agreement with the observation: In the quark-gluon fusion production of the Higgs 
mostly the top ($u_3$) contributes.

In Tables~\ref{Table so13+1.}--\ref{Table so13+1.b}  the phases of all the states are chosen to 
be $1$. Here I use different phases, those 
which enable the usual
presentation of fermions under the change of spin and under 
$\mathbb{C}_{{\cal N}}$ $\cdot {\cal P}_{{\cal N}}$.

\begin{figure}
\centering
\subfigure[]{
\includegraphics{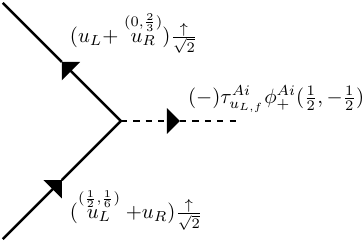}} \quad
\subfigure[]{
\includegraphics{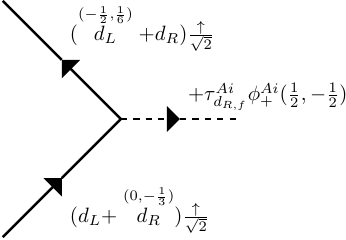}}\\
\vspace{4mm}
\subfigure[]{\includegraphics{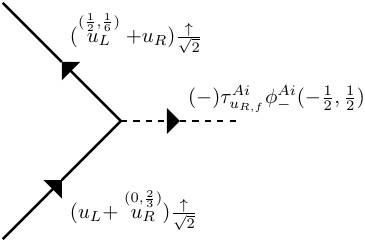}}\quad
\subfigure[]{\includegraphics{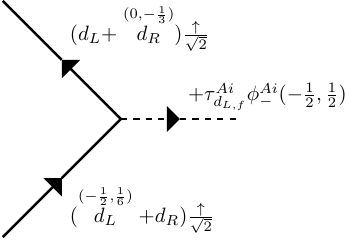}}
\caption{\label{Fig.1} %
The contributions of $u$ and $d$ quarks to the production of the scalar
fields $\Phi^{Ai}_{-}$ and  $\Phi^{Ai}_{+}$, when $\tau^{Ai}$ represent the family 
quantum numbers (which are for the lower four families $\tilde{\tau}^{1i}$ and 
$\tilde{N}^{i}_{L}$) or the family members quantum numbers ($Q, Q', Y'$), are presented:
 (a) the $u$-quark contribution to the scalar fields $\Phi^{Ai}_{+}$,
 (b) the $d$-quark contribution to $\Phi^{Ai}_{+}$,
 (c) the $u$-quark contribution to $\Phi^{Ai}_{-}$,
 (d) the $d$-quark contribution to  $\Phi^{Ai}_{-}$. 
$\tau^{Ai}_{u_{(L,R)},f}$ and $\tau^{Ai}_{d_{(L,R)},f}$ denote the application values of the 
operators $\tilde{\tau}^{1i}$, $\tilde{N}^{i}_{L}$ and of  ($Q, Q', Y'$) on the states. While 
$\tilde{\tau}^{1i}$ and $\tilde{N}^{i}_{L}$ do not distinguish among family members $u$ and
$d$ so that in this case the contribution of $u$ and $d$ have opposite signs, ($Q, Q', Y'$) do,
 influencing the signs and the values. 
}
\end{figure}

In Table~\ref{udprop.} the properties of $u$ and $d$ quarks (and their antiquarks) 
 needed in  Fig.~\ref{Fig.1}  are presented. 
\begin{table}
\caption{
The weak, hyper and electromagnetic charges for quarks 
 in their massless basis are presented. The colour charge is not shown. These and other properties
 of quarks  and 
 leptons 
can be read from Tables~\ref{Table so13+1.}--\ref{Table so13+1.b}. }
 \begin{center}
{ \begin{tabular}{c| r r r} 
 \hline
 state & $ {\cal \tau}^{13}$ & $Y$ & $Q$ \\
 \hline
 $        u_{Ri}$ & $ 0 $              & $\frac{2}{3}$  & $\frac{2}{3}$ \\ 
 $        u_{Li}$  & $ \frac{1}{2}$ & $ \frac{1}{6}$  & $\frac{2}{3}$\\ 
 $        d_{Ri}$  & $ 0 $             & $-\frac{1}{3}$  & $-\frac{1}{3}$ \\
 $        d_{Li}$  & $-\frac{1}{2}$ & $ \frac{1}{6}$  & $-\frac{1}{3}$ \\
\hline
\end{tabular}
}
 \end{center}
\label{udprop.}
 \end{table}
 Fig.~\ref{Fig.1} presents the properties of  $u$ and $d$ quarks, contributing in the quark-gluon fusion to 
the production of the scalar fields 
$\Phi^{Ai}_{\stackrel{78}{(\pm)}}$,~Eq.(\ref{commonAi}), ($A^{Ai}_{\stackrel{78}{(\pm)}}$
$=$ $\Phi^{Ai}_{\stackrel{78}{(\pm)}}$ +$v^{Ai}_{\stackrel{78}{(\pm)}}$ ).  
One notices the opposite signs of the couplings of $u_{i}$ with respect to $d_{i}$ for ether
$\Phi_{-}$ or for $\Phi_{+}$. Correspondingly the fourth family quarks of almost the same mass 
contribute  very little to the production of any scalar, with the scalar Higgs included, which is in agreement 
with the observation. Also to the decay of the Higgs  can the fourth family quarks 
contribute very little: It is, as observed, the $u_{3}$- quark (the top), which contributes  the most to
 either the production or to the decay of the Higgs.

The fourth family quarks can still contribute to the production and to the decay of the scalars of
the masses of a few TeV, to which they couple stronger than to the Higgs. 

%

The figures are valid for any $Ai$ and correspondingly also for any superposition of 
$\Phi^{Ai}_{\pm}$.\\ 


\vspace{2mm}

\subsection{Anomaly cancellation in the {\it spin-charge-family} theory~\footnote{A more
 elaborated version of this report is in preparation.}}
\label{anomaly}

\vspace{2mm}

In the {\it standard model} the triangle anomalies  "miraculously" disappear due to the fact that the sum 
of all possible traces $Tr [\tau^{Ai}$$ \tau^{Bj}$$ \tau^{Ck}]$, where ($\tau^{Ai}, \tau^{Bi}, 
\tau^{Ck}$) are the generators of one, of two or of three of the groups of $SU(3), SU(2)$ and $U(1)$)
 over the representations of one family of the left handed fermions and their anti-fermions (and 
separately of the right handed fermions and their anti-fermions), contributing to  the triangle currents, 
are equal to zero~\cite{Alvarez,AlvarezWitten,Bilal,AlvarezBondiaMartin}.  

Let me demonstrate that this cancellation of the {\it standard model} triangle anomaly follows 
straightforwardly, if the $SO(3,1), SU(2), U(1)$ and $ SU(3)$  are considered as 
subgroups of the orthogonal group $SO(13,1)$. 

To the  triangle anomaly the right-handed spinors (fermions) and antispinors contribute with the opposite 
sign than the left handed spinors and their antispinors. Their common contribution to anomalies is
proportional to~\cite{Bilal} 
\begin{eqnarray}
\label{anomaly0}
(\sum_{(A,i,B,j,C,k)_{L \,\bar{L}}} Tr [\tau^{Ai} \, \tau^{Bj} \,\tau^{Ck}] -
 \sum_{(A,i,B,j,C,k)_{R \,\bar{R}}} Tr [\tau^{Ai} \, \tau^{Bj} \,\tau^{Ck}]\,) \,,
\end{eqnarray}
where $\tau^{Ai}$ are in the {\it standard model} the generators of the infinitesimal transformation 
of the groups $SU(3), SU(2)$ and $U(1)$, while in the {\it spin-charge-family} theory
$\tau^{Ai}$ are irreducible subgroups of the starting orthogonal group $SO(2(2n+1)-1,1)$, $n=3$. 
The indexes ${}_{L \,\bar{L}}$  (${}_{R \,\bar{R}}$) denote
the left (right) handed  spinors  and their antispinors (right (left)), respectively.

In the first seven columns (up  to $||$) of Table~\ref{Table SMandSCFTspinors.} the 
properties of one family members assumed by the {\it standard 
model}, running in the triangle, are presented. 
The last two columns -  taken from Table~\ref{Table so13+1.}  - describe additional properties which 
quarks and leptons (and antiquarks and antileptons) would have, if the {\it standard model} groups
 $SO(3,1), SU(2), SU(3)$ and $U(1)$ are embedded into the $SO(13,1)$ group. 
To demonstrate that the "miraculous" cancellation of the triangle anomalies 
is  "trivial", one takes into account that the {\it standard model} groups can easily be interpreted 
(unified) by making the next step beyond the {\it standard model}.
\begin{table}
{\tiny%
\begin{center}
\caption{\label{Table SMandSCFTspinors.} Properties  of the left handed quarks and leptons and 
antiqurks and antileptons, and of the right handed quarks and leptons and antiquarks and antileptons,
as assumed by the {\it standard model}, are presented in the first seven columns. In the last two 
columns the two 
quantum numbers are added, which the fermions and anti-fermions would have if the {\it standard 
model} groups $SO(3,1), SU(2), SU(3)$ and $U(1)$ are embedded into the $SO(13,1)$ group. 
The whole quark part appears, due to the colour charges, three times. One can check that the 
hyper charge is the sum of $\tau^{4}_{i_{L,R}} + \tau^{23}_{i_{L,R}}$ (Table~\ref{Table so13+1.}).
The quantum numbers are the same for all the families.
}
\begin{tabular}{|r r | c r r r r r ||r r|}
\hline
           & & hand-   & weak    & hyper  & colour&charge & elm   & $SU(2)_{II}$&$U(1)_{II}$\\
            &&edness  & charge  & charge &        &          &charge&     charge     &  charge     \\
$i_{L}$&name    &$ \Gamma^{(3,1)}$&$ \tau^{13}$  & $ Y$ & $\tau^{33}$&$\tau^{38}$
&$Q$&$\tau^{23}$&$\tau^{4}$\\
\hline
$1_{L}$&$ u_{L} $&$ -1     $&$\frac{1}{2}$ &$ \frac{1}{6}$&$\frac{1}{2} $&$\frac{1}{2\sqrt{3}}$&
$ \frac{2}{3}$&0&$\frac{1}{6}$\\
$2_{L}$&$ d_{L} $&$ -1    $&$ -\frac{1}{2}$ &$ \frac{1}{6}$&$\frac{1}{2}$&$\frac{1}{2\sqrt{3}}$ &
$-\frac{1}{3}$&0&$\frac{1}{6}$\\
$3_{L}$&$ u_{L} $&$ -1     $&$\frac{1}{2}$ &$ \frac{1}{6}$&$-\frac{1}{2} $&$\frac{1}{2\sqrt{3}}$&
$ \frac{2}{3}$&0&$\frac{1}{6}$\\
$4_{L}$&$ d_{L} $&$ -1    $&$ -\frac{1}{2}$ &$ \frac{1}{6}$&$-\frac{1}{2}$&$\frac{1}{2\sqrt{3}}$ &
$-\frac{1}{3}$&0&$\frac{1}{6}$\\
$5_{L}$&$ u_{L} $&$ -1     $&$\frac{1}{2}$ &$ \frac{1}{6}$&$ 0 $&$-\frac{1}{\sqrt{3}}$&
$ \frac{2}{3}$&0&$\frac{1}{6}$\\
$6_{L}$&$ d_{L} $&$ -1    $&$ -\frac{1}{2}$ &$ \frac{1}{6}$&$ 0$&$-\frac{1}{\sqrt{3}}$ &
$-\frac{1}{3}$&0&$\frac{1}{6}$\\
\hline
$7_{L}$&$\nu_{L} $&$ -1  $&$ \frac{1}{2}$&$ -\frac{1}{2}$& 0&0&  0  &$0$&$ -\frac{1}{2}$  \\
$8_{L}$&$ e^{L}  $&$ -1  $&$-\frac{1}{2}$&$ -\frac{1}{2}$& 0&0&$-1$&$0$&$ -\frac{1}{2}$ \\
\hline \hline
$9_{L}$&$ \bar{u}{L} $&$ -1$&$  0 $ &$ -\frac{2}{3}$& $ - \frac{1}{2}$ &$-\frac{1}{2\sqrt{3}}$
&$- \frac{2}{3}$ &$ - \frac{1}{2}$ &$ - \frac{1}{6}$\\
$10_{L}$&$ \bar{d}{L} $&$ -1$&$  0 $ &$  \frac{1}{3}$& $ - \frac{1}{2}$ &$-\frac{1}{2\sqrt{3}}$
&$  \frac{1}{3}$ &$   \frac{1}{2}$ &$ - \frac{1}{6}$\\
$11_{L}$&$ \bar{u}{L} $&$ -1$&$  0 $ &$ -\frac{2}{3}$& $  \frac{1}{2}$ &$-\frac{1}{2\sqrt{3}}$
&$- \frac{2}{3}$ &$ - \frac{1}{2}$ &$ - \frac{1}{6}$\\
$12_{L}$&$ \bar{d}{L} $&$ -1$&$  0 $ &$  \frac{1}{3}$& $  \frac{1}{2}$ &$-\frac{1}{2\sqrt{3}}$
&$  \frac{1}{3}$ &$   \frac{1}{2}$ &$ - \frac{1}{6}$\\
$13_{L}$&$ \bar{u}{L} $&$ -1$&$  0 $ &$ -\frac{2}{3}$& $ 0 $&$ \frac{1}{\sqrt{3}}$
&$- \frac{2}{3}$ &$ - \frac{1}{2}$ &$ - \frac{1}{6}$\\
$14_{L}$&$ \bar{d}{L} $&$ -1$&$  0 $ &$  \frac{1}{3}$& $ 0 $&$\frac{1}{\sqrt{3}}$
&$  \frac{1}{3}$ &$   \frac{1}{2}$ &$ - \frac{1}{6}$\\
\hline
$15_{L}$&$\bar{\nu}_{L}$&$ -1$&$ 0 $&$ 0 $& 0&0&$0$&$- \frac{1}{2}$&$ \frac{1}{2}$  \\
$16_{L}$&$\bar{e}_{L}   $&$ -1$&$ 0 $&$ 1 $& 0&0&$1$&$  \frac{1}{2}$&$ \frac{1}{2}$  \\
\hline \hline\hline
$1_{R}$&$u_{R}   $&$ 1$& $0$ & $  \frac{2}{3}$&$ \frac{1}{2}$ &$\frac{1}{2\sqrt{3}}$
&$ \frac{ 2}{3}$&$ \frac{1}{2}$&$ \frac{1}{6}$ \\
$2_{R}$&$d_{R}   $&$ 1$& $0$ & $ -\frac{1}{3}$&$ \frac{1}{2}$ &$\frac{1}{2\sqrt{3}}$ 
&$-\frac{1}{3}$&$- \frac{1}{2}$&$ \frac{1}{6}$  \\
$3_{R}$&$u_{R}   $&$ 1$& $0$ & $  \frac{2}{3}$&$- \frac{1}{2}$ &$\frac{1}{2\sqrt{3}}$
&$ \frac{ 2}{3}$&$ \frac{1}{2}$&$ \frac{1}{6}$ \\
$4_{R}$&$d_{R}   $&$ 1$& $0$ & $ -\frac{1}{3}$&$- \frac{1}{2}$ &$\frac{1}{2\sqrt{3}}$ 
&$-\frac{1}{3}$&$- \frac{1}{2}$&$ \frac{1}{6}$  \\
$5_{R}$&$u_{R}   $&$ 1$& $0$ & $  \frac{2}{3}$&$ 0  $ &$-\frac{1}{\sqrt{3}}$
&$ \frac{ 2}{3}$&$ \frac{1}{2}$&$ \frac{1}{6}$ \\
$6_{R}$&$d_{R}   $&$ 1$& $0$ & $ -\frac{1}{3}$&$ 0  $ &$-\frac{1}{\sqrt{3}}$ 
&$-\frac{1}{3}$&$- \frac{1}{2}$&$ \frac{1}{6}$  \\
\hline
$7_{R}$&$\nu_{R}$&$ 1$& $0$  & $ 0$&0&0&   0 &$  \frac{1}{2}$&$- \frac{1}{2}$           \\
$8_{R}$&$ e_{R}  $&$ 1$& $0$  & $-1$&0&0&$-1$&$- \frac{1}{2}$&$- \frac{1}{2}$  \\
\hline\hline
$9_{R}$&$\bar{u}_{R}   $&$ 1$& $ -\frac{1}{2}$ & $ - \frac{1}{6}$&$ -\frac{1}{2}$ &$-\frac{1}{2\sqrt{3}}$
&$ - \frac{ 2}{3}$&0& $ - \frac{1}{6}$\\
$10_{R}$&$\bar{d}_{R}   $&$ 1$&$  \frac{1}{2}$ & $  - \frac{1}{6}$&$ -\frac{1}{2}$ &$-\frac{1}{2\sqrt{3}}$
&$    \frac{1}{3}$&0& $ - \frac{1}{6}$\\
$11_{R}$&$\bar{u}_{R}   $&$ 1$& $ -\frac{1}{2}$ & $ -  \frac{1}{6}$&$ \frac{1}{2}$ &$-\frac{1}{2\sqrt{3}}$
&$ - \frac{ 2}{3}$&0& $ - \frac{1}{6}$\\
$12_{R}$&$\bar{d}_{R}   $&$ 1$&$  \frac{1}{2}$ & $ -  \frac{1}{6}$&$  \frac{1}{2}$ &$-\frac{1}{2\sqrt{3}}$
&$    \frac{1}{3}$&0& $ - \frac{1}{6}$\\
$13_{R}$&$\bar{u}_{R}   $&$ 1$& $ -\frac{1}{2}$ & $ -  \frac{1}{6}$&$ 0 $ &$ \frac{1}{\sqrt{3}}$
&$ - \frac{ 2}{3}$&0& $ - \frac{1}{6}$\\
$14_{R}$&$\bar{d}_{R}   $&$ 1$&$  \frac{1}{2}$ & $  -  \frac{1}{6}$&$ 0 $ &$ \frac{1}{\sqrt{3}}$
&$    \frac{1}{3}$&0& $ - \frac{1}{6}$\\
\hline
$15_{R}$&$\bar{\nu}_{R}$&$ 1$& $ -\frac{1}{2}$ & $  \frac{1}{2}$&0&0&$0$&0&$\frac{1}{2}$        \\
$16_{R}$&$\bar{e}_{R}   $&$ 1$& $  \frac{1}{2}$ & $  \frac{1}{2}$&0&0&$1$&0&$\frac{1}{2}$ \\
\hline\hline
\end{tabular}
  \end{center}
%
}
%
\end{table}
In the {\it standard model} the triangle anomaly   for the Feynman triangle diagrams, in which the 
gauge vector fields of the charges 
\begin{eqnarray}
\label{anomalyparticular}
&& U(1)\times U(1) \times U(1)\, ,\nonumber\\ 
&& SU(2)\times SU(2) \times U(1)\, ,\nonumber\\
&& SU(3)\times SU(3) \times SU(3)\, ,\nonumber\\ 
&& SU(3)\times SU(3) \times U(1)\, ,\nonumber\\  
&& U(1) \times {\rm gravitational}\,  
\end{eqnarray}
contribute to the triangle anomaly,  occurs if the traces in Eq.(\ref{anomaly0}) are not
zero for either the left handed quarks and leptons and their left handed antiparticles or the right handed
quarks and leptons and their right handed antiparticles. 

To see that embedding  the {\it standard model} groups into the orthogonal group $SO(13,1)$ makes
the cancellation of the triangle anomalies self evident, let us recognize:
The subgroups of the 
$SO(13,1)$ group are $SO(7,1) $ and  $SO(6)$. The subgroups of  $SO(6)$ are the colour group 
$SU(3)$ with the generators denoted by $\tau^{3i}, \, i=(1,\dots,8)$ and the $U(1)$
 (we shall call it $U(1)_{II}$) 
group with the generator $\tau^{4}$. One sees that all the quarks have $\tau^{4}=\frac{1}{6}$, 
all the antiquarks have $\tau^{4}=-\frac{1}{6}$, while the leptons have $\tau^{4}=- \frac{1}{2}$ and
the antileptons have $\tau^{4}= \frac{1}{2}$. Correspondingly the trace of $\tau^{4}$ over all the 
family members is equal to zero.

The subgroups of the $SO(7,1)$, as seen in Table~\ref{Table so13+1.}, have as subgroups $SO(3,1)$, 
$SU(2)_{I}$ and $SU{2}_{II}$, with the generators $\tau^{1i}$ (representing the weak group
 operators) and $\tau^{2i}$ (representing the generators of the additional $SU(2)$ group), respectively. 
The left handed spinors  are $ SU(2)_{I}$ (weak) doublets and 
 $SU(2)_{II}$ singlets, while the right  handed spinors  are the $ SU(2)_{I}$ (weak) singlets and 
 $SU(2)_{II}$ doublets. Correspondingly are the left handed antispinors  the $ SU(2)_{I}$ (weak) singlets 
and  $SU(2)_{II}$ doublets, while the right  handed antispinors  are the $ SU(2)_{I}$ (weak) doublets 
and the $SU(2)_{II}$ singlets.  

The hypercharge of the {\it standard model}  corresponds to the sum of  $\tau^{4}$ and $\tau^{23}$
\begin{eqnarray}
\label{Y}
Y&=& \tau^{4} + \tau^{23} \,.
\end{eqnarray}

To the triangle Feynman diagram, to which three hyper $U(1)$ boson fields contribute, the 
sum  $\sum_{i} Tr (Y_{i})^3$ runs over all the members ($i$) of the left handed 
spinors and antispinors, and of the right handed spinors and antispinors separately. 
Embedding the {\it standard model} groups into $SO(13,1)$ it follows
\begin{eqnarray}
\label{3Ytriangle}
\sum_{i_{L,R}}\, (Y_{i_{L,R}})^3&=&\sum_{i_{L,R}}\,
 (\tau^{4}_{i_{L,R}} + \tau^{23}_{i_{L,R}})^{3}  \nonumber\\
                           &=&\sum_{i_{L,R}}\, (\tau^{4}_{i_{L,R}})^{3} + 
                                 \sum_{i_{L,R}}\, (\tau^{23}_{i_{L,R}})^{3}\nonumber\\
                           &+&\sum_{i_{L,R}}\, 3 \cdot (\tau^{4}_{i_{L,R}})^{2}  \cdot 
\tau^{23}_{i_{L,R}}
                             +  \sum_{i_{L,R}}\, 3  \cdot \tau^{4}_{i_{L,R}}  \cdot 
 (\tau^{23}_{i_{L,R}})^{2}\,,
\end{eqnarray}
for either the left, $i_{L}$, or the right, $i_{R}$, handed members.
Table~\ref{Table SMandSCFTspinors.}  demonstrates  clearly (last column) that $ \sum_{i_{L}}\,
 (\tau^{4}_{i_{L}})^{3} =0 $, when the contribution of  left (right) handed spinors and antispinors
 are taken into account.

Table~\ref{Table SMandSCFTspinors.}  also demonstrates  (the last but one column) even more trivially
 that $\sum_{i_{L}}\, 3  \cdot (\tau^{4}_{i_{L}})^{2}  \cdot \tau^{23}_{i_{L}} =0$,  
since the contribution of either spinors or antispinors - left or right handed - 
separately are equal to zero.

The easiest is to evaluate $ \sum_{i_{L,R}}\, (\tau^{23}_{i_{L}})^{3} =0 $ and $\sum_{i_{L}}\, 3 . 
(\tau^{4}_{i_{L}})^{2}  \cdot \tau^{23}_{i_{L}} =0$ since, as seen from  
Table~\ref{Table SMandSCFTspinors.},  the summation separately within the quarks and lepton
representations  give zero.

Since all the members belong to one spinor representation, it is straightforwardly that all the triangle 
traces are zero, if the {\it standard model}  groups are the subgroups of the orthogonal group 
$SO(13,1)$. 

From only the {\it standard model} assumptions point of view the cancellation of the triangle anomalies
does look  miraculously. For our $\sum_{i_{L,R}}\, (Y_{i_{L,R}})^3$ one obtains for the left handed
members: [$3  \cdot 2   \cdot (\frac{1}{6})^3 + 2  \cdot (-\frac{1}{2})^3 + 3  \cdot 
((-\frac{2}{3})^3 + (\frac{1}{3})^3) + 1^3)$ ], and for the right handed members:  
[$3  \cdot ((\frac{2}{3})^3 +  (-\frac{1}{3})^3) +
(- 1)^3) +3  \cdot 2  \cdot (-\frac{1}{6})^3 + 2  \cdot (\frac{1}{2})^3 $ ].

\vspace{2mm}

\subsection{Dark matter in the {\it spin-charge-family} theory}
\label{darkmatter}

\vspace{2mm}

As discussed in Sect.~\ref{SCFT} the {\it spin-charge-family} theory~\cite{NBled2013,NBled2012,%
norma92,norma93,norma94,pikanorma,portoroz03,JMP,norma95,gmdn07,gn,gn2013,gn2015,NPLB,%
N2014scalarprop,norma2014MatterAntimatter,JMP2015} predicts in the low energy region two 
decoupled groups of four families. In Ref.~\cite{gn} the possibility that the dark matter consists 
of clusters of the fifth family - the stable heavy family of quarks and leptons (with (amost) zero 
Yukawa couplings to the lower group of four families) - is discussed. 

I review here briefly the estimation done in Ref.~\cite{gn}.

We used in Ref.~\cite{gn} the simple hydrogen-like model to evaluate properties of the fifth family
heavy baryons, taking into account that for masses of the order of a few TeV or larger the one
gluon exchange determines the force among the constituents of the fifth family baryons (\cite{gn}, 
Sect.~II).   We estimated the fifth family neutron as the most stable nucleon.

Due to very large masses  of  the fifth family baryons "the nuclear interaction" among  these baryons 
has very interesting properties. 

We also estimated  the behaviour of the neutral clusters when scattering among themselves and with 
the ordinary matter. We studied possible limitations on the family properties due to the cosmological 
evidences, the direct experimental evidences (\cite{gn}, Sect.~IV) and all 
others known properties of the dark matter.

We followed  the behaviour of the fifth family quarks  and antiquarks in the
 plasma of the expanding universe, through the freezing out procedure, solving the Boltzmann 
equations, through the colour phase  transition, while forming neutrons, up to the present dark 
matter (\cite{gn}, Sect.~III).  

The cosmological evolution suggested the limits for the masses of the fifth family quarks
\begin{eqnarray}
\label{massrange}
 10 \; {\rm TeV}  < m_{q_5} \, c^2 < {\rm a \, few} \cdot 10^2\, {\rm TeV} 
\end{eqnarray}
and for the  scattering cross sections
\begin{eqnarray}
\label{scatteringrange} 
 10^{-8}\, {\rm fm}^2\, < \sigma_{c_5}\, <   10^{-6} \,{\rm fm}^2 \,, 
\end{eqnarray} 
while the measured density of  the  dark matter does not put much limitation on the properties of 
heavy enough clusters.  

The direct measurements limited the fifth family quark mass to (\cite{gn}, Sect.~IV.)
\begin{eqnarray}
\label{direct}  {\rm several}\, 10 \,{\rm TeV} < m_{q_{5}}c^2 < 10^5\, {\rm TeV}\,. 
\end{eqnarray} 

We also find that our fifth family baryons of the mass of a few $10$ TeV/${c^2} $ 
have  for a factor more than $100$ times too small scattering amplitude with the ordinary matter
to cause a measurable heat flux on the Earth's surface.

\vspace{2mm}

\subsection{Matter-antimatter in the {\it spin-charge-family} theory}
\label{matterantimatter}

\vspace{2mm}

I shortly overview in this section the properties, quantum numbers, and  discrete symmetries 
of those scalar and vector gauge fields appearing in the starting action 
(Eqs.~(\ref{wholeaction}, \ref{faction})
which cause transitions of antileptons into quarks and back, and antiquarks into quarks and 
back. The appearance of the condensate breaks this symmetry making possible under non 
thermal conditions the ordinary (mostly made of the first family members) matter-antimatter
asymmetry. The reader can find details in Ref.~\cite{norma2014MatterAntimatter}.

Scalar gauge fields, contributing to matter-antimatter asymmetry and causing also the proton 
decay,  carry the triplet or antitriplet colour charges (see Table~\ref{Table bosons.}) and the 
fractional hyper and electromagnetic charge.

The Lagrange densities from Eqs.~(\ref{wholeaction}, \ref{faction}) manifest 
$\mathbb{C}_{{ \cal N}} \cdot {\cal P}_{{\cal N}}$ invariance~\cite{HNds}.   
All the vector and the spinor gauge fields are massless before the appearance of the condensate 
(Subsect.~\ref{condensate}) 
and  reactions creating particles from antiparticles and back go in both directions equivalently. 
Correspondingly there is no matter-antimatter asymmetry. It is the condensate, which breaks
this symmetry.

Let me analyze the Lagrange density of Eq.~(\ref{faction}) before the appearance of the condensate. 
The term $\gamma^{t}\, \frac{1}{2}\, S^{s' s"} \,\omega_{s' s" t}$  can 
be rewritten as follows
\begin{eqnarray}
\gamma^{t}\,\frac{1}{2}\, S^{s' s"} \,\omega_{s' s" t} &=&
\sum_{+,-}\,\sum_{(t\,t')} \,\,\stackrel{t t'}{(\cpm)}\,\frac{1}{2}\, S^{s' s"} \, 
\omega_{\scriptscriptstyle{s" s" \stackrel{t t'}{(\cpm)}}}\,,\nonumber\\
\omega_{\scriptscriptstyle{s" s" \stackrel{t t'}{(\cpm)}}}: &=& 
\omega_{\scriptscriptstyle{s" s" \stackrel{t t'}{(\pm)}}} = 
(\omega_{s' s" t}\,\mp \,i \,\omega_{s' s" t'})\,, \nonumber\\ 
\stackrel{t t'}{(\cpm)}: &=& \stackrel{t t'}{(\pm)} = \frac{1}{2}\, 
(\gamma^{t} \pm \gamma^{t'})\,, \nonumber\\
(t\,t') &\in& ((9\,10), (11\,12),(13\,14))\,. 
\label{factionMaMpart10}
\end{eqnarray}
I introduced the notations $\stackrel{t t'}{(\cpm)}$ and  
$\omega_{\scriptscriptstyle{s" s" \stackrel{t t'}{(\cpm)}}}$ to distinguish among different superposition 
of states in equations below.

 The expression  $\stackrel{t t'}{(\cpm)}\, \frac{1}{2}\, S^{s' s"} \,$  
$\omega_{\scriptscriptstyle{s" s" \,\stackrel{t t'}{(\cpm)}}}$ can be further rewritten 
as follows
\begin{eqnarray}
&&\stackrel{t t'}{(\cpm)}\,\frac{1}{2}\, S^{s' s"} \,  
\omega_{\scriptscriptstyle{s" s" \,\stackrel{t t'}{(\cpm)}}}=\nonumber\\
&& \stackrel{t t'}{(\cpm)}\,\{\, \tau^{2+}\,A^{2+}_{\scriptscriptstyle{\stackrel{t t'}{(\cpm)}}} 
+ \tau^{2-}\,A^{2-}_{\scriptscriptstyle{\stackrel{t t'}{(\cpm)}}} + \tau^{23}\,
A^{23}_{\scriptscriptstyle{\stackrel{t t'}{(\cpm)}}}
+ \tau^{1+}\,A^{1+}_{\scriptscriptstyle{\stackrel{t t'}{(\cpm)}}} + 
\tau^{1-}\,A^{1-}_{\scriptscriptstyle{\stackrel{t t'}{(\cpm)}}} + \tau^{13}\,
A^{13}_{\scriptscriptstyle{\stackrel{t t'}{(\cpm)}}}\, \} \,,\nonumber\\
A^{2\spm}_{\scriptscriptstyle{\stackrel{t t'}{(\cpm)}}} &=& 
(\omega_{\scriptscriptstyle{58 \stackrel{t t'}{(\cpm)}}}+ 
\omega_{\scriptscriptstyle{67 \stackrel{t t'}{(\cpm)}}})\,\smp \, i 
(\omega_{\scriptscriptstyle{57 \stackrel{t t'}{(\cpm)}}}- 
\omega_{\scriptscriptstyle{68 \stackrel{t t'}{(\cpm)}}})\,,
\quad A^{23}_{\scriptscriptstyle{\stackrel{t t'}{(\cpm)}}}= 
(\omega_{\scriptscriptstyle{56 \stackrel{t t'}{(\cpm)}}}+ 
\omega_{\scriptscriptstyle{78 \stackrel{t t'}{(\cpm)}}})\,, 
\nonumber\\
A^{1\spm}_{\scriptscriptstyle{\stackrel{t t'}{(\cpm)}}} &=& 
(\omega_{\scriptscriptstyle{58 \stackrel{t t'}{(\cpm)}}}- 
\omega_{\scriptscriptstyle{67 \stackrel{t t'}{(\cpm)}}})\smp \, i 
(\omega_{\scriptscriptstyle{57 \stackrel{t t'}{(\cpm)}}}+ 
\omega_{\scriptscriptstyle{68 \stackrel{t t'}{(\cpm)}}})\,,\quad 
A^{13}_{\scriptscriptstyle{\stackrel{t t'}{(\cpm)}}}= 
(\omega_{\scriptscriptstyle{56 \stackrel{t t'}{(\cpm)}}}- 
\omega_{\scriptscriptstyle{78 \stackrel{t t'}{(\cpm)}}})\,.
\label{factionMaMpart11}
\end{eqnarray}
Equivalently one  expresses the term $\gamma^{t}\,$
$\frac{1}{2}\, \tilde{S}^{ab} \,\tilde{\omega}_{ab t}$ in Eq.~(\ref{faction}) with 
$\tilde{S}^{ab}$ as the infinitesimal generators of either $\widetilde{SO}(3,1)$ or 
$\widetilde{SO}(4)$ and $\tilde{\omega}_{ab t}$ belonging to the corresponding gauge fields
 with $t =(9,\dots,14)$, 
by using Eqs.~(\ref{so1+3tilde} - \ref{plusminus}), 
 as
\begin{eqnarray}
&&\gamma^{t} \frac{1}{2}\, \tilde{S}^{ab} \,\tilde{\omega}_{ab t} = 
\stackrel{t t'}{(\cpm)} \, \frac{1}{2}\, \tilde{S}^{a b} \, 
\tilde{\omega}_{\scriptscriptstyle{ab \,\stackrel{t t'}{(\cpm)}}}= \nonumber\\
&&\stackrel{t t'}{(\cpm)} \,
\{\,\tilde{\tau}^{2+}\,\tilde{A}^{2+}_{\scriptscriptstyle{\stackrel{t t'}{(\cpm)}}} + 
\tilde{\tau}^{2-}\,\tilde{A}^{2-}_{\scriptscriptstyle{\stackrel{t t'}{(\cpm)}}} + 
\tilde{\tau}^{23}\,\tilde{A}^{23}_{\scriptscriptstyle{\stackrel{t t'}{(\cpm)}}} + \nonumber\\
& & \tilde{\tau}^{1+}\,\tilde{A}^{1+}_{\scriptscriptstyle{\stackrel{t t'}{(\cpm)}}} + 
\tilde{\tau}^{1-}\,\tilde{A}^{1-}_{\scriptscriptstyle{\stackrel{t t'}{(\cpm)}}} + 
\tilde{\tau}^{13}\,\tilde{A}^{13}_{\scriptscriptstyle{\stackrel{t t'}{(\cpm)}}} + \nonumber\\
& & \tilde{N}^{+}_{R}\,\tilde{A}^{N_{R}+}_{\scriptscriptstyle{\stackrel{t t'}{(\cpm)}}} + 
\tilde{N}^{-}_{R}\,\tilde{A}^{N_{R}-}_{\scriptscriptstyle{\stackrel{t t'}{(\cpm)}}} + 
\tilde{N}^{3}_{R}\,\tilde{A}^{N_{R}3}_{\scriptscriptstyle{\stackrel{t t'}{(\cpm)}}} + \nonumber\\
& &
\tilde{N}^{+}_{L}\,\tilde{A}^{N_{L}+}_{\scriptscriptstyle{\stackrel{t t'}{(\cpm)}}} +
\tilde{N}^{-}_{L}\,\tilde{A}^{N_{L}-}_{\scriptscriptstyle{\stackrel{t t'}{(\cpm)}}} + 
\tilde{N}^{3}_{L}\,\tilde{A}^{N_{L}3}_{\scriptscriptstyle{\stackrel{t t'}{(\cpm)}}}\,\}
\,,\nonumber\\ 
\tilde{A}^{N_{R}\spm}_{\scriptscriptstyle{\stackrel{t t'}{(\cpm)}}} &=& 
(\tilde{\omega}_{\scriptscriptstyle{23 \stackrel{t t'}{(\cpm)}}}- i\,
 \tilde{\omega}_{\scriptscriptstyle{01 \stackrel{t t'}{(\cpm)}}})\smp \, i 
(\tilde{\omega}_{\scriptscriptstyle{31 \stackrel{t t'}{(\cpm)}}}- i\,
 \tilde{\omega}_{\scriptscriptstyle{02 \stackrel{t t'}{(\cpm)}}})\,,\quad 
\tilde{A}^{N_{R}3}_{\scriptscriptstyle{\stackrel{t t'}{(\cpm)}}}= 
(\tilde{\omega}_{\scriptscriptstyle{12 \,\stackrel{t t'}{(\cpm)}}}- i\,
\tilde{\omega}_{\scriptscriptstyle{03 \stackrel{t t'}{(\cpm)}}})\,, \nonumber\\
\tilde{A}^{N_{L}\spm}_{\scriptscriptstyle{\stackrel{t t'}{(\cpm)}}} &=& 
(\tilde{\omega}_{\scriptscriptstyle{23 \stackrel{t t'}{(\cpm)}}}+ i\,
\tilde{\omega}_{\scriptscriptstyle{01 \stackrel{t t'}{(\cpm)}}})\smp \, i 
(\tilde{\omega}_{\scriptscriptstyle{31 \stackrel{t t'}{(\cpm)}}}+ i\,
 \tilde{\omega}_{\scriptscriptstyle{02 \stackrel{t t'}{(\cpm)}}})\,,\quad 
\tilde{A}^{N_{R}3}_{\scriptscriptstyle{\stackrel{t t'}{(\cpm)}}}= 
(\tilde{\omega}_{\scriptscriptstyle{12 \stackrel{t t'}{(\cpm)}}}+ i\,
 \tilde{\omega}_{\scriptscriptstyle{03 \stackrel{t t'}{(\cpm)}}})\,. 
\label{factionMaMpart20}
\end{eqnarray}
The expressions for $\tilde{A}^{2\spm}_{\scriptscriptstyle{\stackrel{t t'}{(\cpm)}}}$, 
$\tilde{A}^{23}_{\scriptscriptstyle{\stackrel{t t'}{(\cpm)}}}$, 
$\tilde{A}^{1\spm}_{\scriptscriptstyle{\stackrel{t t'}{(\cpm)}}}$ and 
$\tilde{A}^{1 3}_{\scriptscriptstyle{\stackrel{t t'}{(\cpm)}}}$ can easily be obtained 
from Eq.(\ref{factionMaMpart11}) by replacing in expressions  for 
$A^{2\spm}_{\scriptscriptstyle{\stackrel{t t'}{(\cpm)}}}$, 
$A^{23}_{\scriptscriptstyle{\stackrel{t t'}{(\cpm)}}}$, 
$A^{1\spm}_{\scriptscriptstyle{\stackrel{t t'}{(\cpm)}}}$ 
and $A^{1 3}_{\scriptscriptstyle{\stackrel{t t'}{(\cpm)}}}$, respectively, 
$\omega_{\scriptscriptstyle{s' s" \,\stackrel{t t'}{(\cpm)}}}$ by 
$\tilde{\omega}_{\scriptscriptstyle{s' s" \,\stackrel{t t'}{(\cpm)}}}$. 

The term $\gamma^t$ $\frac{1}{2}\,S^{t' t"}\,\omega_{t' t" t}$ in Eq.~(\ref{faction}) 
%
can be rewritten  with respect to the generators $S^{t' t"}$
and the corresponding gauge fields $\omega_{s' s" t}$  
as one colour octet scalar field and one $U(1)_{II}$ singlet scalar field (Eq.~\ref{so64})
\begin{eqnarray}
\gamma^{t}\,\frac{1}{2}\, S^{t" t'"} \,\omega_{t" t'" t} &=&
\sum_{+,-}\,\sum_{(t\,t')}  \, \stackrel{t t'}{(\cpm)}\, 
\{\,\vec{\tau}^{3}\cdot \vec{A}^{3}_{\scriptscriptstyle{\stackrel{t t'}{(\cpm)}}}\, + \tau^{4}\cdot
A^{4}_{\scriptscriptstyle{\stackrel{t t'}{(\cpm)}}}\,\}\,, \nonumber\\
(t\,t') &\in& ((9\,10), 11\,12),13\,14))\,. 
\label{factionMaMpart21}
\end{eqnarray}

 Considering all the above equations~(\ref{factionMaMpart10} - 
 \ref{factionMaMpart21}),  
and leaving out $p_{\scriptscriptstyle{\stackrel{t t'}{(\cpm)}}}$ 
since in the low energy limit the momentum does not play any role, the ${\mathcal L}_{f" }$  follows
\begin{eqnarray}
{\mathcal L}_{f" }&=&  \psi^{\dagger} \,\gamma^0 (-)\,\{  
\sum_{+,-}\,\sum_{(t\,t')} \,\stackrel{t t'}{(\cpm)}\,{\bf \cdot}\nonumber\\
&&[\,
\tau^{2+}\,A^{2+}_{\scriptscriptstyle{\stackrel{t t'}{(\cpm)}}} 
+ \tau^{2-}\,A^{2-}_{\scriptscriptstyle{\stackrel{t t'}{(\cpm)}}} + \tau^{23}\,
A^{23}_{\scriptscriptstyle{\stackrel{t t'}{(\cpm)}}}  \nonumber\\
&+&
\tau^{1+}\,A^{1+}_{\scriptscriptstyle{\stackrel{t t'}{(\cpm)}}} + 
\tau^{1-}\,A^{1-}_{\scriptscriptstyle{\stackrel{t t'}{(\cpm)}}} + \tau^{13}\,
A^{13}_{\scriptscriptstyle{\stackrel{t t'}{(\cpm)}}} \nonumber\\
&+& 
\tilde{\tau}^{2+}\,\tilde{A}^{2+}_{\scriptscriptstyle{\stackrel{t t'}{(\cpm)}}} + 
\tilde{\tau}^{2-}\,\tilde{A}^{2-}_{\scriptscriptstyle{\stackrel{t t'}{(\cpm)}}} + 
\tilde{\tau}^{23}\,\tilde{A}^{23}_{\scriptscriptstyle{\stackrel{t t'}{(\cpm)}}}\nonumber\\
&+& 
\tilde{\tau}^{1+}\,\tilde{A}^{1+}_{\scriptscriptstyle{\stackrel{t t'}{(\cpm)}}} + 
\tilde{\tau}^{1-}\,\tilde{A}^{1-}_{\scriptscriptstyle{\stackrel{t t'}{(\cpm)}}} + 
\tilde{\tau}^{13}\,\tilde{A}^{13}_{\scriptscriptstyle{\stackrel{t t'}{(\cpm)}}} \nonumber\\
&+&
\tilde{N}^{+}_{R}\,\tilde{A}^{N_{R}+}_{\scriptscriptstyle{\stackrel{t t'}{(\cpm)}}} + 
\tilde{N}^{-}_{R}\,\tilde{A}^{N_{R}-}_{\scriptscriptstyle{\stackrel{t t'}{(\cpm)}}} + 
\tilde{N}^{3}_{R}\,\tilde{A}^{N_{R}3}_{\scriptscriptstyle{\stackrel{t t'}{(\cpm)}}} \nonumber\\
&+&
\tilde{N}^{+}_{L}\,\tilde{A}^{N_{L}+}_{\scriptscriptstyle{\stackrel{t t'}{(\cpm)}}} +
\tilde{N}^{-}_{L}\,\tilde{A}^{N_{L}-}_{\scriptscriptstyle{\stackrel{t t'}{(\cpm)}}} + 
\tilde{N}^{3}_{L}\,\tilde{A}^{N_{L}3}_{\scriptscriptstyle{\stackrel{t t'}{(\cpm)}}} \nonumber\\
&+& 
\sum_{i} \tau^{3i}\,A^{3i}_{\scriptscriptstyle{\stackrel{t t'}{(\cpm)}}}\, + 
\tau^{4}\, A^{4}_{\scriptscriptstyle{\stackrel{t t'}{(\cpm)}}} \nonumber\\
&+& 
\sum_{i} \tilde{\tau}^{3i}\,\tilde{A}^{3i}_{\scriptscriptstyle{\stackrel{t t'}{(\cpm)}}}\, + 
\tilde{\tau}^{4}\, \tilde{A}^{4}_{\scriptscriptstyle{\stackrel{t t'}{(\cpm)}}} \,] 
%
%
\,\}\, \psi\,,
%
%
\label{factionMaM1}
\end{eqnarray}
where $(t,t')$ run in pairs over $[(9,10),\dots(13,14)]$ and the summation must go over $+$ and $-$ 
of ${}_{\scriptscriptstyle{\stackrel{t t'}{(\cpm)}}}$.

On Table~\ref{Table bosons.}, taken from Ref.~\cite{norma2014MatterAntimatter}, the quantum
numbers of the scalar and vector gauge fields, appearing in Eq.~(\ref{faction}), are presented, where
is taken into account that the spin of gauge fields is determined according to Eq.~(\ref{bosonspin}),
\begin{eqnarray}
\label{bosonspin}
(S^{ab})^{c}{}_{e} \,A^{d\dots e \dots g} &=& i\, (\eta^{ac}\delta^{b}_{e}- 
\eta^{bc}\delta^{a}_{e})\, A^{d\dots e \dots g}\,,
\end{eqnarray}
for each index ($ \in (d \dots g)$) of a bosonic field $A^{d\dots g}$ separately.  We must take 
into account also the relation among $S^{ab}$ and the charges (the relations are, of course, the 
same for bosons and fermions), presented in Eqs.~(\ref{so1+3}, \ref{so42}, \ref{so64})).
 \begin{table}
 \begin{tiny}
 \begin{center}
\caption{\label{Table bosons.}%
 Quantum numbers of the scalar gauge fields carrying the space index $t =(9,10,\cdots,14)$, 
 appearing in  Eq.~(\ref{faction}), are presented. The space degrees  of freedom contribute 
one of the triplets values  to the colour charge of all these scalar 
 fields. These scalars are  with respect to the two $SU(2)$ charges, ($\tau^{13}$  and 
$\vec{\tau}^2$), and the two  
 $\widetilde{SU}(2)$  charges, ($\vec{\tilde{\tau}}^1$  and $\vec{\tilde{\tau}}^2$), triplets 
 (that is in the adjoint representations of the  corresponding groups), and they all carry twice the 
 "spinor" number ($\tau^{4}$) of the quarks.  The quantum numbers of the two vector gauge fields, 
 the colour and the  $U(1)_{II}$ ones, are added. 
 }
 \begin{tabular}{|c|c|c|c|c|c|c|c|c|c|c|c|c|c|}
 \hline
 ${\rm field}$&prop. & $\tau^4$&$\tau^{13}$&$\tau^{23}$&($\tau^{33},\tau^{38}$)&$Y$&$Q$&$\tilde{\tau}^4$ 
 &$\tilde{\tau}^{13}$&$\tilde{\tau}^{23}$&$\tilde{N}_{L}^{3}$ &$\tilde{N}_{R}^{3}$  \\
 \hline
 $A^{1\spm}_{\scriptscriptstyle{\stackrel{9\,10}{(\cpm)}}}$& scalar&  $\cmp \frac{1}{3}$&$\spm 1$&$0$ 
 & ($\cpm\frac{1}{2},$ $\cpm \frac{1}{2\sqrt{3}}$)& $\cmp \frac{1}{3}$&$\cmp \frac{1}{3}+ \smp 1$&$0$
 &$0$&$0$&$0$&$0$\\ 
 $A^{13}_{\scriptscriptstyle{\stackrel{9\,10}{(\cpm)}}}$   & scalar&  $\cmp \frac{1}{3}$&$0$&$0$ 
 & ($\cpm\frac{1}{2},$ $\cpm \frac{1}{2\sqrt{3}}$)& $\cmp \frac{1}{3}$&$\cmp \frac{1}{3}$&$0$&$0$&$0$&$0$&$0$\\ 
 $A^{1\spm}_{\scriptscriptstyle{\stackrel{11\,12}{(\cpm)}}}$& scalar&  $\cmp \frac{1}{3}$&$\smp 1$&$0$ 
  & ($\cmp\frac{1}{2},$ $\cpm \frac{1}{2\sqrt{3}}$)& $\cmp \frac{1}{3}$&$\cmp \frac{1}{3}+ \smp 1$&$0$
  &$0$&$0$&$0$&$0$\\ 
  $A^{13}_{\scriptscriptstyle{\stackrel{11\,12}{(\cpm)}}}$   & scalar&  $\cmp \frac{1}{3}$&$0$&$0$ 
 & ($\cmp\frac{1}{2},$ $\cpm \frac{1}{2\sqrt{3}}$)& $\cmp \frac{1}{3}$&$\cmp \frac{1}{3}$&$0$&$0$&$0$&$0$&$0$\\ 
 $A^{1\spm}_{\scriptscriptstyle{\stackrel{13\,14}{(\cpm)}}}$& scalar&  $\cmp \frac{1}{3}$&$\smp 1$&$0$ 
   & ($0,$ $\cmp \frac{1}{\sqrt{3}}$)& $\cmp \frac{1}{3}$&$\cmp \frac{1}{3}+ \smp 1$&$0$
   &$0$&$0$&$0$&$0$\\ 
   $A^{13}_{\scriptscriptstyle{\stackrel{13\,14}{(\cpm)}}}$   & scalar&  $\cmp \frac{1}{3}$&$0$&$0$ 
 & ($0,$ $\cmp \frac{1}{\sqrt{3}}$)& $\cmp \frac{1}{3}$&$\cmp \frac{1}{3}$&$0$&$0$&$0$&$0$&$0$\\ 
 \hline
 $A^{2\spm}_{\scriptscriptstyle{\stackrel{9\,10}{(\cpm)}}}$& scalar&  $\cmp \frac{1}{3}$&$0$&$\spm 1$ 
  & ($\cpm\frac{1}{2},$ $\cpm \frac{1}{2\sqrt{3}}$)& $\cmp \frac{1}{3}+ \smp 1$&$\cmp \frac{1}{3}+ \smp 1$&$0$
  &$0$&$0$&$0$&$0$\\  
  $A^{23}_{\scriptscriptstyle{\stackrel{9\,10}{(\cpm)}}}$   & scalar&  $\cmp \frac{1}{3}$&$0$&$0$ 
 & ($\cpm\frac{1}{2},$ $\cpm \frac{1}{2\sqrt{3}}$)& $\cmp \frac{1}{3}$&$\cmp \frac{1}{3}$&$0$&$0$&$0$&$0$&$0$\\ 
 $\cdots$&&&&&&&&&&&&\\
 \hline
 $\tilde{A}^{1 \spm}_{\scriptscriptstyle{\stackrel{9 10}{(\cpm)}}}$& scalar& $\cmp \frac{1}{3}$&$0$&$0$ 
 & ($\cpm\frac{1}{2},$ $\cpm \frac{1}{2\sqrt{3}}$)& $\cmp \frac{1}{3}$&$\cmp \frac{1}{3}$&$0$
 &$\spm 1$&$0$&$0$&$0$\\ 
 $\tilde{A}^{13}_{\scriptscriptstyle{\stackrel{9 10}{(\cpm)}}}$& scalar& $\cmp \frac{1}{3}$&$0$&$0$ 
  & ($\cpm\frac{1}{2},$ $\cpm \frac{1}{2\sqrt{3}}$)& $\cmp \frac{1}{3}$&$\cmp \frac{1}{3}$&$0$
 &$0$&$0$&$0$&$0$\\ 
  $\cdots$&&&&&&&&&&&&\\
 \hline
 $\tilde{A}^{2 \spm}_{\scriptscriptstyle{\stackrel{9 10}{(\cpm)}}}$& scalar& $\cmp \frac{1}{3}$&$0$&$0$ 
  & ($\cpm\frac{1}{2},$ $\cpm \frac{1}{2\sqrt{3}}$)& $\cmp \frac{1}{3}$&$\cmp \frac{1}{3}$&$0$
  &$0$&$\spm 1$&$0$&$0$\\ 
  $\tilde{A}^{23}_{\scriptscriptstyle{\stackrel{9 10}{(\cpm)}}}$& scalar& $\cmp \frac{1}{3}$&$0$&$0$ 
   & ($\cpm\frac{1}{2},$ $\cpm \frac{1}{2\sqrt{3}}$)& $\cmp \frac{1}{3}$&$\cmp \frac{1}{3}$&$0$
  &$0$&$0$&$0$&$0$\\ 
  $\cdots$&&&&&&&&&&&&\\
 \hline
 $\tilde{A}^{N_{L} \spm}_{\scriptscriptstyle{\stackrel{9 10}{(\cpm)}}}$& scalar& $\cmp \frac{1}{3}$&$0$&$0$ 
   & ($\cpm\frac{1}{2},$ $\cpm \frac{1}{2\sqrt{3}}$)& $\cmp \frac{1}{3}$&$\cmp \frac{1}{3}$&$0$
   &$0$&$0$&$\spm 1$&$0$\\ 
   $\tilde{A}^{N_{L} 3}_{\scriptscriptstyle{\stackrel{9 10}{(\cpm)}}}$& scalar& $\cmp \frac{1}{3}$&$0$&$0$ 
    & ($\cpm\frac{1}{2},$ $\cpm \frac{1}{2\sqrt{3}}$)& $\cmp \frac{1}{3}$&$\cmp \frac{1}{3}$&$0$
   &$0$&$0$&$0$&$0$\\ 
   $\cdots$&&&&&&&&&&&&\\
 \hline
 $\tilde{A}^{N_{R} \spm}_{\scriptscriptstyle{\stackrel{9 10}{(\cpm)}}}$& scalar& $\cmp \frac{1}{3}$&$0$&$0$ 
    & ($\cpm\frac{1}{2},$ $\cpm \frac{1}{2\sqrt{3}}$)& $\cmp \frac{1}{3}$&$\cmp \frac{1}{3}$&$0$
    &$0$&$0$&$0$&$\spm 1$\\ 
    $\tilde{A}^{N_{R} 3}_{\scriptscriptstyle{\stackrel{9 10}{(\cpm)}}}$& scalar& $\cmp \frac{1}{3}$&$0$&$0$ 
     & ($\cpm\frac{1}{2},$ $\cpm \frac{1}{2\sqrt{3}}$)& $\cmp \frac{1}{3}$&$\cmp \frac{1}{3}$&$0$
    &$0$&$0$&$0$&$0$\\ 
    $\cdots$&&&&&&&&&&&&\\
 \hline
 $A^{3 i}_{\scriptscriptstyle{\stackrel{9\,10}{(\cpm)}}}$& scalar&  $\cmp \frac{1}{3}$&$0$&$0$ 
  & ($\spm 1 + \cpm\frac{1}{2},$ $\cpm \frac{1}{2\sqrt{3}}$)& $\cmp \frac{1}{3}$&$\cmp \frac{1}{3}$&$0$
 &$0$&$0$&$0$&$0$\\ 
    $\cdots$&&&&&&&&&&&&\\
  \hline
 $A^{4}_{\scriptscriptstyle{\stackrel{9\,10}{(\cpm)}}}$& scalar&  $\cmp \frac{1}{3}$&$0$&$0$ 
  & ($ \cpm\frac{1}{2},$ $\cpm \frac{1}{2\sqrt{3}}$)& $\cmp \frac{1}{3}$&$\cmp \frac{1}{3}$&$0$
 &$0$&$0$&$0$&$0$\\ 
    $\cdots$&&&&&&&&&&&&\\   
\hline 
$\vec{A}^{3}_{m}$& vector&  $0$&$0$&$0$ 
  & octet& $0$&$0$&$0$
 &$0$&$0$&$0$&$0$\\ 
\hline
 $A^{4}_{m}$& vector&  $0$&$0$&$0$ 
  & $0$& $0$&$0$&$0$
 &$0$&$0$&$0$&$0$\\ 
\hline   
 \end{tabular}
 \end{center}
 \end{tiny}
 %
  \end{table}
The scalar fields  with the scalar index $s=(9,10,\cdots,14)$, presented in Table~\ref{Table bosons.}, 
carry one of the triplet colour charges and the "spinor" charge equal to twice the quark "spinor" charge,
or the antitriplet colour charges and the anti "spinor" charge. 
They carry in addition the quantum numbers of the adjoint representations originating in $S^{ab}$ 
or in $\tilde{S}^{ab}$~\footnote{Although carrying the colour charge in  one of the triplet or 
antitriplet states, these fields can not be interpreted as superpartners of the quarks since they do
 not have quantum numbers as required by, let say, the $N=1$ supersymmetry. The hyper 
charges  and the electromagnetic charges are namely not those required by the 
supersymmetric partners to the family members.}. 

Let us choose the $57^{th}$ line of Table~\ref{Table so13+1.}, which represents in the 
spinor technique the left handed positron,~$\bar{e}^{\,\,+}_{L}$, to see what do the scalar fields, 
appearing in Eq.~(\ref{factionMaM1}) and in 
Table~\ref{Table bosons.}, do when applying on the left handed members of the Weyl 
representation presented on Table~\ref{Table so13+1.}, 
containing quarks and leptons and antiquarks and antileptons~\cite{pikanorma,portoroz03,HNds}. 

If we make, let say, the choice of the 
term ($\gamma^0 \stackrel{9 10}{(+)}\, \tau^{2 \sminus}\,$)  
 $A^{2\sminus}_{\scriptscriptstyle{\stackrel{9\,10}{(\oplus)}}} $ (the scalar field 
$ A^{2\sminus}_{\scriptscriptstyle{\stackrel{9\,10}{(\oplus)}}} $ is presented  in the $7^{th}$ 
line in Table~\ref{Table bosons.} and in the second line of Eq.~(\ref{factionMaM1})), 
the family quantum numbers will not be affected and 
they can be any. The state carries the "spinor" (lepton) 
number  $\tau^{4}=\frac{1}{2}$, the weak charge $\tau^{13} =0$, the second $SU(2)_{II}$ 
charge $\tau^{23} =\frac{1}{2}$ and the colour charge $(\tau^{33},\tau^{38})=(0,0)$. 
Correspondingly, its hyper charge ($Y(=\tau^{4}+\tau^{23})$) is $1$ and the electromagnetic 
charge $Q(=Y + \tau^{13})$  is $1$. 

So, what does the term $\gamma^0 \stackrel{9 10}{(+)}\, \tau^{2 \sminus}\,$  
 $A^{2\sminus}_{\scriptscriptstyle{\stackrel{9\,10}{(\oplus)}}} $ make on this spinor 
$\bar{e}^{+}_{L}$? Making use of Eqs.~(\ref{snmb:gammatildegamma}, \ref{graphbinoms}, 
\ref{plusminus})  of~\ref{technique} one easily finds that operator 
$\gamma^0 \stackrel{9 10}{(+)}\,$ 
$ \tau^{2 -}\,$ transforms the left handed  positron  into $\stackrel{03}{(+i)}\,\stackrel{12}{(+)}|
\stackrel{56}{[-]}\,\stackrel{78}{[-]}||\stackrel{9 \;10}{(+)}\;\;
\stackrel{11\;12}{(-)}\;\;\stackrel{13\;14}{(-)} $, which is $d_{R}^{c1} $, presented on line $3$
of  Table~\ref{Table so13+1.}. Namely, $\gamma^0$ transforms $\stackrel{03}{[-i]}$ into 
$\stackrel{03}{(+i)}$, $\stackrel{9 10}{(+)}$ transforms $\stackrel{9 \;10}{[-]}$ into 
$\stackrel{9 \;10}{(+)}$, while $\tau^{2 -}$ ($= -\stackrel{56}{(-)}$ $\stackrel{78}{(-)}$)
transforms $\stackrel{56}{(+)}$ $\stackrel{78}{(+)}$ into $\stackrel{56}{[-]}$ $\stackrel{78}{[-]}$. 
The state $d_{R}^{c1} $ carries the "spinor" (quark) number  $\tau^{4}=\frac{1}{6}$, the weak charge 
$\tau^{13} =0$, the second $SU(2)_{II}$ charge $\tau^{23} =-\frac{1}{2}$ and the colour charge 
$(\tau^{33},\tau^{38})=(\frac{1}{2},\frac{1}{2\sqrt{3}})$. Correspondingly its hyper charge  is
($Y=\tau^{4}+\tau^{23}=$) $-\frac{1}{3}$ and the electromagnetic charge ($Q=Y + \tau^{13}=$) 
$ -\frac{1}{3}$. The scalar field $A^{2\sminus}_{\scriptscriptstyle{\stackrel{9\,10}{(\oplus)}}} $
carries just  the needed quantum numbers as we can see in the $7^{th}$ line of Table~\ref{Table bosons.}.

\begin{figure}
\includegraphics{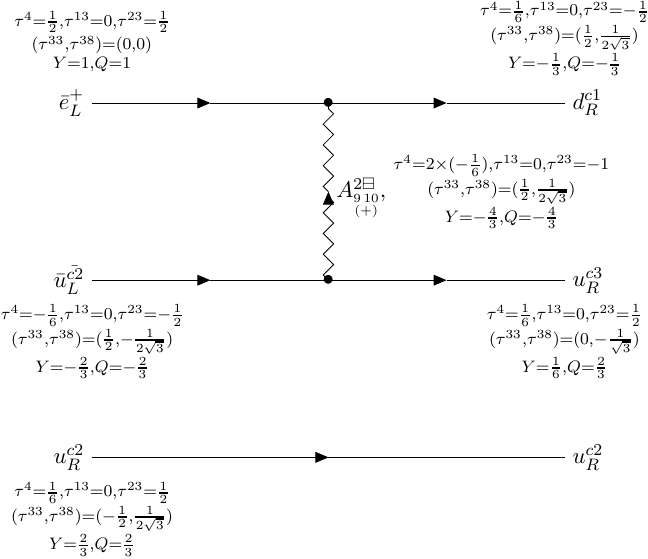}
\centering
\caption{\label{proton is born1.} The birth of a "right handed proton" out of an positron%
~$\bar{e}^{\,\,+}_{L}$, 
antiquark $\bar{u}_L^{\bar{c2}}$ and quark (spectator) $u_{R}^{c2}$.  
The family quantum number can be any.} 
\end{figure}

If the antiquark $ \bar{u}_{L}^{\bar{c2}}$, from the line $43$ (it is not presented, but one can 
very easily construct it) in Table~\ref{Table so13+1.},  with
the "spinor" charge $\tau^{4}=-\frac{1}{6}$, the weak charge $\tau^{13} =0$, 
the second $SU(2)_{II}$ charge $\tau^{23} =-\frac{1}{2}$, the colour charge 
$(\tau^{33},\tau^{38})=(\frac{1}{2},-\frac{1}{2\sqrt{3}})$, the hyper charge 
$Y(=\tau^{4}+\tau^{23}=$) $-\frac{2}{3}$ and the electromagnetic charge $Q (\,=Y + \tau^{13}=$) 
$ -\frac{2}{3}$ submits the $A^{2 \sminus}_{\scriptscriptstyle{\stackrel{9\,10}{(\oplus)}}} $ scalar 
field, it transforms into $u_{R}^{c3}$ from the line $17$ of Table~\ref{Table so13+1.}, carrying the 
quantum numbers $\tau^{4}=\frac{1}{6}$, $\tau^{13} =0$, $\tau^{23} =\frac{1}{2}$, 
$(\tau^{33},\tau^{38})=(0,-\frac{1}{\sqrt{3}})$, $Y=\frac{2}{3}$ and $Q=\frac{2}{3} $ . 
These two quarks, $d_{R}^{c1} $ and $u_{R}^{c3}$ can bind  together 
with $u_{R}^{c2}$ from the $9^{th}$ line of the same table (at low enough energy, after the electroweak 
transition, and if they belong to a superposition with the left handed partners to the first family)
into the colour chargeless baryon - a proton.
This transition is presented in Figure~\ref{proton is born1.}.

The opposite transition at low energies would make the proton decay.

Similar transitions go also with other scalars from  Eq.~(\ref{factionMaM1}) and 
Table~\ref{Table bosons.}. 
The $\vec{\tilde{A}}^{1}_{\stackrel{t'\,t"}{(+)}}$, $\vec{\tilde{A}}^{2}_{\stackrel{t'\,t"}{(+)}}$,
$\vec{\tilde{A}}^{N_{L}}_{\stackrel{t'\,t"}{(+)}}$ and $\vec{\tilde{A}}^{N_{L}}_{\stackrel{t'\,t"}{(+)}}$
fields cause transitions among the family members, changing a particular member into the antimember 
of another colour and of another family. The term $\gamma^0 \stackrel{9 10}{(+)}\, \tilde{N}^{-}_{R}\,$  
$A^{\tilde{N}_{R}-}_{\scriptscriptstyle{\stackrel{9\,10}{(\oplus)}}} $  transforms $\bar{e}^{+}_{R}$ 
into $u^{c1}_{L}$, changing the family quantum numbers.

The action from Eqs.~(\ref{wholeaction}, \ref{faction}) 
manifests 
$\mathbb{C}_{{ \cal N}} \cdot {\cal P}_{{\cal N}}$ invariance.  All the vector and 
the spinor gauge fields are massless. 

Since none of the scalar fields from Table~\ref{Table bosons.} have been observed and also no  
vector gauge fields like $\vec{A}^{2}_{m}$, $A^{4}_{m}$ and other scalar and vector fields, 
there must exist a mechanism, which 
makes the non observed scalar and vector gauge fields massive enough~\footnote{I expect that 
the condensate  (Table~\ref{Table con.}) appears on the scale of unification - $\ge 
\approx 10^{16}$ GeV. Interacting with these vector gauge fields the condensate make them 
massive.}.

Scalar fields from Table~\ref{Table bosons.} carry the colour and the electromagnetic charge. 
Therefore their nonzero vacuum expectation values would not be in agreement with the observed 
phenomena. 
One, however, notices that all the scalar gauge fields from Table~\ref{Table bosons.} and several 
other scalar and vector gauge fields 
couple to the condensate with the nonzero quantum number $\tau^{4}$ and $\tau^{23}$ and 
nonzero family quantum numbers. 

It is not difficult to recognize that the desired condensate must have spin zero, $Y=\tau^4 + $ 
$\tau^{23}=0$, $Q=Y + \tau^{13}=0$ and $\vec{\tau}^{1}=0$ in order that in the low energy 
limit the {\it spin-charge-family} theory would manifest effectively as the {\it standard model}.

I make a choice of the two right handed neutrinos of the $VIII^{th}$ family 
coupled into a scalar, with $\tau^4=-1$, $\tau^{23}=1$, correspondingly $Y=0$, $Q=0$ and 
$\vec{\tau}^{1}=0$, and with family quantum numbers~(Eqs. (\ref{so42tilde}, \ref{so1+3tilde})) 
$\tilde{\tau}^4=-1 $, $\tilde{\tau}^{23}=1 $, $\tilde{N}_{R}^{3}=1$, and correspondingly with 
$\tilde{Y}=$ $\tilde{\tau}^4 + \tilde{\tau}^{23}=0$, $\tilde{Q}=\tilde{Y} + \tilde{\tau}^{13}=0$, 
and $\vec{\tilde{\tau}}^{1}=0$. 
The condensate carries the family quantum numbers of the upper four families, 
see Subsect.~\ref{condensate}.

The condensate made out of spinors couples to spinors differently than to antispinors -  
"anticondensate" would namely carry $\tau^4=1$, and $\tau^{23}=-1$ - breaking 
correspondingly the $\mathbb{C}_{{ \cal N}} \cdot $ ${\cal P}_{{\cal N}}$ symmetry:  
The reactions creating particles from antiparticles  are not any longer symmetric to those 
creating antiparticle from particles.
 
Such  a condensate leaves the hyper field $A^{Y}_{m}$ ($= \sin \vartheta_{2}\,A^{23}_{m} +
\cos \vartheta_{2}\,A^{4}_{m}$) (for the choice  that $\sin \vartheta_{2}= \cos \vartheta_{2}$ 
and $g^{4}= g^{2}$, there is no justification for such a choice, $A^{Y}_{m} = \frac{1}{\sqrt{2}}\,
(A^{23}_{m} + A^{4}_{m})$ follows) massless, while it gives masses to $A^{2\pm}_m$  and $A^{Y'}_m$ 
($=\frac{1}{\sqrt{2}}\,(A^{4}_{m} - A^{23}_{m})$ for $\sin \vartheta_{2}= \cos \vartheta_{2}$) 
and it gives masses also to all the scalar gauge fields from Table~\ref{Table bosons.}, since they all couple 
to the condensate through $\tau^{4}$.

The weak vector gauge fields, $\vec{A}^{1}_m$, the hyper charge vector gauge fields, 
$A^{Y}_{m}$, and the colour vector gauge fields, $\vec{A}^{1}_m$, remain massless. 

The scalar fields with the scalar space index $s=(7,8)$ (there are three singlets which couple to all eight 
families,  two triplets which couple only to the upper four families and another two triplets  which
couple only to the lower four families) -  carrying the weak and the hyper charges of the 
Higgs's scalar - wait for gaining 
nonzero vacuum  expectation values  to change their masses while causing the electroweak break. 

The condensate does  what is needed so that in the low energy regime
the {\it spin-charge-family} manifests as an effective theory. This effective theory agrees with the 
{\it standard model} to such an extent that it is in agreement with the observed phenomena, 
explaining the {\it standard model} assumptions and predicting new fermion and boson 
fields. 

It also may hopefully explain  the observed matter-antimatter asymmetry if the 
conditions in the expanding universe would be appropriate, Ref.~(\cite{norma2014MatterAntimatter},
Sect. VI.). The work needed 
to check these conditions in the expanding universe within the {\it spin-charge-family} theory 
is very demanding. Although we do have some experience with following the history of the 
expanding universe~\cite{gn}, this study needs much more efforts, not only in  calculations, 
but also in understanding the mechanism of the condensate appearance, relations among the velocity 
of the expansion, the temperature and the dimension of space-time in the period of the appearance 
of the condensate.
This study has not yet been really started.

\vspace{3mm}

\section{Conclusions}
\label{conclusion}

\vspace{3mm}

To better understand the history of the universe and also to make next step in understanding 
the dynamics of the elementary fermion fields and boson (vector and scalar) gauge fields it is 
needed to explain the assumptions of the {\it standard model}, as well as the phenomena like 
the existence of the dark matter, matter-antimatter asymmetry and dark energy.

We must understand the origin of: {\bf  A.} the family members quantum numbers, 
 {\bf  B.}  the family quantum numbers,  {\bf  C.} the origin of vector gauge fields,  {\bf  D.}
the origin of the Higgs and Yukawa couplings.
 
 One of the most urgent questions in the elementary particle physics is: {\it Where do 
the families originate?} Explaining the origin of families would answer the question about the 
number of families which are possibly observable at the low energy regime, about the origin 
of the scalar field(s) and the Yukawa couplings (the couplings of fermions to the scalar field(s)), 
about the differences in the fermions properties - the differences in the masses and mixing 
matrices among family members --
quarks and leptons, as well as about the hierarchy in quark and lepton masses.

I demonstrated in this talk, that the {\it spin-charge-family} theory - starting with the simple action
in $d=(13+1)$ for fermions and bosons - offers the explanation for all the assumptions of the
 {\it standard model}:  

\noindent 
{\bf a.} $\;\,$ The theory explains all the properties of the family members - quarks and leptons,
left and right handed, and their right and left handed antiquarks and antileptons~\footnote{One
Weyl representation of $SO(13+1)$ contains, if analyzed with respect to the {\it standard model }
groups, all the members of one family, the coloured quarks and colourless leptons, and the 
anticoloured antiquarks and (anti)colourless antileptons, with the left handed spinors  
carrying the weak charge and the right handed ones weak chargeless, while the left handed 
antispinors are weak chargeless and the right handed ones carry the weak charge.}, explaining why
 the left handed spinors carry the weak charge while the right handed do not (the right handed 
neutrino is the regular member of each family).\\ 
{\bf b.} $\;\,$  It explains the appearance and the properties of the families of family members. \\
{\bf c.} $\;\,$ It explains the existence of the gauge vector fields of the family members charges. \\
{\bf d.} $\;\,$ It explains the appearance and the properties of the scalar field (the Higgs) and the 
Yukawa couplings.\\

All the gauge fields, vector and scalar in $d=(3+1)$, origin in vielbeins and the two kinds of spin
connection fields in $d=(13+1)$ -  the gravity.  The two spin connection fields are uniquely
expressible  with the vielbeins, if there are no spinors present~\cite{DNproof}.

\noindent
It also offers the explanation for the phenomena, which are not part of the 
{\it standard model}, like:\\ 
{\bf e.} $\;\,$ It explains the existence of the dark matter.\\
{\bf f.} $\;\,$ It explain the origin of the (ordinary) matter-antimatter asymmetry.

\vspace{2mm}
\subsection{Predictions for the future experiments}
\label{predictions}

\vspace{2mm}

\noindent
The theory predicts:\\
{\bf g.} $\;\,$ There are twice two groups of four families of quarks and leptons at low energies.\\  
{\bf g.i.} The fourth family  with masses above $1$ TeV, weakly coupled to the observed
three families, will be measured at the LHC.\\ 
{\bf g.ii.} The quarks and leptons of the fifth family - that is of the stable one of the upper four 
families - form the dark matter. The  family members, which form the chargeless clusters, manifest,
 due to their very heavy masses, a "new nuclear force".\\
{\bf h.} $\;\,$ The predicted scalar fields with the space index $(7,8)$ manifest in $(d=(3+1))$  as
the weak and hyper charges doublets  (as required by the {\it standard model} Higgs) with respect
to the space index.
These scalars carry in addition:\\ 
 {\bf h.i.} Either they carry one of the three family members quantum numbers, 
$(Q,Q',Y')$ - belonging correspondingly  to one of three singlets. \\
 {\bf h.ii.} Or they carry family quantum numbers - belonging correspondingly  to one of the
 twice two triplets.\\
{\bf h.iii.} The  three singlets and the two triplets determine mass matrices of the lower 
four families, contributing to masses of the heavy vector bosons. \\ 
{\bf h.iv.} These scalars determine the observed Higgs and the Yukawa couplings.\\
{\bf i.} $\;\,$ The predicted scalar fields with the space index $(9,10,..,14)$ are triplets
with respect to the space index. They cause the transitions from antileptons into quarks and antiquarks 
into quarks and back. \\
{\bf i.i.}  The condensate breaks the matter-antimatter symmetry, causing the asymmetry
in the (ordinary) matter with respect to antimatter. \\ 
{\bf i.ii.} These condensate is responsible also for the proton decay.\\
{\bf j.} $\;\,$ The condensate is a scalar of the two right handed neutrinos with the family quantum 
numbers of the upper four families. \\
{\bf k.} $\;\,$ The condensate gives masses to all the gauge fields with which it 
interacts.\\
{\bf k.i.}  It gives masses to all scalar fields and to vector fields, leaving massless only the colour, 
the weak, the hyper vector gauge fields and the gravity in ($(3+1)$). \\
{\bf l.} $\;\,$ There is the $SU(2)$ (belonging together with the weak $SU(2)$ to $SO(4)$ 
gauge  fields included in $SO(7,1)$) vector gauge field, which gain masses of the order of the 
appearance of the condensate.\\ 
{\bf m.} $\;\,$ At the electroweak break the scalar fields with the space index $(7,8)$ change their
 mutual interaction, and gaining nonzero vacuum expectation values, break the weak and the 
hyper charges and correspondingly the mass protection of fermions, making them massive.\\
{\bf n.} $\;\,$ The symmetry of mass matrices allow, in the case that the experimental data 
for the mixing submatrix $3 \times 3$ of the $4 \times 4$ mixing matrix would be accurate, to 
determine the mixing matrix and the masses of the fourth family quarks. The accuracy, with 
which the masses of the six lower families are measured so far, does not influence the results
appreciably. Due to uncertainty of the experimental data for the $3\times 3$ mixing submatrix 
we are only able to determine the $4 \times 4$ quark mixing matrix for a chosen masses of the
the fourth family quarks. However, we also predict how will the $3 \times 3$ submatrix of the 
mixing matrix change with more accurate measurements.\\
{\bf n.i.} The fourth family quarks mass matrices are for masses above $1$ TeV closer and 
closer to the democratic matrices. The less the scalars with the family members quantum
numbers contribute to masses of the fourth family quarks, the closer is $m_{u_4}$ to
 $m_{d_4}$.\\
{\bf n.ii.} The large contribution of the scalars with the family members quantum numbers 
($Q,Q',Y'$) to the masses of the lower four families manifests in the large differences 
of quarks masses of the lower four families.\\
{\bf n.iii.} Although we have done calculations also for leptons, further analyses of their 
properties must  wait for more accurate experimental data.\\
{\bf o.} $\;\,$ In the case that  the $u_{4}$ and $d_{4}$ quarks have similar masses -
determined mostly by the scalar fields carrying the family quantum numbers - they contribute
mostly to the production of these scalars, while their contribution to the 
production of those scalars which carry the family members quantum numbers - to the Higgs in 
particular - is much weaker, which is in 
agreement with the experiment~\footnote{The 
coupling constants of the singlet scalar fields differ among themselves and also from the coupling 
constants of the two triplet scalar fields.}.\\
{\bf p.} $\;\,$ All the degrees of freedom discussed in this talk are already a part of the simple 
starting action~Eq.(\ref{wholeaction}).\\
{\bf p.i.} The way of breaking symmetries (ordered by the conditions determining the history 
of our universe) is assumed so that it leads in $d=(3+1)$ to the observable symmetries, although
we could in principle derive it from the starting action and boundary conditions.\\
{\bf p.ii.} Also the effective interaction among scalar fields is assumed, although we could 
derive it in principle from the starting action and the boundary conditions.\\
{\bf r.} The {\it spin-charge-family} theory easily explains what in the {\it standard model}
seems like a miracle: no triangle anomalies.\\
{\bf s.} A lot of efforts has been put in this theory to show that it could work as a next step 
below the {\it standard model} proving like:\\
{\bf s.i.} There is possibility in the Kaluza-Klein-like theory that breaking symmetries can leave 
fermions massless.\\
{\bf s.ii.} That vielbeins in the Kaluza-Klein theories and spin connections in the {\it spin-charge-family}
theory represent the same vector gauge fields in $d=(3+1)$.

\vspace{2mm}

\subsection{Open questions in the {\it spin-charge-family} theory}
\label{answers}

\vspace{2mm}

\noindent
There are several open problems in the {\it spin-charge-family} theory:\\
{\bf t.} $\;\,$ Since this theory is, except that fermions carry two kinds of spins - one
kind taking  care of spin and charges, the second one taking care of families - a kind of 
the Kaluza-Klein theories, it shares at very high energy with these theories the quantization
problem. \\  
{\bf u.} $\;\,$ The dimension of space-time, $d=(13+1)$, is in the {\it spin-charge-family}
theory chosen, since $SO(13,1)$ contains all the members, assumed in the {\it standard 
model}.\\  
It contains also the right handed neutrino (which carries the $Y'$ quantum number). \\
{\bf u.i.} It should be shown, however, how has nature "made the decision" in evolution 
to go through this dimension and what is indeed the dimension of space-time (infinite?).\\
{\bf v.} $\;\,$ There are many other open question, like: \\
{\bf v.i.}  What is the reason for the (so small) dark energy?\\ 
{\bf v.ii.} At what energy the electroweak phase transition occurs?
(Let me add that there is the answer within the {\it spin-charge-family} theory to this open problem.)\\
{\bf v.iii.} Why do we have fermions and bosons?~\cite{hnfermionization2015}\\

It is encouraging that the more work is done on this theory the more answers to the open 
questions is found.

\appendix

\vspace{3mm}

\section{Short presentation of spinor technique~\cite{JMP2015,norma93,hn02,hn03}}
\label{technique}

\vspace{3mm}

This appendix is a short review (taken from~\cite{JMP}) of the technique~\cite{norma93,DKhn,%
hn02,hn03}, 
initiated and developed in Ref.~\cite{norma93}, while  proposing the {\it spin-charge-family} 
theory~\cite{NBled2013,NBled2012,norma92,norma93,norma94,%
pikanorma,portoroz03,JMP,norma95,gmdn07,gn,gn2013,gn2015,NPLB,N2014scalarprop,%
norma2014MatterAntimatter,JMP2015}.
All the internal degrees of freedom of spinors, with 
family quantum numbers included, are describable in the space of $d$-anticommuting (Grassmann) 
coordinates~\cite{norma93}, if the dimension of ordinary space is also $d$. 
There are two kinds of operators in the Grassmann space fulfilling the Clifford algebra and 
anticommuting with one another~Eq.(\ref{gammatildegamma}). The technique  was further 
developed in the present 
shape together with H.B. Nielsen~\cite{DKhn,hn02,hn03}. 

In this last stage we  rewrite a spinor basis, written in Ref.~\cite{norma93} as products of polynomials 
of Grassmann coordinates of odd and even Grassmann character, chosen to be eigenstates of the Cartan 
subalgebra defined by the two kinds of the Clifford algebra objects, as products of nilpotents and 
projections, formed as odd and even objects of $\gamma^a$'s, respectively, and  chosen to be eigenstates 
of a Cartan subalgebra of the Lorentz groups defined by $\gamma^a$'s and $\tilde{\gamma}^a$'s.   

The technique can be used to construct a spinor basis for any dimension $d$
and any signature in an easy and transparent way. Equipped with the graphic presentation of basic states,  
the technique offers an elegant way to see all the quantum numbers of states with respect to the two 
Lorentz groups, as well as transformation properties of the states under any Clifford algebra object. 

Ref.~\cite{JMP2015}, App.~B, 
briefly represents the starting point~\cite{norma93} of this technique. 
There are two kinds of the Clifford algebra objects, 
$\gamma^a$'s and $\tilde{\gamma}^a$'s.

These objects 
have properties,
\begin{eqnarray}
\label{gammatildegamma}
&& \{ \gamma^a, \gamma^b\}_{+} = 2\eta^{ab}\,, \quad\quad    
\{ \tilde{\gamma}^a, \tilde{\gamma}^b\}_{+}= 2\eta^{ab}\,, \quad,\quad
\{ \gamma^a, \tilde{\gamma}^b\}_{+} = 0\,.
\end{eqnarray}

If $B$ is a Clifford algebra object, let say a polynomial of $\gamma^a$,
$B=a_{0} + a_{a}\,\gamma^a + a_{a b}\, \gamma^a \gamma^b + \cdots + 
a_{a_1 a_2 \dots a_d}\, 
\gamma^{a_1}\gamma^{a_2}\dots \gamma^{a_d}$, 
one finds 
\begin{eqnarray}
(\tilde{\gamma}^a B : &=& i(-)^{n_B} \,B \gamma^a \,) \;|\psi_0 \,>, \nonumber\\
B &=& a_0 + a_{a_0} \gamma^{a_0} + a_{a_1 a_2} \gamma^{a_1} \gamma^{a_2} + \cdots + 
a_{a_1 \cdots a_d} \gamma^{a_1}\cdots \gamma^{a_d} \,,
\label{tildegcliffordappendix}
\end{eqnarray}
where $|\psi_{0}>$ is a vacuum state, defined in Eq.~(\ref{graphherscal}) and 
$(-)^{n_B} $ is equal to $1$ for the term in the polynomial which has an even number of $\gamma^b$'s, 
and to $-1$ for the term with an odd number of  $\gamma^b$'s,
for any $d$, even or odd, and  $I$ is the unit element in the 
Clifford algebra. 

It follows from Eq.~(\ref{tildegcliffordappendix}) that the two kinds of the 
Clifford algebra objects are connected with
the left and the right multiplication of any Clifford algebra objects $B$
 (Eq.~(\ref{tildegcliffordappendix})).


The Clifford algebra objects $S^{ab}$ and $\tilde{S}^{ab}$ close the algebra of the Lorentz 
group 
\begin{eqnarray}
\label{sabtildesab}
S^{ab}: &=& (i/4) (\gamma^a \gamma^b - \gamma^b \gamma^a)\,, \nonumber\\
\tilde{S}^{ab}: &=& (i/4) (\tilde{\gamma}^a \tilde{\gamma}^b 
- \tilde{\gamma}^b \tilde{\gamma}^a)\,,
\end{eqnarray}
$ \{S^{ab}, \tilde{S}^{cd}\}_{-}= 0\,$, 
$\{S^{ab},S^{cd}\}_{-} = $ $ i(\eta^{ad} S^{bc} + \eta^{bc} S^{ad} - \eta^{ac} S^{bd} - \eta^{bd} S^{ac})\,$,
$\{\tilde{S}^{ab},\tilde{S}^{cd}\}_{-} $ $= i(\eta^{ad} \tilde{S}^{bc} + \eta^{bc} \tilde{S}^{ad} 
- \eta^{ac} \tilde{S}^{bd} - \eta^{bd} \tilde{S}^{ac})\,$.
%

We assume  the ``Hermiticity'' property for $\gamma^a$'s  
\begin{eqnarray}
\gamma^{a\dagger} = \eta^{aa} \gamma^a\,,
\label{cliffher}
\end{eqnarray}
in order that 
$\gamma^a$ 
are compatible with (\ref{gammatildegamma}) and formally unitary, 
i.e. $\gamma^{a \,\dagger} \,\gamma^a=I$. 

One finds from Eq.~(\ref{cliffher}) that $(S^{ab})^{\dagger} = \eta^{aa} \eta^{bb}S^{ab}$.

Recognizing from Eq.(\ref{sabtildesab})  that the two Clifford algebra objects 
$S^{ab}, S^{cd}$ with all indices different commute, and equivalently for 
$\tilde{S}^{ab},\tilde{S}^{cd}$, we  select  the Cartan subalgebra of the algebra of the 
two groups, which  form  equivalent representations with respect to one another 
\begin{eqnarray}
S^{03}, S^{12}, S^{56}, \cdots, S^{d-1\; d}, \quad {\rm if } \quad d &=& 2n\ge 4,
\nonumber\\
S^{03}, S^{12}, \cdots, S^{d-2 \;d-1}, \quad {\rm if } \quad d &=& (2n +1) >4\,,
\nonumber\\
\tilde{S}^{03}, \tilde{S}^{12}, \tilde{S}^{56}, \cdots, \tilde{S}^{d-1\; d}, 
\quad {\rm if } \quad d &=& 2n\ge 4\,,
\nonumber\\
\tilde{S}^{03}, \tilde{S}^{12}, \cdots, \tilde{S}^{d-2 \;d-1}, 
\quad {\rm if } \quad d &=& (2n +1) >4\,.
\label{choicecartan}
\end{eqnarray}

The choice for  the Cartan subalgebra in $d <4$ is straightforward.
It is  useful  to define one of the Casimirs of the Lorentz group -  
the  handedness $\Gamma$ ($\{\Gamma, S^{ab}\}_- =0$) in any $d$ 
\begin{eqnarray}
\Gamma^{(d)} :&=&(i)^{d/2}\; \;\;\;\;\;\prod_a \quad (\sqrt{\eta^{aa}} \gamma^a), \quad {\rm if } \quad d = 2n, 
\nonumber\\
\Gamma^{(d)} :&=& (i)^{(d-1)/2}\; \prod_a \quad (\sqrt{\eta^{aa}} \gamma^a), \quad {\rm if } \quad d = 2n +1\,.
\label{hand}
\end{eqnarray}
One proceeds equivalently for $\tilde{\Gamma}^{(d)} $, substituting $\gamma^a$'s by $\tilde{\gamma}^a$'s.
We understand the product of $\gamma^a$'s in the ascending order with respect to 
the index $a$: $\gamma^0 \gamma^1\cdots \gamma^d$. 
It follows from Eq.(\ref{cliffher})
for any choice of the signature $\eta^{aa}$ that
$\Gamma^{\dagger}= \Gamma,\;
\Gamma^2 = I.$
We also find that for $d$ even the handedness  anticommutes with the Clifford algebra objects 
$\gamma^a$ ($\{\gamma^a, \Gamma \}_+ = 0$) , while for $d$ odd it commutes with  
$\gamma^a$ ($\{\gamma^a, \Gamma \}_- = 0$). 

To make the technique simple we introduce the graphic presentation 
as follows 
\begin{eqnarray}
\stackrel{ab}{(k)}:&=& 
\frac{1}{2}(\gamma^a + \frac{\eta^{aa}}{ik} \gamma^b)\,,\quad \quad
\stackrel{ab}{[k]}:=
\frac{1}{2}(1+ \frac{i}{k} \gamma^a \gamma^b)\,,
\label{signature}
\end{eqnarray}
where $k^2 = \eta^{aa} \eta^{bb}$.
It follows then 
\begin{eqnarray}
\gamma^{a}&=& \stackrel{ab}{(k)} + \stackrel{ab}{(-k)}\,, \quad \quad 
\gamma^{b} = ik\eta^{aa}\,(\stackrel{ab}{(k)} - \stackrel{ab}{(-k)})\,,\nonumber\\
S^{ab}    &=& \frac{k}{2} (\stackrel{ab}{[k]}- \stackrel{ab}{[-k]})\,.
\label{signaturegamma}
\end{eqnarray}
One can easily check by taking into account the Clifford algebra relation 
(Eq.~(\ref{gammatildegamma})) and the
definition of $S^{ab}$ and $\tilde{S}^{ab}$ (Eq.~(\ref{sabtildesab}))
that the nilpotent $\stackrel{ab}{(k)}$ and the projector $\stackrel{ab}{[k]}$ are "eigenstates" of
$S^{ab}$ and $\tilde{S}^{ab}$ 
\begin{eqnarray}
        S^{ab}\, \stackrel{ab}{(k)}= \frac{1}{2}\,k\, \stackrel{ab}{(k)}\,,\quad \quad 
        S^{ab}\, \stackrel{ab}{[k]}= \frac{1}{2}\,k \,\stackrel{ab}{[k]}\,,\nonumber\\
\tilde{S}^{ab}\, \stackrel{ab}{(k)}= \frac{1}{2}\,k \,\stackrel{ab}{(k)}\,,\quad \quad 
\tilde{S}^{ab}\, \stackrel{ab}{[k]}=-\frac{1}{2}\,k \,\stackrel{ab}{[k]}\,,
\label{grapheigen}
\end{eqnarray}
which means that we get the same objects back multiplied by the constant $\frac{1}{2}k$ in the case 
of $S^{ab}$, while $\tilde{S}^{ab}$ multiply $\stackrel{ab}{(k)}$ by $k$ and $\stackrel{ab}{[k]}$ 
by $(-k)$ rather than $(k)$. 
This also means that when 
$\stackrel{ab}{(k)}$ and $\stackrel{ab}{[k]}$ act from the left hand side on  a
vacuum state $|\psi_0\rangle$ the obtained states are the eigenvectors of $S^{ab}$.
We further recognize 
that $\gamma^a$ 
transform  $\stackrel{ab}{(k)}$ into  $\stackrel{ab}{[-k]}$, never to $\stackrel{ab}{[k]}$, 
while $\tilde{\gamma}^a$ transform  $\stackrel{ab}{(k)}$ into $\stackrel{ab}{[k]}$, never to 
$\stackrel{ab}{[-k]}$ 
\begin{eqnarray}
&&\gamma^a \stackrel{ab}{(k)}= \eta^{aa}\stackrel{ab}{[-k]},\; 
\gamma^b \stackrel{ab}{(k)}= -ik \stackrel{ab}{[-k]}, \; 
\gamma^a \stackrel{ab}{[k]}= \stackrel{ab}{(-k)},\; 
\gamma^b \stackrel{ab}{[k]}= -ik \eta^{aa} \stackrel{ab}{(-k)}\,,\nonumber\\
&&\tilde{\gamma^a} \stackrel{ab}{(k)} = - i\eta^{aa}\stackrel{ab}{[k]},\;
\tilde{\gamma^b} \stackrel{ab}{(k)} =  - k \stackrel{ab}{[k]}, \;
\tilde{\gamma^a} \stackrel{ab}{[k]} =  \;\;i\stackrel{ab}{(k)},\; 
\tilde{\gamma^b} \stackrel{ab}{[k]} =  -k \eta^{aa} \stackrel{ab}{(k)}\,. 
\label{snmb:gammatildegamma}
\end{eqnarray}
From Eq.(\ref{snmb:gammatildegamma}) it follows
\begin{eqnarray}
\label{stildestrans}
S^{ac}\stackrel{ab}{(k)}\stackrel{cd}{(k)} &=& -\frac{i}{2} \eta^{aa} \eta^{cc} 
\stackrel{ab}{[-k]}\stackrel{cd}{[-k]}\,,\,\quad\quad
\tilde{S}^{ac}\stackrel{ab}{(k)}\stackrel{cd}{(k)} = \frac{i}{2} \eta^{aa} \eta^{cc} 
\stackrel{ab}{[k]}\stackrel{cd}{[k]}\,,\,\nonumber\\
S^{ac}\stackrel{ab}{[k]}\stackrel{cd}{[k]} &=& \frac{i}{2}  
\stackrel{ab}{(-k)}\stackrel{cd}{(-k)}\,,\,\quad\quad
\tilde{S}^{ac}\stackrel{ab}{[k]}\stackrel{cd}{[k]} = -\frac{i}{2}  
\stackrel{ab}{(k)}\stackrel{cd}{(k)}\,,\,\nonumber\\
S^{ac}\stackrel{ab}{(k)}\stackrel{cd}{[k]}  &=& -\frac{i}{2} \eta^{aa}  
\stackrel{ab}{[-k]}\stackrel{cd}{(-k)}\,,\,\quad\quad
\tilde{S}^{ac}\stackrel{ab}{(k)}\stackrel{cd}{[k]} = -\frac{i}{2} \eta^{aa}  
\stackrel{ab}{[k]}\stackrel{cd}{(k)}\,,\,\nonumber\\
S^{ac}\stackrel{ab}{[k]}\stackrel{cd}{(k)} &=& \frac{i}{2} \eta^{cc}  
\stackrel{ab}{(-k)}\stackrel{cd}{[-k]}\,,\,\quad\quad
\tilde{S}^{ac}\stackrel{ab}{[k]}\stackrel{cd}{(k)} = \frac{i}{2} \eta^{cc}  
\stackrel{ab}{(k)}\stackrel{cd}{[k]}\,. 
\end{eqnarray}
From Eq.~(\ref{stildestrans}) we conclude that $\tilde{S}^{ab}$ generate the 
equivalent representations with respect to $S^{ab}$ and opposite. 

Let us deduce some useful relations

\begin{eqnarray}
\stackrel{ab}{(k)}\stackrel{ab}{(k)}& =& 0\,, \quad \quad \stackrel{ab}{(k)}\stackrel{ab}{(-k)}
= \eta^{aa}  \stackrel{ab}{[k]}\,, \quad \stackrel{ab}{(-k)}\stackrel{ab}{(k)}=
\eta^{aa}   \stackrel{ab}{[-k]}\,,\quad
\stackrel{ab}{(-k)} \stackrel{ab}{(-k)} = 0\,, \nonumber\\
\stackrel{ab}{[k]}\stackrel{ab}{[k]}& =& \stackrel{ab}{[k]}\,, \quad \quad
\stackrel{ab}{[k]}\stackrel{ab}{[-k]}= 0\,, \;\;\quad \quad  \quad \stackrel{ab}{[-k]}\stackrel{ab}{[k]}=0\,,
 \;\;\quad \quad \quad \quad \stackrel{ab}{[-k]}\stackrel{ab}{[-k]} = \stackrel{ab}{[-k]}\,,
 \nonumber\\
\stackrel{ab}{(k)}\stackrel{ab}{[k]}& =& 0\,,\quad \quad \quad \stackrel{ab}{[k]}\stackrel{ab}{(k)}
=  \stackrel{ab}{(k)}\,, \quad \quad \quad \stackrel{ab}{(-k)}\stackrel{ab}{[k]}=
 \stackrel{ab}{(-k)}\,,\quad \quad \quad 
\stackrel{ab}{(-k)}\stackrel{ab}{[-k]} = 0\,,
\nonumber\\
\stackrel{ab}{(k)}\stackrel{ab}{[-k]}& =&  \stackrel{ab}{(k)}\,,
\quad \quad \stackrel{ab}{[k]}\stackrel{ab}{(-k)} =0,  \quad \quad 
\quad \stackrel{ab}{[-k]}\stackrel{ab}{(k)}= 0\,, \quad \quad \quad \quad
\stackrel{ab}{[-k]}\stackrel{ab}{(-k)} = \stackrel{ab}{(-k)}\,.\nonumber\\
\label{graphbinoms}
\end{eqnarray}
We recognize in Eq.~(\ref{graphbinoms}) 
the demonstration of the nilpotent and the projector character of the Clifford algebra objects 
$\stackrel{ab}{(k)}$ and $\stackrel{ab}{[k]}$, respectively. 
Defining
\begin{eqnarray}
\stackrel{ab}{\tilde{(\pm i)}} = 
\frac{1}{2} \, (\tilde{\gamma}^a \mp \tilde{\gamma}^b)\,, \quad
\stackrel{ab}{\tilde{(\pm 1)}} = 
\frac{1}{2} \, (\tilde{\gamma}^a \pm i\tilde{\gamma}^b)\,, 
\label{deftildefun}
\end{eqnarray}
one recognizes that
\begin{eqnarray}
\stackrel{ab}{\tilde{( k)}} \, \stackrel{ab}{(k)}& =& 0\,, 
\quad \;
\stackrel{ab}{\tilde{(-k)}} \, \stackrel{ab}{(k)} = -i \eta^{aa}\,  \stackrel{ab}{[k]}\,,
\quad\;
\stackrel{ab}{\tilde{( k)}} \, \stackrel{ab}{[k]} = i\, \stackrel{ab}{(k)}\,,
\quad\;
\stackrel{ab}{\tilde{( k)}}\, \stackrel{ab}{[-k]} = 0\,.
\label{graphbinomsfamilies}
\end{eqnarray}
Recognizing that
\begin{eqnarray}
\stackrel{ab}{(k)}^{\dagger}=\eta^{aa}\stackrel{ab}{(-k)}\,,\quad
\stackrel{ab}{[k]}^{\dagger}= \stackrel{ab}{[k]}\,,
\label{graphherstr}
\end{eqnarray}
we define a vacuum state $|\psi_0>$ so that one finds
\begin{eqnarray}
< \;\stackrel{ab}{(k)}^{\dagger}
 \stackrel{ab}{(k)}\; > = 1\,, \nonumber\\
< \;\stackrel{ab}{[k]}^{\dagger}
 \stackrel{ab}{[k]}\; > = 1\,.
\label{graphherscal}
\end{eqnarray}

Taking into account the above equations it is easy to find a Weyl spinor irreducible representation
for $d$-dimensional space, with $d$ even or odd.

For $d$ even we simply make a starting state as a product of $d/2$, let us say, only nilpotents 
$\stackrel{ab}{(k)}$, one for each $S^{ab}$ of the Cartan subalgebra  elements (Eq.(\ref{choicecartan})),  
applying it on an (unimportant) vacuum state. 
For $d$ odd the basic states are products
of $(d-1)/2$ nilpotents and a factor $(1\pm \Gamma)$.  
Then the generators $S^{ab}$, which do not belong 
to the Cartan subalgebra, being applied on the starting state from the left, 
 generate all the members of one
Weyl spinor.  
\begin{eqnarray}
\stackrel{0d}{(k_{0d})} \stackrel{12}{(k_{12})} \stackrel{35}{(k_{35})}\cdots \stackrel{d-1\;d-2}{(k_{d-1\;d-2})}
|\psi_0 \,>\nonumber\\
\stackrel{0d}{[-k_{0d}]} \stackrel{12}{[-k_{12}]} \stackrel{35}{(k_{35})}\cdots \stackrel{d-1\;d-2}{(k_{d-1\;d-2})}
|\psi_0 \,>\nonumber\\
\stackrel{0d}{[-k_{0d}]} \stackrel{12}{(k_{12})} \stackrel{35}{[-k_{35}]}\cdots \stackrel{d-1\;d-2}{(k_{d-1\;d-2})}
|\psi_0 \,>\nonumber\\
\vdots \nonumber\\
\stackrel{0d}{[-k_{0d}]} \stackrel{12}{(k_{12})} \stackrel{35}{(k_{35})}\cdots \stackrel{d-1\;d-2}{[-k_{d-1\;d-2}]}
|\psi_0 \,>\nonumber\\
\stackrel{od}{(k_{0d})} \stackrel{12}{[-k_{12}]} \stackrel{35}{[-k_{35}]}\cdots \stackrel{d-1\;d-2}{(k_{d-1\;d-2})}
|\psi_0\,> \nonumber\\
\vdots 
\label{graphicd}
\end{eqnarray}
All the states have the same handedness $\Gamma $, since $\{ \Gamma, S^{ab}\}_{-} = 0$. 
States, belonging to one multiplet  with respect to the group $SO(q,d-q)$, that is to one
irreducible representation of spinors (one Weyl spinor), can have any phase. We made a choice
of the simplest one, taking all  phases equal to one.

The above graphic representation demonstrates that for $d$ even 
all the states of one irreducible Weyl representation of a definite handedness follow from a starting state, 
which is, for example, a product of nilpotents $\stackrel{ab}{(k_{ab})}$, by transforming all possible pairs
of $\stackrel{ab}{(k_{ab})} \stackrel{mn}{(k_{mn})}$ into $\stackrel{ab}{[-k_{ab}]} \stackrel{mn}{[-k_{mn}]}$.
There are $S^{am}, S^{an}, S^{bm}, S^{bn}$, which do this.
The procedure gives $2^{(d/2-1)}$ states. A Clifford algebra object $\gamma^a$ being applied from the left hand side,
transforms  a 
Weyl spinor of one handedness into a Weyl spinor of the opposite handedness. Both Weyl spinors form a Dirac 
spinor.


We shall speak about left handedness when $\Gamma= -1$ and about right
handedness when $\Gamma =1$ for either $d$ even or odd.

While $S^{ab}$ which do not belong to the Cartan subalgebra (Eq.~(\ref{choicecartan})) generate 
all the states of one representation,  $\tilde{S}^{ab}$ which do not belong to the 
Cartan subalgebra (Eq.~(\ref{choicecartan}))  generate the states of $2^{d/2-1}$ equivalent representations.

Making a choice of the Cartan subalgebra set (Eq.~(\ref{choicecartan})) 
of the algebra $S^{ab}$ and 
$\tilde{S}^{ab}$  
%
$S^{03}, S^{12}, S^{56}, S^{78}, S^{9 \;10}, S^{11\;12}, S^{13\; 14}\,$, 
$\tilde{S}^{03}, \tilde{S}^{12}, \tilde{S}^{56}, \tilde{S}^{78}, \tilde{S}^{9 \;10}, 
\tilde{S}^{11\;12}, \tilde{S}^{13\; 14}\,$,
%
a left handed ($\Gamma^{(13,1)} =-1$) eigenstate of all the members of the 
Cartan  subalgebra, representing a weak chargeless  $u_{R}$-quark with spin up, hyper charge ($2/3$) 
and  colour ($1/2\,,1/(2\sqrt{3})$), for example, can be written as 
\begin{eqnarray}
&& \stackrel{03}{(+i)}\stackrel{12}{(+)}|\stackrel{56}{(+)}\stackrel{78}{(+)}
||\stackrel{9 \;10}{(+)}\stackrel{11\;12}{(-)}\stackrel{13\;14}{(-)} |\psi_{0} \rangle = \nonumber\\
&&\frac{1}{2^7} 
(\gamma^0 -\gamma^3)(\gamma^1 +i \gamma^2)| (\gamma^5 + i\gamma^6)(\gamma^7 +i \gamma^8)||
\nonumber\\
&& (\gamma^9 +i\gamma^{10})(\gamma^{11} -i \gamma^{12})(\gamma^{13}-i\gamma^{14})
|\psi_{0} \rangle \,.
\label{start}
\end{eqnarray}
This state is an eigenstate of all $S^{ab}$ and $\tilde{S}^{ab}$ which are members of the Cartan 
subalgebra (Eq.~(\ref{choicecartan})). 

The operators $ \tilde{S}^{ab}$, which do not belong to the Cartan subalgebra (Eq.~(\ref{choicecartan})),  
generate families from the starting $u_R$ quark, transforming the $u_R$ quark from Eq.~(\ref{start}) 
to the $u_R$ of another family,  keeping all of the properties with respect to $S^{ab}$ unchanged.
In particular, $\tilde{S}^{01}$ applied on a right handed $u_R$-quark 
from Eq.~(\ref{start}) generates a 
state which is again  a right handed $u_{R}$-quark,  weak chargeless,  with spin up,
hyper charge ($2/3$)
and the colour charge ($1/2\,,1/(2\sqrt{3})$)
\begin{eqnarray}
\tilde{S}^{01}\;
\stackrel{03}{(+i)}\stackrel{12}{(+)}| \stackrel{56}{(+)} \stackrel{78}{(+)}||
\stackrel{9 10}{(+)} \stackrel{11 12}{(-)} \stackrel{13 14}{(-)}= -\frac{i}{2}\,
&&\stackrel{03}{[\,+i]} \stackrel{12}{[\,+\,]}| \stackrel{56}{(+)} \stackrel{78}{(+)}||
\stackrel{9 10}{(+)} \stackrel{11 12}{(-)} \stackrel{13 14}{(-)}\,.\nonumber\\
\label{tildesabfam}
\end{eqnarray}

Below some useful relations~\cite{pikanorma} are presented 

%
\begin{eqnarray}
\label{so1+3}
\vec{N}_{\pm}(= \vec{N}_{(L,R)}): &=& \,\frac{1}{2} (S^{23}\pm i S^{01},S^{31}\pm i S^{02}, 
S^{12}\pm i S^{03} )\,,
\end{eqnarray}
 \begin{eqnarray}
 \label{so42}
 \vec{\tau}^{1}:&=&\frac{1}{2} (S^{58}-  S^{67}, \,S^{57} + S^{68}, \,S^{56}-  S^{78} )\,,\nonumber\\
 \vec{\tau}^{2}:&=& \frac{1}{2} (S^{58}+  S^{67}, \,S^{57} - S^{68}, \,S^{56}+  S^{78} )\,,
 \end{eqnarray}
 \begin{eqnarray}
 \label{so64}
 \vec{\tau}^{3}: = &&\frac{1}{2} \,\{  S^{9\;12} - S^{10\;11} \,,
  S^{9\;11} + S^{10\;12} ,\, S^{9\;10} - S^{11\;12} ,\nonumber\\
 && S^{9\;14} -  S^{10\;13} ,\,  S^{9\;13} + S^{10\;14} \,,
  S^{11\;14} -  S^{12\;13}\,,\nonumber\\
 && S^{11\;13} +  S^{12\;14} ,\, 
 \frac{1}{\sqrt{3}} ( S^{9\;10} + S^{11\;12} - 
 2 S^{13\;14})\}\,,\nonumber\\
 \tau^{4}: = &&-\frac{1}{3}(S^{9\;10} + S^{11\;12} + S^{13\;14})\,,
 \end{eqnarray}
\begin{eqnarray}
\label{so1+3tilde}
\vec{\tilde{N}}_{L,R}:&=& \,\frac{1}{2} (\tilde{S}^{23}\pm i \tilde{S}^{01},
\tilde{S}^{31}\pm i \tilde{S}^{02}, \tilde{S}^{12}\pm i \tilde{S}^{03} )\,,
\end{eqnarray}
 \begin{eqnarray}
 \label{so42tilde}
 \vec{\tilde{\tau}}^{1}:&=&\frac{1}{2} (\tilde{S}^{58}-  \tilde{S}^{67}, \,\tilde{S}^{57} + 
 \tilde{S}^{68}, \,\tilde{S}^{56}-  \tilde{S}^{78} )\,,\;\;\nonumber\\
 \vec{\tilde{\tau}}^{2}:&=&\frac{1}{2} (\tilde{S}^{58}+  \tilde{S}^{67}, \,\tilde{S}^{57} - 
 \tilde{S}^{68}, \,\tilde{S}^{56}+  \tilde{S}^{78} )\,, 
 \end{eqnarray}
 \begin{eqnarray}
 \label{so64tilde}
 \tilde{\tau}^{4}: = &&-\frac{1}{3}(\tilde{S}^{9\;10} + \tilde{S}^{11\;12} + \tilde{S}^{13\;14})\,.
 \end{eqnarray}
\begin{eqnarray}
\label{plusminus}
N^{\pm}_{+}         &=& N^{1}_{+} \pm i \,N^{2}_{+} = 
 - \stackrel{03}{(\mp i)} \stackrel{12}{(\pm )}\,, \quad N^{\pm}_{-}= N^{1}_{-} \pm i\,N^{2}_{-} = 
  \stackrel{03}{(\pm i)} \stackrel{12}{(\pm )}\,,\nonumber\\
\tilde{N}^{\pm}_{+} &=& - \stackrel{03}{\tilde{(\mp i)}} \stackrel{12}{\tilde{(\pm )}}\,, \quad 
\tilde{N}^{\pm}_{-}= 
  \stackrel{03} {\tilde{(\pm i)}} \stackrel{12} {\tilde{(\pm )}}\,,\nonumber\\ 
\tau^{1\pm}         &=& (\mp)\, \stackrel{56}{(\pm )} \stackrel{78}{(\mp )} \,, \quad   
\tau^{2\mp}=            (\mp)\, \stackrel{56}{(\mp )} \stackrel{78}{(\mp )} \,,\nonumber\\ 
\tilde{\tau}^{1\pm} &=& (\mp)\, \stackrel{56}{\tilde{(\pm )}} \stackrel{78}{\tilde{(\mp )}}\,,\quad   
\tilde{\tau}^{2\mp}= (\mp)\, \stackrel{56}{\tilde{(\mp )}} \stackrel{78}{\tilde{(\mp )}}\,.
\end{eqnarray}

\vspace{3mm}

\section{{\it Standard model} assumptions}
\label{sm}

\vspace{3mm}

More than $40 $ years ago the {\it standard model} offered an  {\it elegant new step} in
 understanding 
the origin of fermions and bosons by postulating:  
\begin{itemize}
\item The existence  of  {\it massless family members}: coloured quarks and colourless leptons, both 
left and right handed, the  left handed members distinguishing from the right handed ones in the weak 
and hyper charges and correspondingly mass protected, Table~\ref{Table SMfermions.}. 
\item The existence of the {\it vector gauge fields } (massless before the electroweak break) to
the observed  charges of the family  members, Table~\ref{Table SMvectors.}.  
\item The existence of a massive {\it scalar field} carrying the weak charge ($\pm \frac{1}{2}$)
and the hyper charge ($\mp \frac{1}{2}$), which by its "nonzero vacuum expectation values" 
breaks the weak and the hyper charge and correspondingly breaks the mass protection of 
fermions and those vector bosons which interact with this vacuum, Table~\ref{Table SMscalar.}.
\item The existence of the {\it Yukawa couplings} of fermions,  which together
with (the  gluons and)  the  scalar take care of the properties of fermions after the electroweak
break. 
\end{itemize}
\begin{table}
{\tiny%
\begin{center}
\caption{
\label{Table SMfermions.}
Table represents the {\it standard model} assumptions for each of  the three so far observed 
($i=1,2,3$)  families of quarks and leptons, massless before the electroweak break.
Each family contains the left handed weak charged quarks and right handed weak chargeless 
quarks, each quark belonging to the colour triplet $(1/2,1/(2\sqrt{3}))$, $(-1/2,1/(2\sqrt{3}))$, 
$(0,-1/(\sqrt{3})) $, and the left handed weak charged  and right handed weak chargeless
colourless leptons.  $\tau^{13}$ defines the third component of the weak charge, $Y$ is  the hyper 
charge determining the electromagnetic charge $Q= Y + \tau^{13}$. The  {\it standard model} 
assumes to each family member of each family the corresponding anti-fermions. }
\begin{tabular}{|r c r r r r |}
\hline
$\alpha$&{\bf { hand-}} & {\bf { weak}}  &{\bf { hyper}} & 
{\bf { colour}} & elm \\
        &{\bf { edness}}& {\bf { charge}}& {\bf { charge}} &  
         {\bf { charge}} & charge\\

name    &${\bf { -4i S^{03} S^{12}}}$&${\bf { \tau^{13}}}$  &
${\bf { Y}}$       & &$Q$\\
\hline
&&&&&\\
${\bf {  u^{i}_{L}}}$&$ {\bf { -1}}$&$ {\bf { \frac{1}{2}}}$&
${\bf { \frac{1}{6}}}$& {\bf { colour triplet}}&$\frac{2}{3}$\\
&&&&&\\
${\bf {  d^{i}_{L}}}$&$ {\bf { -1}}$&${\bf { -\frac{1}{2}}}$&
${\bf { \frac{1}{6}}}$& {\bf { colour triplet}}&$-\frac{1}{3}$\\
&&&&&\\
\hline
${\bf {\nu^{i}_{L}}}$&$ {\bf { -1}}$&$ {\bf { \frac{1}{2}}}$&  
${\bf { -\frac{1}{2}}}$&{\bf {  colourless}}&$0$  \\
&&&&&\\
${\bf {  e^{i}_{L}}}$&$ {\bf { -1}}$ &${\bf { -\frac{1}{2}}}$&  
${\bf { -\frac{1}{2}}}$& {\bf { colourless}}&$-1$ \\
&&&&&\\
\hline
&&&&&\\
${\bf {   u^{i}_{{\bf { R}}}}}$&$ {\bf { 1}}$&{\bf { weakless }} & 
${\bf {  \frac{2}{3}}}$&{\bf { colour triplet}}&$\frac{ 2}{3}$\\
&&&&&\\
${\bf {   d^{i}_{{\bf { R}}}}}$&$ {\bf { 1}}$& {\bf { weakless }}&  
${\bf { -\frac{1}{3}}}$&{\bf { colour triplet}}&$-\frac{1}{3}$\\
&&&&&\\
\hline
${\bf { \nu^{i}_{{\bf { R}}}}}$&$ {\bf { 1}}$& {\bf { weakless}}& 
${\bf { 0 }}          $& {\bf { colourless}}&$0$           \\
&&&&&\\
${\bf {   e^{i}_{{\bf { R}}}}}$&$ {\bf { 1}}$& {\bf { weakless}}&  
${\bf { -1 }} $& {\bf { colourless}}&$-1$ \\
\hline\hline
\end{tabular}
  \end{center}  
}
\end{table}
\begin{table}
{\small%
\begin{center}
\caption{\label{Table SMvectors.} Vector fields, the gauge fields of the hyper, weak 
and colour charges, all massless before the electroweak break. They all are vectors in the adjoint 
representations with respect to the weak, colour and hyper charges.}
\begin{tabular}{|r|c|r|r|r|c|}
\hline
name &{\bf { hand- }}        & {\bf {  weak }}&
{\bf { hyper}}  &{\bf {  colour }}& elm \\
     &{\bf { edness }}& {\bf {  charge}}&{\bf { charge}}
     & {\bf {   charge}}& charge\\  
\hline
{\bf {  hyper photon}}&$ 0$              &$ 0$   &  $0$      & colourless&$0$\\
&&&&&\\
{\bf {  weak bosons }}&$ 0$              &{\bf { triplet}}&  $0$ & colourless&
{\bf { triplet }}\\
&&&&&\\
{\bf {  gluons }}     & $0$              &$0$    &  $0$   & {\bf { colour octet}}&$0$ \\
\hline
\end{tabular}
  \end{center}
}
\end{table}
\begin{table}
{\small
\begin{center}
\caption{\label{Table SMscalar.} 
Higgs is the scalar field  with the weak charge and the hyper charge $\pm\frac{1}{2}$ and 
$\mp\frac{1}{2}$, respectively. The $0.\,$ Higgs$_{d} $ and $0.\,$ Higgs$_{u}$ are not assumed. 
In Table these two are added to manifest the fundamental representation of the charge groups. 
The two components, $< $Higgs$_{u}>$ and$< $Higgs$_{d}>$, "dress" in the {\it standard model}
the right handed $u$-quarks and $d$-quarks, respectively, giving them the charges of the left 
partners.} 
\begin{tabular}{|r|c|r|r|r|c|}
\hline
name &{\bf { hand- }}        & {\bf {  weak }}&
{\bf { hyper}}  &{\bf {  colour }}& {\bf { elm}} \\
     &{\bf { edness }}& {\bf {  charge}}&{\bf { charge}}
     & {\bf {   charge}}& {\bf { charge}}\\  
\hline
&&&&&\\
 $ 0 \cdot$ Higgs$_{u}$ & $ 0$ & $\frac{1}{2}$ & $ \frac{1}{2}$ & colourless& $ 1$\\                                        
                                        &&&&&\\
 $< $Higgs$_{d}>$&$ 0$&$- \frac{1}{2}$& $\frac{1}{2}$& colourless& $ 0$\\                                        
\hline
\hline
&&&&&\\
 $<$Higgs$_{u}>$&$ 0$&$\frac{1}{2}$& $-\frac{1}{2}$&  colourless &$  0$\\
                                        &&&&&\\
$ 0 \cdot$ Higgs$_{d}$&$ 0$&$ - \frac{1}{2}$&$ -\frac{1}{2}$& colourless&$ -1$\\                                        
\hline
\end{tabular}
  \end{center}
}
\end{table}

%


The {\it standard model} assumptions have been confirmed without offering surprises.  The last 
unobserved field, the Higgs, detected in June  2012, was confirmed in March 2013.

\ack
The author thanks the organizers of IARD - Martin Land and Matej Pav\v si\v c - to give
her opportunity to present her work, which might have a real chance to show the next step
beyond the standard models.


\end{document}